\documentclass[trackchanges,twocolumn]{aastex701}
\usepackage{placeins}

\usepackage{booktabs}
\usepackage{multirow}
\usepackage{amsmath}
\usepackage{tabularx}
\usepackage{float}
\usepackage{hyperref}
\begin{document}

\title{\galsyn\ I: A Forward-Modeling Code for Synthetic Galaxy Observations \\ from Hydrodynamical Simulations and First Data Release from IllustrisTNG}

\author[0000-0002-5258-8761]{Abdurro'uf}\email{fnuabdur@iu.edu}
\affiliation{Department of Astronomy, Indiana University, 727 East Third Street, Bloomington, IN 47405, USA}

\author[0000-0001-7113-2738]{Henry C. Ferguson}\email{ferguson@stsci.edu}
\affiliation{Space Telescope Science Institute (STScI), 3700 San Martin Drive, Baltimore, MD 21218, USA}

\author[0000-0003-2342-7501]{Samir Salim}\email{salims@iu.edu}
\affiliation{Department of Astronomy, Indiana University, 727 East Third Street, Bloomington, IN 47405, USA}

\author[0000-0001-9298-3523]{Kartheik G. Iyer}\email{kartheikiyer@gmail.com}
\affiliation{Flatiron Institute, 162 5th Avenue. New York, NY 10010, USA}

\author[0000-0002-7908-9284]{Larry D. Bradley}\email{
lbradley@stsci.edu}
\affiliation{Space Telescope Science Institute (STScI), 3700 San Martin Drive, Baltimore, MD 21218, USA}

\author[0000-0001-7410-7669]{Dan Coe}\email{dcoe@stsci.edu}
\affiliation{Space Telescope Science Institute (STScI), 
3700 San Martin Drive, Baltimore, MD 21218, USA}
\affiliation{Association of Universities for Research in Astronomy (AURA), Inc.
for the European Space Agency (ESA)}
\affiliation{Department of Physics and Astronomy, The Johns Hopkins University, 3400 N Charles St. Baltimore, MD 21218, USA}

\author[0009-0009-3404-5673]{Novan Saputra Haryana}
\email{novan.haryana@astr.tohoku.ac.jp}
\affiliation{Astronomical Institute, Tohoku University, 6-3, Aramaki, Aoba-ku, Sendai, Miyagi, 980-8578, Japan}

\author[0000-0002-1050-7572]{Sultan Hassan}\email{shassan@stsci.edu}
\affiliation{Space Telescope Science Institute (STScI), 3700 San Martin Drive, Baltimore, MD 21218, USA}

\author[0000-0003-1187-4240]{Intae Jung}
\email{}
\affiliation{Department of Astronomy and Space Science, Chungbuk National University, Cheongju, 28644, Republic of Korea}

\author[0000-0002-3475-7648]{Gourav Khullar}\email{gkhullar@uw.edu}
\affiliation{Department of Astronomy, University of Washington, Physics-Astronomy Building, Box 351580, Seattle, WA 98195-1700, USA}
\affiliation{eScience Institute, University of Washington, Physics-Astronomy Building, Box 351580, Seattle, WA 98195-1700, USA}

\author[0000-0002-8512-1404]{Takahiro Morishita}\email{takahiro.morishita.d8@tohoku.ac.jp}
\affiliation{Astronomical Institute, Tohoku University, 6-3, Aramaki, Aoba-ku, Sendai, Miyagi, 980-8578, Japan}

\author[0000-0002-8530-9765]{Lamiya Mowla}\email{lmowla@wellesley.edu}
\affiliation{Department of Physics and Astronomy, Wellesley College, Wellesley, MA 02481, USA}


\newcommand{\LCDM}{$\Lambda$CDM}

\newcommand{\red}[1]{{\color{red} #1}}
\newcommand{\redss}[1]{{\color{red} ** #1}}
\newcommand{\redbf}[1]{{\color{red}\bf #1 \color{black}}}

\newcommand{\ny}{$\tilde {\rm n}$}
\newcommand{\about}{$\sim$}
\newcommand{\appr}{$\approx$}
\newcommand{\gt}{$>$}
\newcommand{\uJy}{$\mu$Jy}
\newcommand{\sig}{$\sigma$}
\newcommand{\Lya}{Lyman-$\alpha$}
\renewcommand{\th}{$^{\rm th}$}
\newcommand{\lam}{$\lambda$}

\newcommand{\tentothe}[1]{$10^{#1}$}
\newcommand{\tentotheminus}[1]{$10^{-#1}$}
\newcommand{\e}[1]{$\times 10^{#1}$}
\newcommand{\en}[1]{$\times 10^{-#1}$}
\newcommand{\cgsfluxunits}{erg$\,$s$^{-1}\,$cm$^{-2}$}
\newcommand{\cgsfluxdensityunits}{erg$\,$s$^{-1}\,$cm$^{-2}$\,\AA$^{-1}$}
\newcommand{\linefluxunits}{\tentotheminus{20} \cgsfluxunits}

\newcommand{\logU}{$\log(U)$}
\newcommand{\logOH}{12+log(O/H)}

\newcommand{\sinv}{s$^{-1}$}

\newcommand{\footnoteurl}[1]{\footnote{\url{#1}}}

\newcommand{\tnm}[1]{\tablenotemark{#1}}
\newcommand{\super}[1]{$^{\rm #1}$}
\newcommand{\supa}{$^{\rm a}$}
\newcommand{\supb}{$^{\rm b}$}
\newcommand{\supc}{$^{\rm c}$}
\newcommand{\supd}{$^{\rm d}$}
\newcommand{\supe}{$^{\rm e}$}
\newcommand{\supf}{$^{\rm f}$}
\newcommand{\supg}{$^{\rm g}$}
\newcommand{\suph}{$^{\rm h}$}
\newcommand{\supi}{$^{\rm i}$}
\newcommand{\supj}{$^{\rm j}$}
\newcommand{\supk}{$^{\rm k}$}
\newcommand{\supl}{$^{\rm l}$}
\newcommand{\supm}{$^{\rm m}$}
\newcommand{\supn}{$^{\rm n}$}
\newcommand{\supo}{$^{\rm o}$}

\newcommand{\sqarcmin}{arcmin\squared}

\newcommand{\supcomma}{$^{\rm ,}$}

\newcommand{\rhalf}{$r_{1/2}$}

\newcommand{\chisq}{$\chi^2$}

\newcommand{\Zgas}{$Z_{\rm gas}$}  
\newcommand{\Zstar}{$Z_*$}  

\newcommand{\inv}{\per}
\newcommand{\Mstar}{$M^*$}
\newcommand{\Lstar}{$L^*$}
\newcommand{\phistar}{$\phi^*$}

\newcommand{\sigmaMsunpc}{$M_{\odot}\,\rm{pc}^{-2}$}
\newcommand{\sigmaMsunkpc}{$M_{\odot}\,\rm{kpc}^{-2}$}
\newcommand{\mutan}{$\mu_{\rm tan}$}

\newcommand{\logM}{log($M_*$/\Msun)}

\newcommand{\LUV}{$L_{UV}$}
\newcommand{\MUV}{$M_{UV}$}

\newcommand{\Msun}{$M_\odot$}
\newcommand{\Lsun}{$L_\odot$}
\newcommand{\Zsun}{$Z_\odot$}

\newcommand{\Mvir}{$M_{vir}$}
\newcommand{\Mt}{$M_{200}$}
\newcommand{\Mf}{$M_{500}$}

\newcommand{\Ndotion}{$\dot{N}_{\rm ion}$}
\newcommand{\xiion}{$\xi_{\rm ion}$}
\newcommand{\logxiion}{log(\xiion)}
\newcommand{\fesc}{$f_{\rm esc}$}

\newcommand{\XHI}{$X_{\rm HI}$}
\newcommand{\XHII}{$X_{\rm HII}$}
\newcommand{\RHII}{$R_{\rm HII}$}

\newcommand{\Halpha}{H$\alpha$}
\newcommand{\Hbeta}{H$\beta$}
\newcommand{\Hgamma}{H$\gamma$}
\newcommand{\Hdelta}{H$\delta$}
\newcommand{\Halphaw}{\Halpha\,$\lambda$6563}
\newcommand{\Hbetaw}{\Hbeta\,$\lambda$4861}
\newcommand{\Hgammaw}{H$\gamma$\,$\lambda$4342}
\newcommand{\Hdeltaw}{H$\delta$\,$\lambda$4103}
\newcommand{\Ha}{\Halpha}
\newcommand{\Hb}{\Hbeta}

\newcommand{\I}{\,{\sc i}}
\newcommand{\II}{\,{\sc ii}}
\newcommand{\III}{\,{\sc iii}}
\newcommand{\IV}{\,{\sc iv}}
\newcommand{\VI}{\,{\sc vi}}
\newcommand{\VII}{\,{\sc vii}}
\newcommand{\VIII}{\,{\sc viii}}

\newcommand{\HI}{H\,{\sc i}}
\newcommand{\HII}{H\,{\sc ii}}
\newcommand{\HeI}{He\,{\sc i}}
\newcommand{\HeII}{He\,{\sc ii}}

\newcommand{\CII}{[C\,{\sc ii}]}
\newcommand{\CIIw}{\CII\,$\lambda$2325 (blend)}
\newcommand{\CIII}{C\,{\sc iii}]}
\newcommand{\CIIIwa}{\CIII\,$\lambda$1907}
\newcommand{\CIIIwb}{\CIII\,$\lambda$1909}
\newcommand{\CIIId}{C\,{\sc iii}]}
\newcommand{\CIIIdw}{C\,{\sc iii}]\,$\lambda\lambda$1907,1909}
\newcommand{\CIV}{C\,{\sc iv}}
\newcommand{\CIVw}{\CIV\,$\lambda$1549}
\newcommand{\OII}{[O\,{\sc ii}]}
\newcommand{\OIIwa}{\OII\,$\lambda$3726}
\newcommand{\OIIwb}{\OII\,$\lambda$3729}
\newcommand{\OIIdw}{\OII\,$\lambda\lambda$3726,3729}
\newcommand{\OIII}{[O\,{\sc iii}]}
\newcommand{\OIIIw}{\OIII\,$\lambda$5007}
\newcommand{\OIIIww}{\OIII\,$\lambda\lambda$4959,5007}
\newcommand{\OIIIwa}{\OIII\,$\lambda$4363}
\newcommand{\OIIIwc}{\OIII\,$\lambda$4959}
\newcommand{\NeIII}{[Ne\,{\sc iii}]}
\newcommand{\NeIIIw}{\NeIII\,$\lambda$3870}
\newcommand{\NeIIIwb}{\NeIII\,$\lambda$3969}
\newcommand{\HeIw}{HeI\,$\lambda$3890}
\newcommand{\HeIwa}{HeI\,$\lambda$4473}
\newcommand{\HeIIw}{HeII\,$\lambda$1640}
\newcommand{\NII}{[N\,{\sc ii}]}
\newcommand{\NIII}{N\,{\sc iii}]}
\newcommand{\NIV}{N\,{\sc iv}]}
\newcommand{\NIIIw}{\NIII\,$\lambda$1748}
\newcommand{\NIVw}{\NIV\,$\lambda$1486}
\newcommand{\MgII}{Mg\,{\sc ii}}
\newcommand{\MgIIw}{\MgII\,$\lambda$2800}

\newcommand{\SII}{[S\,{\sc ii}]}
\newcommand{\SIIwa}{\SII\,$\lambda$6716}
\newcommand{\SIIwb}{\SII\,$\lambda$6731}
\newcommand{\SIIdw}{\SII\,$\lambda\lambda$6716,6731}

\newcommand{\Lyaw}{Ly$\alpha$\,$\lambda$1216}

\newcommand{\OIIIwfivem}{\OIII\,$\lambda$52$\mu$m}
\newcommand{\OIIIweightm}{\OIII\,$\lambda$88$\mu$m}



\newcommand{\Om}{\Omega_{\rm M}}
\newcommand{\OL}{\Omega_\Lambda}

\newcommand{\etal}{et al.}

\newcommand{\citeps}{\citep}

\newcommand{\HST}{{\em HST}}
\newcommand{\SST}{{\em SST}}
\newcommand{\Hubble}{{\em Hubble}}
\newcommand{\Spitzer}{{\em Spitzer}}
\newcommand{\Chandra}{{\em Chandra}}
\newcommand{\JWST}{{\em JWST}}
\newcommand{\Planck}{{\em Planck}}

\newcommand{\Bradac}{{Brada\v{c}}}

\newcommand{\citepeg}[1]{\citep[e.g.,][]{#1}}

\newcommand{\range}[2]{\! \left[ _{#1} ^{#2} \right] \!}  

\newcommand{\galsyn}{\texttt{GalSyn}}
\newcommand{\fsps}{\texttt{FSPS}}
\newcommand{\grizli}{\textsc{grizli}}
\newcommand{\eazypy}{\textsc{eazypy}}
\newcommand{\msaexp}{\textsc{msaexp}}
\newcommand{\trilogy}{\textsc{trilogy}}
\newcommand{\bagpipes}{\texttt{bagpipes}}
\newcommand{\beagle}{\textsc{beagle}}
\newcommand{\photutils}{\textsc{photutils}}
\newcommand{\SEP}{\textsc{sep}}
\newcommand{\piXedfit}{\textsc{piXedfit}}
\newcommand{\pyneb}{\textsc{pyneb}}
\newcommand{\HIIC}{\textsc{hii-chi-mistry}}
\newcommand{\astropy}{\textsc{astropy}}
\newcommand{\astrodrizzle}{\textsc{astrodrizzle}}
\newcommand{\multinest}{\textsc{multinest}}
\newcommand{\cloudy}{\textsc{Cloudy}}
\newcommand{\jdaviz}{\textsc{Jdaviz}}
\newcommand{\emcee}{\textsc{emcee}}
\newcommand{\galfit}{\textsc{GALFIT}}
\newcommand{\prospector}{\textsc{prospector}}
\newcommand{\db}{\textsc{Dense Basis}}

\renewcommand{\tt}[1]{\texttt{#1}}

\newcommand{\SE}{\tt{SourceExtractor}}

\newcommand{\PD}[1]{\textcolor{blue}{[PD: #1\;]}}

\newcommand{\JD}{MACS0647$-$JD}

\newcommand{\edense}{$n_{e}$}
\newcommand{\ROII}{$R_{[\rm{OII}]}$}
\newcommand{\OIIratio}{\OII\,$\lambda$3729/$\lambda$3726}

\newcommand{\RSII}{$R_{[\rm{SII}]}$}
\newcommand{\SIIratio}{\SII\,$\lambda$6716/$\lambda$6731}
\newcommand{\CIIIratio}{C\,{\sc iii}]\,$\lambda$1909/$\lambda$1907}

\newcommand{\lya}{\hbox{Ly$\alpha$}}        
\newcommand{\nv}{\hbox{\sc N\,v}}           
\newcommand{\niv}{\hbox{\sc N\,iv]}}      
\newcommand{\ariv}{\hbox{\sc Ar\,iv]}}      
\newcommand{\civ}{\hbox{\sc C\,iv}}         
\newcommand{\heii}{\hbox{He\,{\sc ii}}}     
\newcommand{\cii}{\hbox{\sc C\,ii]}}      
\newcommand{\hei}{\hbox{He\,{\sc i}}}     
\newcommand{\oiiisemi}{\hbox{\sc O\,iii]}}  
\newcommand{\ciii}{\hbox{\sc C\,iii]}}      
\newcommand{\siiii}{\hbox{Si\,{\sc iii]}}}  
\newcommand{\mgii}{\hbox{Mg\,{\sc ii}}}     
\newcommand{\oi}{\hbox{O\,{\sc i}}}     
\newcommand{\oii}{\hbox{\sc [O\,ii]}}     
\newcommand{\hb}{\hbox{\sc H$\beta$}}       
\newcommand{\oiii}{\hbox{\sc [O\,iii]}}     
\newcommand{\ha}{\hbox{\sc H$\alpha$}}      
\newcommand{\hd}{\hbox{\sc H$\delta$}}      
\newcommand{\hg}{\hbox{\sc H$\gamma$}}      
\newcommand{\he}{\hbox{\sc H$\epsilon$}}      
\newcommand{\het}{\hbox{\sc H$\eta$}}      
\newcommand{\sii}{\hbox{[S\,{\sc ii}]}}     
\newcommand{\siii}{\hbox{[S\,{\sc iii}]}}   
\newcommand{\pab}{\hbox{Pa$\beta$}}      
\newcommand{\pag}{\hbox{Pa$\gamma$}}
\newcommand{\neiii}{\hbox{[Ne\,{\sc iii]}}}  
\newcommand{\siiiIR}{\hbox{[S\,{\sc iii]}}}  
\newcommand{\feii}{\hbox{[Fe\,{\sc ii]}}}  
\newcommand{\nev}{\hbox{[Ne\,{\sc v]}}}  
\newcommand{\oiv}{\hbox{O\,{\sc iv]}}}  
\newcommand{\siv}{\hbox{[S\,{\sc iv]}}}  

\newcommand{\kms}{km\,s$^{-1}$}

\begin{abstract}

We present \galsyn\ (Galaxy Synthesizer), a modular and flexible Python package for generating synthetic observations from hydrodynamical galaxy simulations. \galsyn\ generates synthetic spectrophotometric data cubes for individual galaxies from simulation cutouts, employing a particle-by-particle spectral modeling approach that enables the rapid production of large synthetic datasets required for statistical population studies, offering a computationally efficient alternative to full radiative transfer codes. Users have full control over the spectral modeling choices, including the stellar population synthesis engine, stellar isochrones, spectral libraries, and initial mass functions. Dust attenuation is modeled at spatially resolved scales using a line-of-sight column-density method, with a comprehensive suite of fixed and adaptive attenuation laws. A decoupled kinematics model independently Doppler-shifts the stellar and nebular components, enabling realistic synthetic integral field unit data cubes. It also provides features to add observational realism, including PSF convolution and noise simulation. Beyond generating imaging and spectroscopic data cubes, \texttt{GalSyn} reconstructs spatially resolved physical property maps and star formation histories (SFHs) of a galaxy. Alongside this paper, we present the first data release of synthetic imaging observations and resolved SFHs generated from the IllustrisTNG simulations. This release includes four mock extragalactic survey fields (spanning $5$, $8$, $137$, and $365$ arcmin$^{2}$) together with data cubes of the individual galaxies within them, data cubes of 290 local massive galaxies and their progenitors tracked across $0<z<5$, and 259 major-merger systems. Each galaxy data cube contains imaging across 47 filters spanning HST, JWST, Euclid, Rubin/LSST, and the Roman Space Telescope. \galsyn\ is publicly available at \url{https://github.com/aabdurrouf/GalSyn}.

\end{abstract}



\section{Introduction} \label{sec:introduction} 

The current era of astronomical research has seen an unprecedented influx of observational data. The advent of state-of-the-art telescopes and surveys, including the \textit{James Webb Space Telescope (JWST)} \citep{Gardner2023,Rigby2023}, the \textit{Euclid} mission \citep{EuclidSkyOverview}, the \textit{Vera C. Rubin Observatory} \citep{Ivezic2019}, and the upcoming \textit{Nancy Grace Roman Space Telescope} \citep{2015Spergel_roman}, has opened a new paradigm of ``big data'' in astronomy. These powerful instruments are capable of observing galaxies with unprecedented depth, spatial resolution, and field of view, generating datasets that are orders of magnitude larger and more complex than those previously available. However, this wealth of observational data presents both opportunities and challenges. While it enables detailed studies of galaxy properties and their evolution, it also demands sophisticated theoretical frameworks and computational tools capable of bridging the gap between numerical simulations and observational results.

Forward modeling and synthetic observations have emerged as an important component in this endeavor, serving as essential intermediaries between theoretical models and real observations. By creating mock observations from simulations, we can directly compare theoretical predictions with observational data, validate galaxy formation models, and interpret complex observational signatures. This approach is particularly crucial in the field of galaxy evolution, where the interplay between various physical processes (e.g.,~star formation, feedback processes, and chemical enrichment) produces complex observational signatures that could be better understood through detailed forward modeling.

A crucial component in the forward modeling is computing the emergent spectrum from a collection of stars, gas, and dust in the interstellar medium (ISM) of a galaxy, which is very complex. Radiative transfer (RT) techniques, exemplified by codes such as \texttt{SKIRT} \citep{Camps2015,Camps2020}, \texttt{SUNRISE} \citep{Jonsson2006}, and \texttt{POWDERDAY} \citep{Robitaille2011,Narayanan2021}, can be applied to track the propagation of photons through the ISM, accounting for absorption, scattering, and re-emission. However, these codes do not themselves model the emergent spectrum of a given stellar population. Furthermore, the emergent spectrum of the ``star particles'' in a simulation ought to, in principle, include both a composite population of stars and the effects of the local sub-grid ISM on this emergent spectrum. 

The primary advantage of RT codes lies in their physical accuracy and self-consistency. They naturally capture phenomena such as dust heating, cooling, and scattered light contributions. However, this accuracy comes at a substantial computational cost. Radiative transfer calculations are computationally intensive, often requiring significant computational resources and time, particularly for high-resolution simulations or large galaxy samples. This computational burden can limit the exploration of parameter space and the generation of large synthetic datasets required for statistical studies.

An alternative approach has emerged in the form of particle-by-particle spectral modeling, as implemented in recent codes such as \texttt{Synthesizer} \citep{Lovell2025,Roper2025,Harvey2026}, \texttt{PyMGAL} \citep{Janulewicz2025}, and \texttt{FORECAST} \citep{Fortuni2023}. Briefly speaking, these methods assign spectral energy distributions (SEDs) directly to star particles based on their stellar masses, ages, and metallicities, subsequently applying dust attenuation prescriptions and summing the contributions to produce synthetic galaxy imaging and spectra data cubes.

While this approach may sacrifice some of the geometric complexity captured by full radiative transfer, it offers significant advantages in terms of computational efficiency and methodological flexibility. Particle-based spectral modeling enables rapid generation of synthetic observations for large galaxy samples, making it particularly suitable for population studies and statistical analyses of galaxy properties. With this flexibility and efficiency, this tool could be used to thoroughly study the impact of specific choices in forward modeling on the resulting observational properties.

The particle-based spectral modeling approach, as implemented in our newly developed \galsyn\ (Galaxy Synthesizer) package, provides a comprehensive control over each physical process involved in the synthesis procedure. Users can select from multiple stellar population synthesis models, customize isochrone grids and spectral libraries, experiment with different initial mass functions, and implement flexible dust attenuation laws. By treating these ingredients as customizable building blocks rather than fixed inputs, \galsyn\ enables data-driven Bayesian model selection across diverse observational regimes, from imaging to IFU spectroscopy. This framework allows us to evaluate competing physical prescriptions directly against observational data, quantitatively testing and relaxing traditionally held assumptions in galaxy evolution modeling.

Such flexibility is crucial for advancing our understanding of galaxy evolution in several ways. First, it enables systematic exploration of degeneracies among different physical processes and their observational signatures. This can be done, such as by studying the effects of varying physical models on the observational properties. Second, it enables optimization studies aimed at identifying the combination of physical models that best reproduce observed galaxy properties. Third, it facilitates the development and testing of new theoretical prescriptions within a consistent framework. This capability is particularly valuable for identifying optimal configurations of physical models that can generate synthetic observations, whether broadband images or integral field unit (IFU) spectroscopic data, that closely match real observational datasets. 

In this paper, which is the first in a series, we describe the forward-modeling algorithms implemented in \galsyn\ and demonstrate its robustness and core capabilities. We also present the initial release (DR1) of synthetic data cubes generated from the IllustrisTNG simulation suites \citep{Nelson2018, Nelson2019, Marinacci2018, Naiman2018, Springel2018, Pillepich2018, Pillepich2019, Nelson2019_b}. This release includes four mock survey fields covering areas of 5 (TNG50-1), 8 (TNG50-1), 137 (TNG100-1), and 365 (TNG300-1) arcmin$^2$, imaging data cubes of individual galaxies within those fields, and data cubes for two specific TNG50 samples: progenitors of local massive galaxies ($\log(M_{*}/M_{\odot}) > 10.5$) and 200 major-merger systems. Each galaxy data cube contains imaging data across multiple filters (HST, JWST, Roman, LSST, and Euclid) and a wide range of physical property maps. The software is available on GitHub\footnote{\galsyn\ codebase: \url{https://github.com/aabdurrouf/GalSyn}.} and archived in Zenodo (\dataset[doi: 10.5281/zenodo.21498033]{https://doi.org/10.5281/zenodo.21498033}). The DR1 dataset is available on GitHub\footnote{DR1 link: \url{https://github.com/aabdurrouf/GalSyn_dr1}.} and archived in Zenodo (\dataset[doi: 10.5281/zenodo.21498294]{https://doi.org/10.5281/zenodo.21498294}). A forthcoming paper will provide a comprehensive exploration of how various dust attenuation laws impact observable properties.
  
This paper is structured as follows. In Section~\ref{sec:galsyn_design}, we describe the \galsyn\ design and physical ingredients within it. Subsequently, in Section~\ref{sec:datacube_observational_effects}, we describe the data products and further processing of the data cubes by adding observational effects. The first set of data release is presented in Section~\ref{sec:data_release}, while the summary and outlook are given in Section~\ref{sec:summary}.   

\begin{figure*}[ht]
\centering
\includegraphics[width=0.8\textwidth]{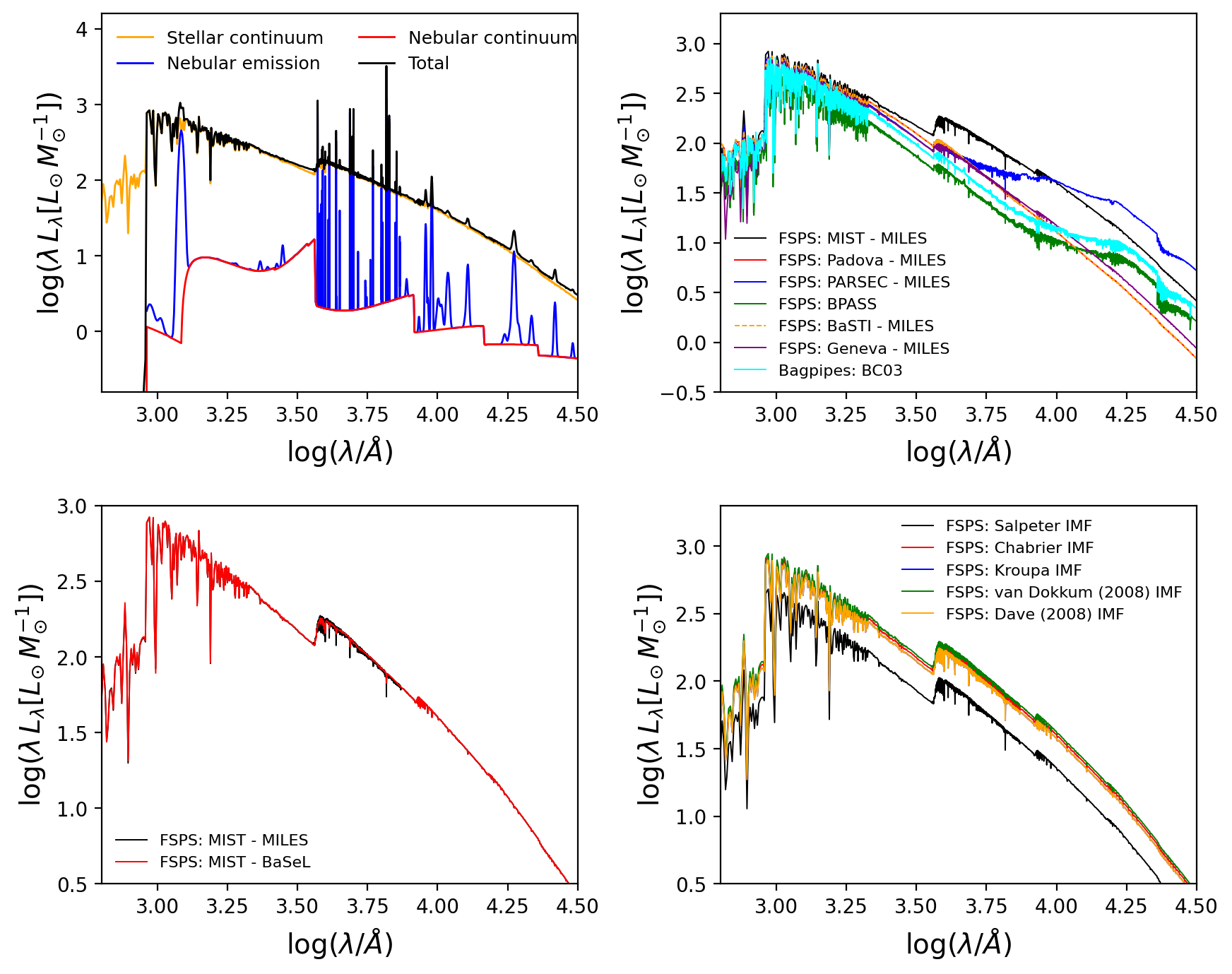}
\caption{Comparison of SSP spectra available in \galsyn. \textit{Top left}: an example of SSP spectrum broken down into its components: stellar continuum, nebular continuum, and nebular emission. \textit{Top right}: comparison of SSP spectra for various isochrones using the MILES spectral library and \citet{2003Chabrier} IMF, calculated using the FSPS code. A BC03 spectrum generated with the Bagpipes code is also shown for reference. \textit{Bottom left}: comparison of SSP spectra for different spectral libraries, but the same isochrone (MIST) and IMF (\citealt{2003Chabrier}). \textit{Bottom right}: SSP spectra with various IMFs, but the same isochrone (MIST) and spectral library (MILES). All of the spectra are for stellar population with an age of $5$ Myr and a solar metallicity.}
\label{fig:ssp_spectra}
\end{figure*}

\section{\galsyn\ Design} \label{sec:galsyn_design}

\galsyn\ is designed with a modular structure, allowing flexible customization of modeling ingredients. The overall analysis workflow of \galsyn\ is described below.

\subsection{Preparation Stage} \label{sec:preparation}

\subsubsection{Input Simulation Data} \label{sec:input_sim_data}

The first step in the \galsyn\ workflow is the preparation of the input simulation data. \galsyn\ is designed to be agnostic to the specific hydrodynamical simulation data used, enabling its application to a wide range of simulations such as IllustrisTNG \citep{Nelson2019}, EAGLE \citep{Schaye2015}, and others. This flexibility is achieved by first converting the raw simulation output into a standardized HDF5 file format, which serves as the input for all subsequent synthesis tasks. This step ensures that the core \galsyn\ routines operate on a consistent data structure, regardless of the data's origin. 

The resulting standardized file must contain the datasets listed in Table~\ref{tab:sim_params}, including the coordinates, velocities, and physical properties (e.g.,~mass and metallicity) of the star and gas particles. Some of the gas parameters are derivable from the other parameters. The gas temperature can be derived from the internal energy and electron abundance assuming the effective equation of state model \citep{Springel2003}\footnote{Gas temperature in IllustrisTNG data can be calculated following procedure described in \protect\url{https://www.tng-project.org/data/docs/faq/\#gen6}}. The hydrogen mass can be derived from the total gas mass assuming a constant hydrogen fraction ($X_{\rm H}$).

The instantaneous SFR of a gas cell exceeding the star formation threshold ($\simeq 0.1\,\rm{cm}^{-3}$ in IllustrisTNG) is calculated by dividing its estimated cold-phase mass by a density-dependent star formation timescale. The star-forming gas is modeled as a two-phase medium where overall energy is a mass-weighted average of cold dense clouds and a hot ambient gas. Within this model, the cell's cold-phase mass is obtained by multiplying the total cell mass by a cold fraction determined by local thermal equilibrium based on the local gas density \citep{Springel2003}.

\begin{table*}[ht] 
\centering 
\caption{List of Physical Parameters in the Standardized Input Simulation Data} \label{tab:sim_params} 
\begin{tabular}{@{}lll@{}} 
\toprule Parameter & Description & Units \\ 
\midrule \multicolumn{3}{c}{\textbf{Properties for both Star and Gas Particles}} \\ 
\midrule \texttt{coords} & 3D particle positions & kpc \\ 
\texttt{vel} & 3D particle velocities & km/s \\ 
\texttt{mass} & Particle mass (surviving mass for stars) & M$_{\odot}$ \\ 
\texttt{zmet} & Particle metallicity (mass fraction) & Dimensionless \\ 
\midrule \multicolumn{3}{c}{\textbf{Star-Specific Properties}} \\ 
\midrule \texttt{init\_mass} & Stellar mass at the time of formation & M$_{\odot}$ \\ \texttt{form\_z} & Redshift at which the star particle formed & \\ 
\midrule \multicolumn{3}{c}{\textbf{Gas-Specific Properties}} \\ 
\midrule \texttt{sfr\_inst} & Instantaneous star formation rate & M$_{\odot}$/yr \\ 
\texttt{temp} & Gas temperature & Kelvin \\ 
\texttt{mass\_H} & Hydrogen mass of the gas particle & M$_{\odot}$ \\ 
\bottomrule 
\end{tabular} 
\end{table*}

The \texttt{simutils\_tng} module provides a streamlined set of tools for handling data from the IllustrisTNG simulation. The procedure involves two main steps: data acquisition with functions for downloading a cutout for a specific subhalo or parent halo at a given snapshot, and conversion to a standardized data format that includes the necessary unit conversion and calculations to transform the data into the physical units required by \galsyn. New modules for handling data from other simulations will be added to \galsyn\ in future updates.

\subsubsection{SSP and Nebular Emission Grids} \label{sec:ssp_grids}

The foundational element of the light synthesis in \galsyn\ is the assignment of a Simple Stellar Population (SSP) spectrum, including nebular emissions, to each star particle. \galsyn\ offers a highly flexible framework for this process, allowing users to either generate SSPs on-the-fly or, for significantly improved computational efficiency, use pre-computed grids of SSP models. These pre-computed grids, which store spectra across a range of ages, metallicities, and ionization parameters ($U$), can be generated using dedicated modules (\texttt{ssp\_generator\_fsps} and \texttt{ssp\_generator\_bagpipes}). The ionization parameter, which describes the intensity of the ionizing radiation field relative to the density of the gas, is a parameter controlling the nebular emission component. 

\galsyn\ provides two options for SPS codes: the Flexible Stellar Population Synthesis \citep[\fsps;][]{Conroy2009,Conroy2010} and \bagpipes\ \citep{Carnall2018}. When using \fsps, users can customize a wide array of model components, including the choice of stellar isochrones (MIST, \citealt{Choi2016}; Padova, \citealt{Girardi2000}; PARSEC, \citealt{Bressan2012}; BaSTI, \citealt{Pietrinferni2004}; and Geneva, \citealt{Schaller1992}), stellar spectral libraries (MILES, \citet{Sanchez-Blazquez2006}, and BaSeL, \citealt{Lejeune1997}), and a comprehensive selection of Initial Mass Functions (IMFs), including \citet{Salpeter1955}, \citet{2003Chabrier}, \citet{Kroupa2001}, \citet{vanDokkum2008}, \citet{Dave2008}, and tabulated piece-wise power law IMF (specified in \texttt{imf.dat} file in \fsps\ data directory). Furthermore, FSPS implementation includes the option of the BPASS model \citet{Eldridge2017}, which incorporates the effects of binary stellar evolution. Additionally, \galsyn\ also provides the 2016 version of the \citet{Bruzual2003} isochrones called through \bagpipes. This implements \citet{Kroupa2002} IMF. 

By providing access to these different SPS engines and their various internal model choices, \galsyn\ serves as a powerful laboratory. This flexibility allows us to systematically investigate how different assumptions about stellar evolution and atmospheres impact the emergent light from galaxies. The resulting SSP spectra are normalized to $1\,M_{\odot}$, with the surviving stellar mass recorded in the output HDF5 file. During the synthesis process, the ratio between this surviving mass and the mass of the star particles is used to correctly scale the final spectra.

\galsyn's nebular emission modeling is physically coupled to the young stellar populations (with age below $10$ Myr) whose ionizing photon production is intrinsically calculated by the chosen SPS code (either \fsps\ or \bagpipes). Both codes use the photoionization code \texttt{CLOUDY} \citep{Ferland1998, Ferland2013} to self-consistently model the resulting nebular continuum and emission lines. There are generally two parameters in the nebular emission modeling: $U$ and gas-phase metallicity. While \fsps\ treats gas-phase metallicity as an independent parameter, \bagpipes\ inherently ties it to have the same value as the stellar metallicity. To maintain computational efficiency in \fsps\ without introducing an additional free parameter (which will add a dimension in SSP grids), we allow users to tie the gas-phase metallicity to the stellar metallicity using a fixed input ratio. The calculation of $U$ from simulation data is described in Section~\ref{sec:ionization_parameter}. Figure~\ref{fig:ssp_spectra} shows a comparison of SSP spectra generated with various IMF, stellar isochrones, and spectral libraries.

\begin{table*}[ht] 
\centering 
\caption{Stellar Population Synthesis (SPS) Modeling Options in GalSyn} \label{tab:ssp_options} 
\begin{tabular}{@{}lll@{}} \toprule \textbf{Model Component} & \textbf{FSPS Options} & \textbf{Bagpipes Options} \\ 
\midrule \multirow{2}{*}{\textbf{SPS Engine}} & Flexible Python wrapper for the Fortran FSPS code. & Built upon the Bruzual \& Charlot (2003) models. \\ 
& (Conroy et al. 2009; Conroy \& Gunn 2010) & (Carnall et al. 2018) \\ 
\midrule \textbf{Isochrones} & MIST, BaSTI, Padova, PARSEC, Geneva & \multirow{2}{*}{Bruzual \& Charlot (2003) models} \\ 
\textbf{Spectral Library} & MILES, BaSeL \\ 
\midrule \textbf{Initial Mass Function} & Chabrier (2003), Kroupa (2001), Salpeter (1955), & Kroupa (2001) \\ 
\textbf{(IMF)} & Custom power-law, van Dokkum (2008), Dave (2008) & \\ 
\midrule \textbf{Binary Evolution} & BPASS models can be integrated. & Not included by default. \\ 
\bottomrule 
\end{tabular} 
\end{table*}

\begin{figure*}[ht]
\centering
\includegraphics[width=0.7\linewidth]{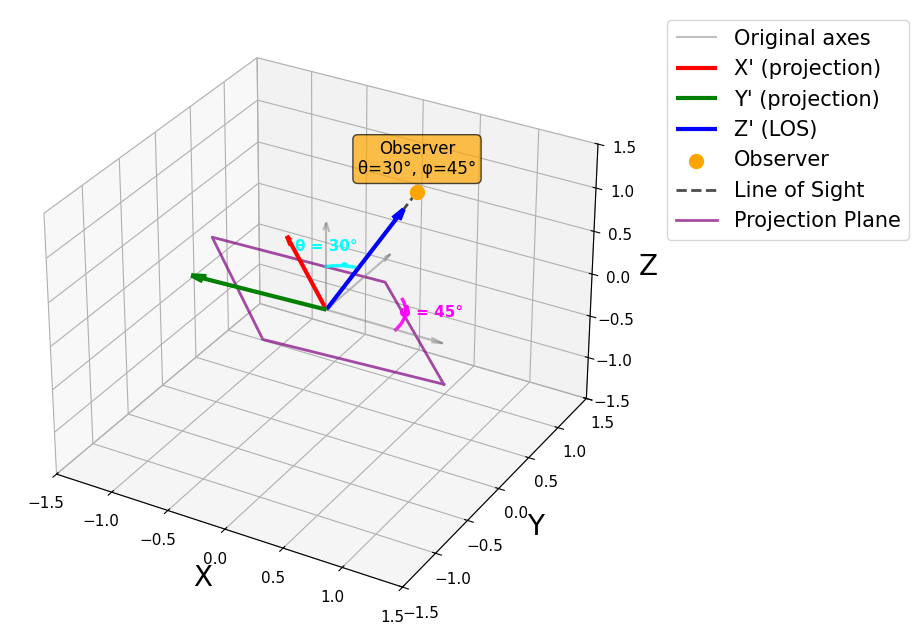}
\caption{Visualization of two-dimensional projection from three-dimensional coordinates given a viewing angle specified by the polar ($\theta$) and azimuth angles ($\phi$). The new coordinate system (x', y', z') obtained from the transformation is shown by the red, green, and blue arrows. In this new frame, the observer's LOS is aligned with the z'-axis, and the x'-y' plane represents the plane of the sky.}
\label{fig:visualize_projection}
\end{figure*}

\subsection{Synthesis Process} \label{sec:synthesis}

The \galsyn\ synthesis pipeline begins by performing a spatial projection, transforming the 3D coordinates of star and gas particles from a simulation into a 2D image plane based on a specified viewing angle. The core of the synthesis is then executed in parallel (employing multiple CPUs) for each pixel. The initial pixel size (i.e.,~sampling) is defined by the input smoothing length, which is assumed constant. For every star particle within a pixel's line-of-sight, the code assigns an SSP spectrum, applies a kinematic Doppler shift based on particle velocities, and models dust attenuation from both the diffuse interstellar medium and the local birth cloud. The light from all particles in the pixel is then aggregated, and the final spectrum is redshifted, attenuated by the intergalactic medium (IGM), and integrated through user-defined filter transmission functions to produce the observed broadband flux. Finally, the synthetic cube is regridded into the user-desired pixel scale. 

\subsubsection{Spatial Projection} \label{sec:spatial_projection}

The first step in the synthesis process is to transform the three-dimensional distribution of star and gas particles from the simulation into a two-dimensional image plane that represents the observer's view. This process constructs the viewing geometry, projects the particles, and calculates their positions along the line of sight. 

The viewing orientation is defined by two key parameters: the polar angle ($\theta$) and the azimuth angle ($\phi$). These angles specify the observer's position in a spherical coordinate system centered on the galaxy. The polar angle controls the inclination, representing a tilt from the simulation's z-axis, while the azimuth angle controls the rotation of the galaxy in the xy-plane. Based on these angles, the code constructs a rotation matrix that transforms the original (x, y, z) coordinates of every particle into a new coordinate system (x', y', z'). In this new frame, the z'-axis is aligned with the observer's line of sight (LOS), and the x'-y' plane represents the plane of the sky. Figure~\ref{fig:visualize_projection} visualizes the projection of three-dimensional coordinates into a two-dimensional image plane.

After this transformation, the (x', y') coordinates of each particle are used to bin them into a 2D pixel grid, forming the basis of the final image. The initial pixel grid is set by the input smoothing length (\texttt{smoothing\_length}), which can be thought of as the spatial resolution of the simulation. The z' coordinate of each particle becomes its distance along the line of sight. A crucial calculation is performed at this stage: the code determines the minimum z' value among all particles (both stars and gas) and normalizes all z' distances relative to this minimum. This results in a relative line-of-sight distance for every particle, where a value of 0 corresponds to the particle closest to the observer. This relative distance is essential for the subsequent dust attenuation modeling (to be described in Section~\ref{sec:dust_attenuation}), as it allows the code to robustly identify which gas particles are physically located \textit{in front of} any given star particle, and thus contribute to the attenuation of its light.

The central coordinate of the synthetic cutout is determined through a two-pass procedure. In the first pass, all star particles are projected onto a preliminary 2D pixel grid spanning the full extent of the simulation data, using the same rotation matrix described above. A stellar mass density map is then constructed by accumulating the mass of each star particle into its corresponding pixel. The pixel containing the highest total stellar mass is identified as the galaxy center, and its physical coordinate in the projected plane is taken as the central reference point. In the second pass, the final cutout grid is defined such that this most-massive pixel lies at the geometric center of the output image, with the cutout dimensions set by the user-specified output size. The physical size of the cutout can either be specified directly by the user in kpc, or determined automatically as the smallest square that encloses a user-defined fraction of the total projected stellar mass (default: 99\%).

\subsubsection{Calculation of Ionization Parameter} \label{sec:ionization_parameter}

To account for the local physical conditions of the ISM in simulated galaxy into the nebular emission modeling, we calculate the ionization parameter ($U$) for each pixel. For an ionization-bounded H~{\sc ii} region, $U$ can be calculated as 
\begin{equation} \label{eq:u}
    U = \frac{Q}{4 \pi R^{2} n_{\text{H}} c},
\end{equation}
where $Q$, $n_{\text{H}}$, $R$, and $c$ are the hydrogen-ionizing photon production rate, hydrogen number density, characteristic size of the H~{\sc ii} region, and the speed of light, respectively \citep[e.g.,][]{Reddy2023a,Reddy2023b}. The radius can be approximated as the Str{\"o}mgren radius:
\begin{equation} \label{eq:stromgen_radius}
    R_{S} = \left( \frac{3Q}{4\pi \alpha_{B} n_{\rm{H}}^{2} \epsilon} \right)^{1/3},
\end{equation}
where $\alpha_{B}$ is the case-B recombination coefficient and $\epsilon$ is the volume filling factor. Here, we assume fully ionized gas ($n_{\rm H} \approx n_{e}$, electron density) for simplicity. Substituting radius in Equation~\ref{eq:u} with $R_{S}$, we obtain 
\begin{equation}
    U = k \left( n_{\text{H}} Q \epsilon^{2} \right)^{1/3}
\end{equation}
with $k=\left(\frac{3\alpha_{B}^{2}}{16\pi c^{3}} \right)^{1/3}$. Assuming $\alpha_{B}=2.59\times 10^{-13}\,\text{cm}^{3}\text{s}^{-1}$ for temperature of $T=10^{4}\,\text{K}$, we obtain $k=5.3\times 10^{-20}\,\text{cm}\,\text{s}^{1/3}$.

The production rate of hydrogen-ionizing photons can be calculated via
\begin{equation}
    Q = \xi_{\rm ion}\,{L_{\rm UV}}, 
\end{equation}
where $\xi_{\rm ion}$ is the ionizing photon production efficiency and $L_{\rm UV}$ is the ultraviolet (UV) luminosity. $\xi_{\rm ion}$ is a free parameter with a default value of $10^{25.39}\,\text{erg}^{-1} \text{Hz}$, while $L_{\rm UV}$ is calculated from pixel's instantaneous SFR assuming \citet{Kennicutt1998}.  

We note that this formulation represents an effective sub-grid approximation for the mean ionization state of the gas within a given pixel. Because simulation star particles represent SSPs rather than individual stars, a pixel generally contains an ensemble of sub-pixel H II regions rather than a single monolithic Str{\"o}mgren sphere. However, adopting an effective Str{\"o}mgren geometry preserves the essential physical coupling between the local ionizing photon budget (driven by young stars) and the surrounding gas density. As shown Figure~\ref{fig:plot_sfr_vs_logu}, this effective treatment naturally reproduces the observed empirical scaling between $U$ and SFR density on spatially resolved scales without requiring additional free parameters.

The calculated $U$ is clipped to a standard physical range of $10^{-4}$ to $10^{-1}$. The derived $U$ of a pixel is then used in assigning spectra of stellar particles in the pixel (to be described in Section~\ref{sec:ssp_assignment}). This approach allows the ionization state of the gas to vary spatially across the simulated galaxy, self-consistently reflecting the relationship between local star formation feedback and the surrounding gas density.

\subsubsection{Spectra Assignment} \label{sec:ssp_assignment}

The core of \galsyn's light generation is assigning an SSP spectrum to each star particle based on its age, metallicity, mass, and the ionization parameter of the ISM where it resides. This is handled in one of two ways to obtain two separate spectral components: the stellar continuum and the nebular emission. 

\begin{itemize}
\item Method 1: Using a pre-computed SSP grid. This is the fastest, default method. The code reads from an HDF5 file containing a pre-calculated grid of spectra. For each star particle, it uses the particle’s age, metallicity, and ionization parameter (of the pixel where the particle resides) to locate the corresponding entry in the SSP grid. The spectrum is then obtained by interpolating between grid points using one of three supported techniques: \texttt{linear}, \texttt{nearest}, or \texttt{cubic}. Interpolation is applied consistently to both the stellar continuum and nebular emission components of the star particle. Finally, the interpolated spectra are rescaled by the particle's mass. The robustness of the interpolation methods are tested and discussed in Appendix~\ref{sec:ssp_interpolation_robustness}.

\item Method 2: On-the-fly SSP generation. In this mode, an SSP instance (either \fsps\ or \bagpipes) is initialized for each parallel worker and configured with the star particle's specific age, metallicity, and ionization parameter. To isolate the nebular emission, the SSP model is run twice: once with nebular emission enabled to get a total spectrum, and a second time with it disabled to get the stellar continuum only. The final nebular spectrum is then obtained by subtracting the stellar-only from the total spectrum, resulting in two spectral components for each star particle.
\end{itemize}

\begin{figure*}[ht]
\centering
\includegraphics[width=0.325\textwidth]{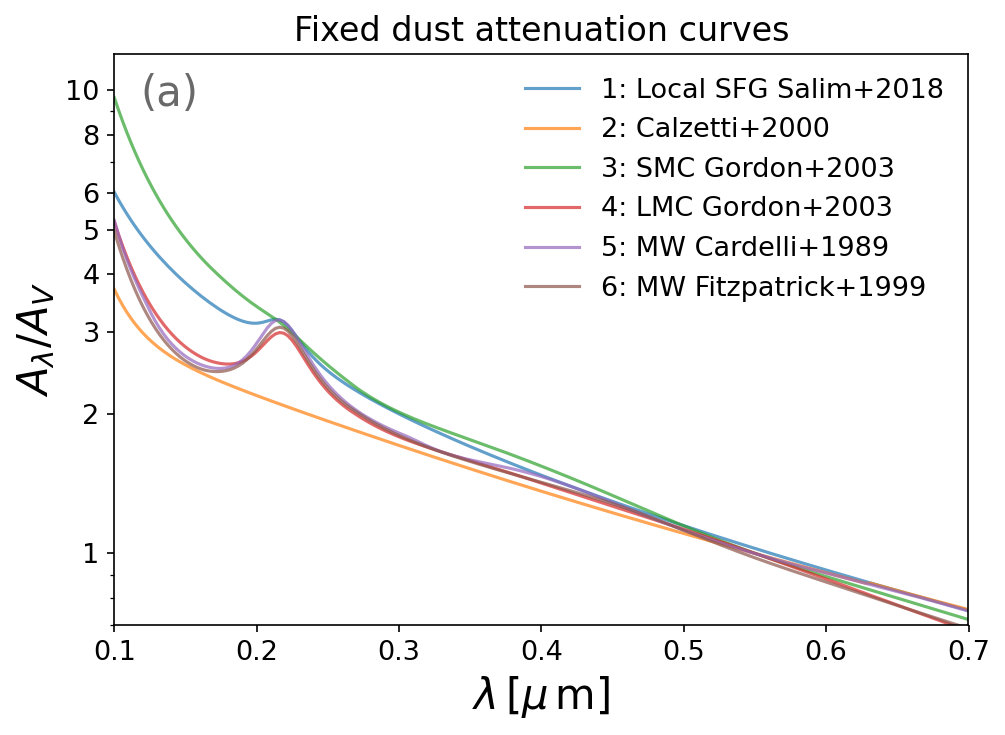}
\includegraphics[width=0.325\textwidth]{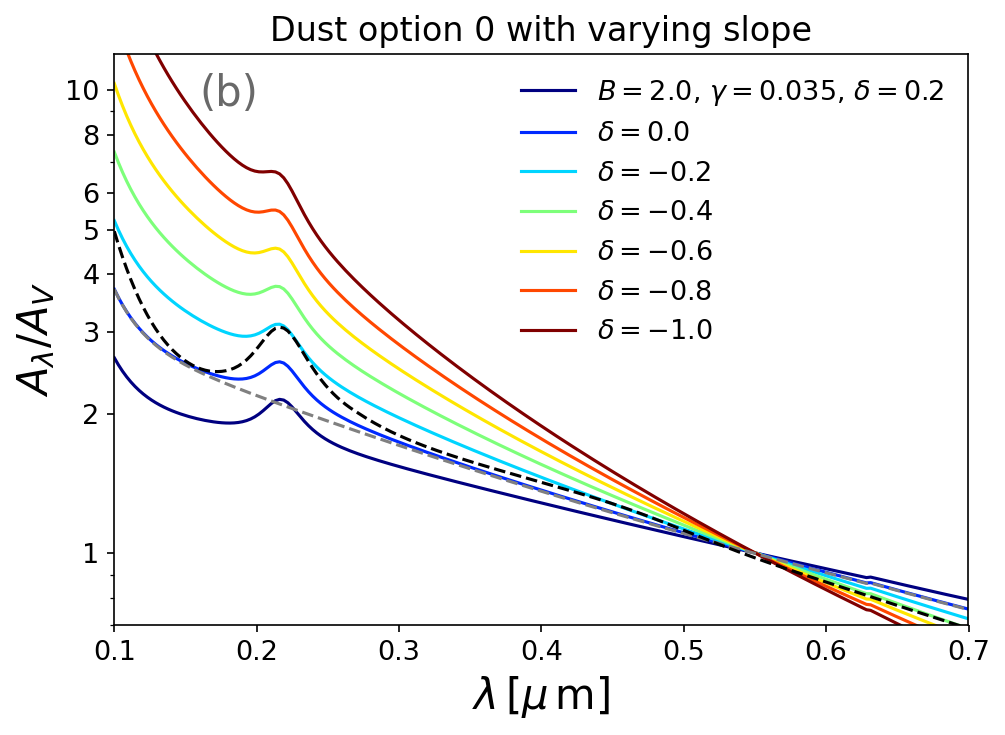}
\includegraphics[width=0.325\textwidth]{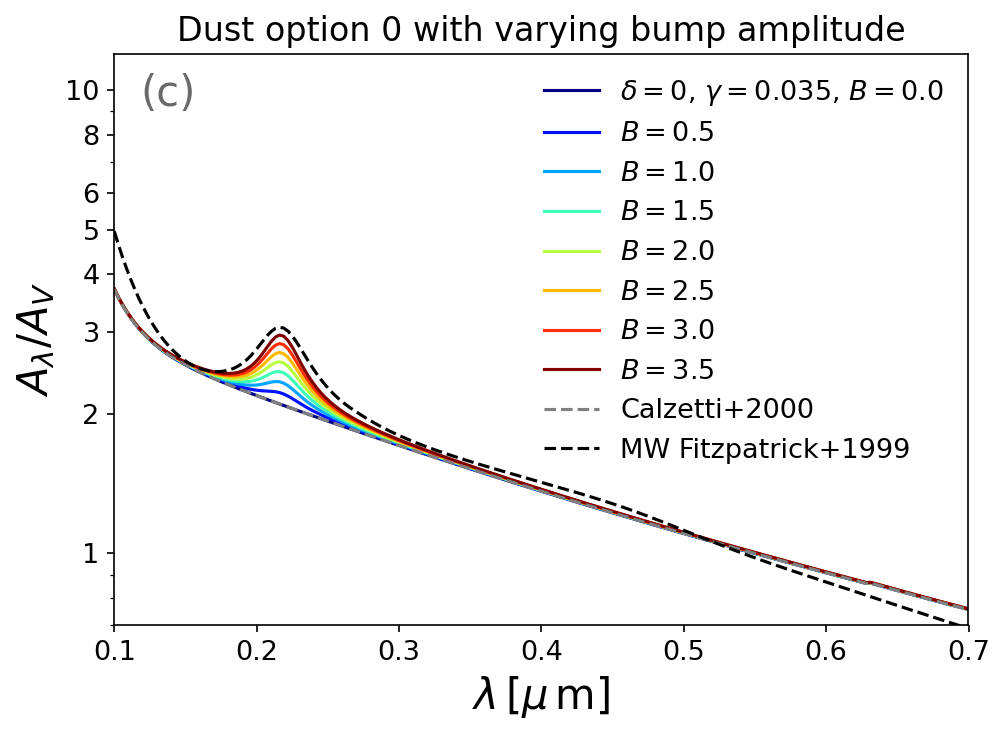}
\includegraphics[width=0.325\textwidth]{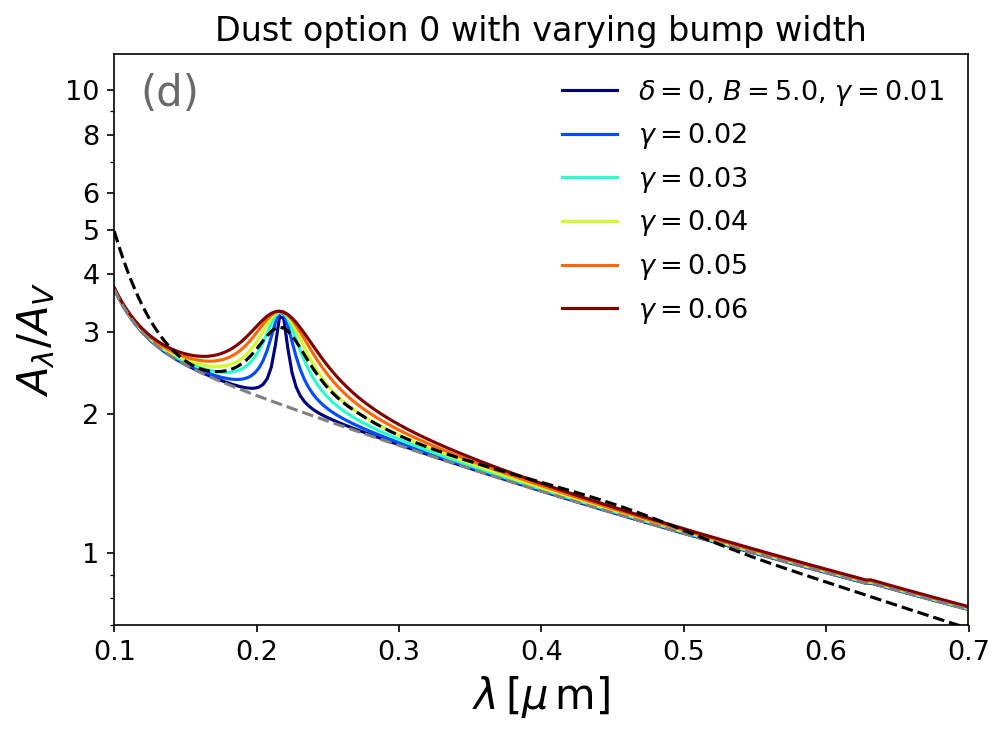}
\includegraphics[width=0.325\textwidth]{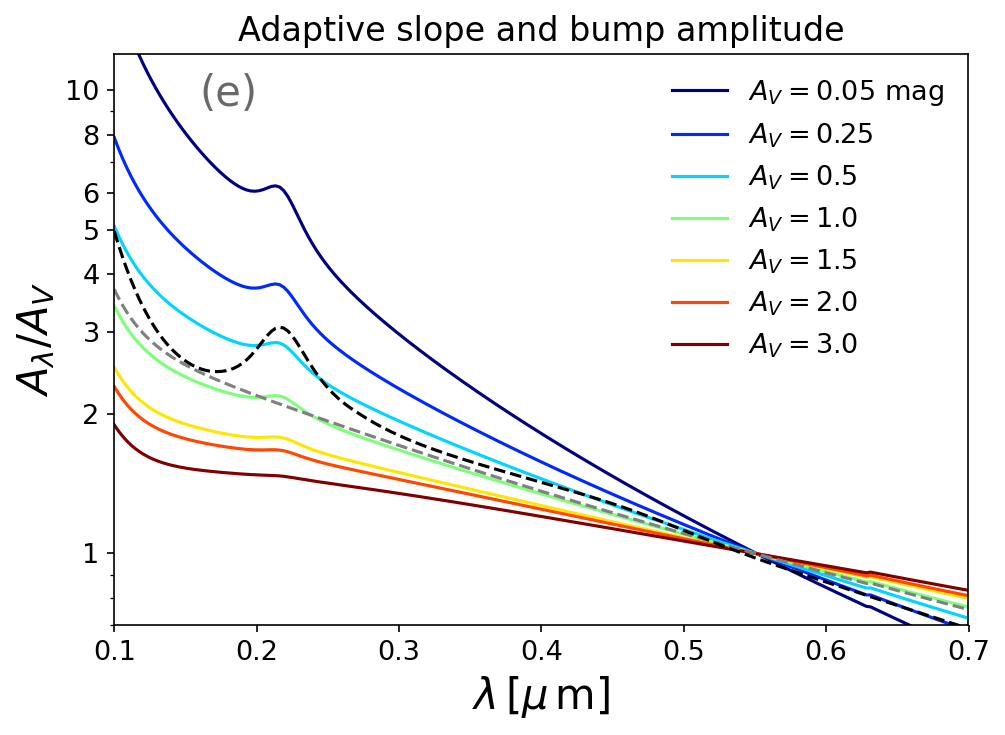}
\includegraphics[width=0.325\textwidth]{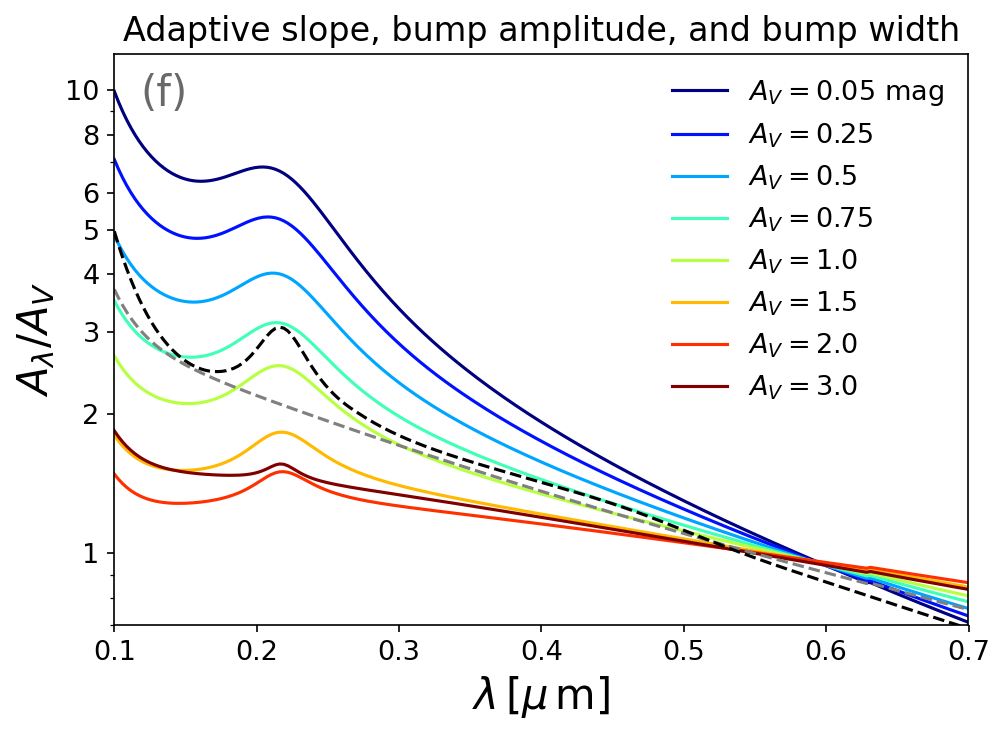}
\caption{Options for the dust attenuation curves of the diffuse ISM in \galsyn. (a) The fixed dust attenuation law options: \citet{Salim2018}, \citet{Calzetti2000}, SMC by \citet{Gordon2003}, LMC by \citet{Gordon2003}, MW by \citet{Cardelli1989}, and MW by \citet{Fitzpatrick1999}. These dust curves have a fixed shape, independent of the local $A_{V}$. (b) The modified Calzetti law (option 0), as defined in Equations~\ref{eq:modif_calzetti}-\ref{eq:drude}. This model is shown with a varying slope ($\delta$) and fixed UV bump parameters ($B=2.0$ and $\gamma=0.035$). (c) The modified Calzetti law shown with a varying bump amplitude but fixed slope ($\delta=0$) and bump width ($\gamma=0.035\,\mu$m). (d) The modified Calzetti law shown with a varying bump width but fixed slope ($\delta=0$) and bump amplitude ($B=5$). (e) The modified Calzetti law is applied in an adaptive mode with slope and bump amplitude changes depending on $A_{V}$. The slope varies with $A_{V}$ following the empirical relation from \citet{Salim2018}, and the bump amplitude then varies depending on slope following the relation from \citet{Kriek2013}. (f) Dust attenuation curve obtained by fitting modified Calzetti law to the integrated $A_{\lambda}$ curves from synthetic IFS cubes of simulated galaxies from the NIHAO-SKIRT catalog \citep{Faucher2023}. Details about this will be presented in Abdurro'uf et al. (in prep.).}
\label{fig:dust_curves}
\end{figure*}

\subsubsection{Kinematic Effects} \label{sec:kinematic_effects}

\galsyn\ includes a treatment of galaxy kinematics, which is designed to realistically model the distinct motions of stellar populations and the interstellar medium (ISM). Rather than assuming that ionized gas moves with the stars that excite it, the code implements a decoupled kinematics model. This ensures that the stellar and nebular components of a star particle's spectrum are Doppler-shifted independently, reflecting their different physical origins.

The stellar continuum, which represents the integrated light from the stars themselves, is shifted based on the line-of-sight velocity of its parent star particle. This directly ties the stellar absorption features in the final spectrum to the underlying stellar dynamics of the simulated galaxy.

In contrast, the nebular emission spectrum is shifted according to the light-weighted average line-of-sight velocity of the local gas particles. These are identified by applying the following criteria. For each star particle, the code searches for nearby gas particles that are located physically within a close 3D physical distance defined by \texttt{max\_dist\_neb} (with a default value of $500$ pc) of the star particle and having positive instantaneous SFR.

This procedure ensures that the emission lines in the final spectrum trace the kinematics of the local ISM, while the stellar absorption features trace the stellar kinematics. This decoupling is essential for creating realistic synthetic Integral Field Unit (IFU) data, as it allows for the independent measurement and analysis of gas and stellar rotation curves, a critical technique in observational studies of galaxy dynamics.

\subsubsection{Dust Attenuation} \label{sec:dust_attenuation}

\galsyn\ aims to realistically model both the production and processing of light in galaxies; therefore, incorporating dust effects is essential. In the current version of \galsyn, however, we only include dust attenuation and omit dust emission, as the main goal of the code is to generate spectrophotometric data cubes covering the rest-frame ultraviolet to near-infrared ($\sim 0.05-2\,\mu$m), where dust emission has a negligible impact. The inclusion of dust emission modeling to obtain full infrared spectra is deferred to future work. 

Most current hydrodynamical simulations (e.g., IllustrisTNG, EAGLE) do not explicitly track dust as a separate component. Instead, dust properties in galaxies can be inferred from the properties of gas particles. The core of dust attenuation modeling involves estimating the total extinction experienced by starlight and determining its wavelength dependence. To this end, \galsyn\ provides two distinct modeling approaches for assigning the level of dust attenuation on spatially resolved scales.

The first method (hereafter called ``line-of-sight'' or LOS) is based on the column density of intervening gas particles positioned along the line of sight to each star particle. In this approach, attenuation is applied individually to the spectrum of each star particle. The second option (hereafter called ``sfr\_AV'') determines the $A_{V}$ for a given pixel based on the instantaneous SFR surface density of the gas cells within that pixel. This $A_{V}$ value is then applied uniformly to all star particles contributing to that pixel. This latter option is inspired by empirical correlations between SFR (or SFR density) and $A_{V}$ \citep[e.g.,][]{Smail2021}, providing users with a simple yet physically motivated alternative for dust modeling. As shown by \citet{Sommovigo2025}, a correlation between $A_{V}$ and SFR surface density naturally emerged from radiative transfer post-processing of TNG galaxies. The specific correlation can be input by the user in the form of a dictionary (\texttt{av\_sfrden\_relation}). In the following, we will describe the LOS method.         

In the LOS method, we model spatially resolved dust attenuation for each stellar particle by estimating the dust optical depth along the line of sight. We also include additional attenuation from birth clouds that affects young stars. The implementation of our dust attenuation model is described in detail below. Our approach is similar to the methods in \citet{Vogelsberger2020} (model B) and \citet{Nelson2018} (model C), but with added flexibility in several aspects, including the choices for dust attenuation (and extinction) models. In the following, we first describe the attenuation modeling for the diffuse ISM, and then for the birth clouds.

The core of the dust model is estimating the amount of dust that starlight must pass through, which is calculated dynamically for each star particle. For a given star particle, the code identifies all gas particles in front of it along the line of sight. Only ``cold'' gas is selected (i.e.,~gas with a nonzero instantaneous SFR and a temperature below 8000 K), as this is where dust is expected to reside.  

The hydrogen column density ($N_{\rm H}$), which is a direct proxy for the amount of intervening material, is then computed by summing the hydrogen mass of all foreground cold gas particles and dividing it by the pixel’s physical area. With this and the gas-phase metallicity ($Z_{\rm g}$), we can then estimate the $V$-band resolved optical depth for the stellar particle via 

\begin{equation}
\tau_{\rm V,ISM} = \tau_{\rm dust}(z) \left(\frac{Z_{\rm g}}{Z_{\odot}} \right)^{\gamma} \left(\frac{N_{\rm H}}{N_{\rm H,0}} \right),
\end{equation}
where $\gamma=1$, $N_{\rm H,0}=2.1\times 10^{21}\rm{cm}^{-2}$, and $\tau_{\rm dust}(z)$ (\texttt{scale\_dust\_tau}) is a redshift-dependent scale factor of the optical depth. This implies that the $V$-band optical depth is assumed to be linearly dependent on the metal mass in the galaxy disc \citep[e.g.,][]{Somerville2012,Yung2019} and the redshift-dependent scale factor scales like the average dust-to-metal ratio. The $\tau_{\rm dust}(z)$ function in \galsyn\ is easily customizable, with the default implementation following that of \citet[][ Table~3 therein]{Vogelsberger2020}. 

Next, we convert the optical depth into the more commonly used $V$-band (0.55 $\mu$m) attenuation ($A_{V}$) using the relation derived from a dust-star geometry model in which ionized gas and dust are co-spatial and uniformly mixed \citep{Calzetti1994, Vogelsberger2020}:
\begin{equation}
A_{V} = -2.5 \log \left(\frac{1-e^{-\tau_{V}}}{\tau_{V}} \right).
\end{equation}
Here, $\tau_{V}$ should be interpreted as an effective optical depth that accounts for both dust absorption and scattering, including photons scattered into the line of sight. Accordingly, we treat this quantity as attenuation rather than extinction.

The $A_{V}$ value provides the overall normalization of dust attenuation, but a functional form is still required to describe how attenuation varies with wavelength (i.e., the dust attenuation curve). In principle, the shape of this curve depends on the physical properties of the dust, such as its spatial distribution, density, grain size distribution, and chemical composition. However, in the absence of detailed knowledge of these quantities, an assumed attenuation curve needs to be adopted. 

In \galsyn, we provide a suite of empirical dust attenuation laws commonly used in galaxy SED modeling. Additionally, we include an analytical attenuation curve extracted from SKIRT synthetic data, which is expected to serve as a representative global attenuation model derived from the radiative transfer simulation. The details of this extraction process will be described in a forthcoming paper (Abdurro'uf et al., in prep.). Within \galsyn, these dust attenuation curves are broadly classified into two categories: fixed and adaptive. 

The fixed dust law has an unchanging shape, independent of the local $A_{V}$. This includes the commonly used empirical law, such as the Milky Way (MW) extinction law \citep[e.g.,][]{Cardelli1989,Fitzpatrick1999}, the Small Magellanic Cloud (SMC) extinction law \citep[e.g.,][]{Prevot1984,Gordon2003}, the Large Magellanic Cloud (LMC) extinction law \citep[e.g.,][]{Nandy1981,Clayton1985,Gordon2003}, the 
\citet{Calzetti2000} dust attenuation law, and the \citet{Salim2018} law, as well as the modified \citet{Calzetti2000} law with fixed parameters, explained below.

In the adaptive dust category, the shape of the dust attenuation curve is adaptive to the amount of dust attenuation ($A_{V}$), which allows the attenuation curve to be steeper (i.e.,~more UV extinction) in regions with less dust and vice versa, as expected from the observations \citep[e.g.,][]{Noll2009,Kriek2013,Chevallard2013,Salmon2016,Salim2018,Salim2020,Nagaraj2022}. The baseline of this attenuation curve is the \citet{Calzetti2000} dust law, which is modified with a variable power-law slope and the addition of a UV bump \citep{Noll2009}.

The modified attenuation curve follows the prescription of \citet{Kriek2013}:
\begin{equation}
\frac{A(\lambda)}{A_V} = \frac{k'(\lambda) + D(\lambda)}{4.05} \left( \frac{\lambda}{\lambda_V} \right)^{\delta}
\label{eq:modif_calzetti}
\end{equation}
where $k'(\lambda)$ represents the original \citet{Calzetti2000} law and $\lambda_{V}=0.55\,\mu\,\mathrm{m}$. Here, $\delta$ represents the power-law slope modification (\texttt{dust\_index}), and $D(\lambda)$ is a Lorentzian-like Drude profile describing the UV bump feature at $2175$\AA:

\begin{equation}
D(\lambda) = \frac{B\,(\lambda \, \gamma)^2}{(\lambda^2 - \lambda_0^2)^2 + (\lambda \, \gamma)^2}
\label{eq:drude}
\end{equation}
with $\lambda$ in microns, $B$ the bump amplitude (\texttt{bump\_amp}), $\lambda_{0}=0.2175\,\mu$m the central wavelength of the bump, and $\gamma$ the bump width (full width at half maximum, FWHM, \texttt{bump\_dwave}). While the bump width is commonly fixed at the empirical value of $0.035\,\mu$m \citep{Noll2009}, we make it a free parameter in \galsyn. As will be shown in Abdurro'uf et al. (in prep.), the bump width tends to be inversely correlated with $A_{V}$ in SKIRT RT simulations. 

The three parameters of the modified \citet{Calzetti2000} dust curve ($\delta$, $B$, and $\gamma$) are fully customizable within \galsyn. Users may choose to hold these parameters constant at specific values or define them as functions of $A_{V}$ by providing the desired correlations in the form of a dictionary. This flexibility allows us to mimic empirical dust attenuation trends, such as the observed correlation between the power-law slope and $A_{V}$ \citep[e.g.,][]{Salim2018, Nagaraj2022, Battisti2020} or the relationship between the UV bump amplitude and the power-law slope, such as $B=0.85-1.9\,\delta$ reported by \citet{Kriek2013}.

Below, we list the dust attenuation law options (\texttt{dust\_law}) in \galsyn:
\begin{itemize}
\item Option 0: The modified Calzetti law as detailed above. The three parameters (\texttt{dust\_index}, \texttt{bump\_amp}, and \texttt{bump\_dwave}) can be fixed at user-defined values or configured to depend on $A_{V}$ by supplying the desired correlations in the form of a dictionary. 
\item Option 1: The \citet{Salim2018} attenuation law.
\item Option 2: The original \citet{Calzetti2000} law.
\item Option 3: The SMC extinction law from \citet{Gordon2003}.
\item Option 4: The LMC extinction law from \citet{Gordon2003}.
\item Option 5: The MW extinction law from \citet{Cardelli1989}.
\item Option 6: The MW extinction law from \citet{Fitzpatrick1999}. 
\end{itemize}

The \citet{Salim2018} dust attenuation law uses a custom polynomial reddening curve $k(\lambda)$ combined with the Drude profile:
\begin{equation}
A(\lambda) = \frac{A_{V}\,k(\lambda)}{R_V}.
\label{eq:salim}
\end{equation}
The reddening curve $k(\lambda)$ is defined as:
\begin{equation}
k(\lambda) = a_0 + \frac{a_1}{\lambda} + \frac{a_2}{\lambda^2} + \frac{a_3}{\lambda^3} + D(\lambda,B) + R_V,
\label{eq:salim_k}
\end{equation}
where $a_0$, $a_1$, $a_2$, and $a_3$ are the polynomial coefficients (\texttt{salim\_a0} to \texttt{salim\_a3}), $D$ is the Drude profile with bump strength B (\texttt{salim\_B}), and $R_V$ is the total-to-selective extinction ratio (\texttt{salim\_RV}).

To illustrate the variation in the shape of the dust attenuation curves provided in \galsyn, we present them in Figure~\ref{fig:dust_curves}. In Abdurro’uf et al. (in prep.), we derive a global dust attenuation law from radiative transfer simulations using synthetic spectroscopic data cubes from the NIHAO-SKIRT catalog \citep{Faucher2023}. These data were generated using the SKIRT code. We fitted the integrated dust attenuation curves using the modified Calzetti formula described above. Our results indicate that all the primary parameters (slope, bump amplitude, and bump width) tend to correlate strongly with $A_{V}$. Specifically, as $A_{V}$ increases, the slope increases, while both the bump amplitude and width decrease. The resulting curve is shown in the bottom right panel (f) of Figure~\ref{fig:dust_curves}. This attenuation curve can be easily implemented in \galsyn\ using option 0.

Finally, the LOS method accounts for the extra attenuation for young stars with an age less than a specified value (\texttt{t\_esc}), typically $10$ Myr. The shape of this curve is defined as a simple power law of 
\begin{equation}
A(\lambda) = A_{V,\rm{BC}}\,\left(\frac{\lambda}{\lambda_{V}} \right)^{\delta_{\rm BC}}
\end{equation}
with $A_{V,\rm{BC}}$ is the $V$-band attenuation from the birth clouds and $\delta_{\rm BC}$ is a power-law slope. The $A_{V,\rm{BC}}$ can be set to relate with the $A_V$ of the diffuse ISM through $\eta$ (\texttt{dust\_eta}), defined as $\equiv A_{V,\rm{BC}}/A_{V}$. The $\delta_{\rm BC}$ is a free parameter with a default of $-0.7$.

\begin{figure*}[ht]
\centering
\includegraphics[width=0.9\textwidth]{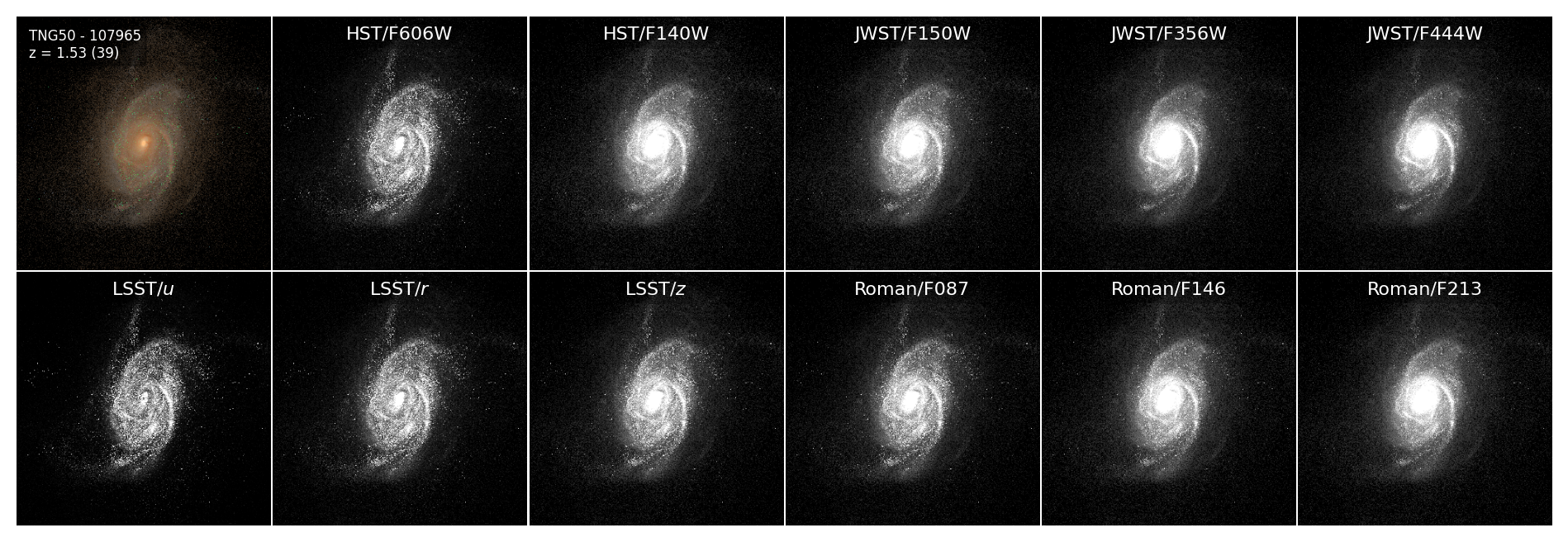}
\caption{Example of synthetic images of a simulated galaxy from TNG50 (subhalo ID=107965) at $z=1.53$ across multiple filters from various telescopes, including HST, JWST, Rubin/LSST, and Roman Space Telescope. The top-left panel shows a color composite image generated from JWST filters: F115W (blue), F150W (green), and F200W (red). These images have a pixel size of $0.03$ arcsec$^{-1}$ and have no observational effects being added.}
\label{fig:idealized_images1}
\end{figure*}

\subsubsection{Cosmological Redshifting} \label{sec:redshifting}

Up to this point, the synthetic pixel spectra are in the rest frame of the galaxy. To transform them into observable quantities, we apply cosmological redshifting. This process accounts for the expansion of the Universe, which both stretches the wavelength of light and diminishes its observed flux. The rest-frame spectrum, with luminosity density $L_{\lambda}$ in units of $L_{\odot}$ \AA$^{-1}$ is converted to an observed-frame flux density, $F_{\lambda}$ in units of erg s$^{-1}$ cm$^{-2}$ \AA$^{-1}$. The transformation is based on the galaxy's redshift, $z$, and the luminosity distance, $D_{L}(z)$, as determined by the chosen cosmology. 

\subsubsection{IGM absorption} \label{sec:igm_absorption}

As light from a distant galaxy travels to an observer, it passes through the Intergalactic Medium (IGM), where intervening clouds of neutral hydrogen absorb photons, primarily at ultraviolet wavelengths. This effect is crucial for accurately modeling the observed spectra of high-redshift galaxies. \galsyn\ applies this absorption to the final, redshifted spectrum of each pixel. The final observed flux is the product of the intrinsic flux and the IGM transmission. \galsyn\ provides two widely used empirical models for IGM transmission: \citet{Madau1995} and \citet{Inoue2014}.

\subsubsection{Integrating Through Filter Transmission Functions} \label{sec:filtering} 

To transform the synthesized spectra into observable photometric fluxes, we simulate the process of observing the object through a set of standard astronomical filters. 
The average flux density from a given spectrum that is observed through a filter is obtained by integrating the spectrum through the transmission function of the filter. This calculation is performed for each pixel in the image grid and for every specified filter, yielding the final multi-band photometric data cubes. The resulting flux is in units of erg s$^{-1}$ cm$^{-2}$ \AA$^{-1}$. This flux is then converted into the user-selected units. The available options for this are: MJy/sr, nJy, AB magnitude, and erg s$^{-1}$ cm$^{-2}$ \AA$^{-1}$.

In the generation of a photometric datacube, \galsyn\ also calculates the pivot wavelength from the filter transmission function, which is useful for converting flux densities between different unit options.

\subsection{Physical Property Maps Reconstruction} \label{sec:property_maps}

\begin{figure*}[ht]
\centering
\includegraphics[width=0.95\linewidth]{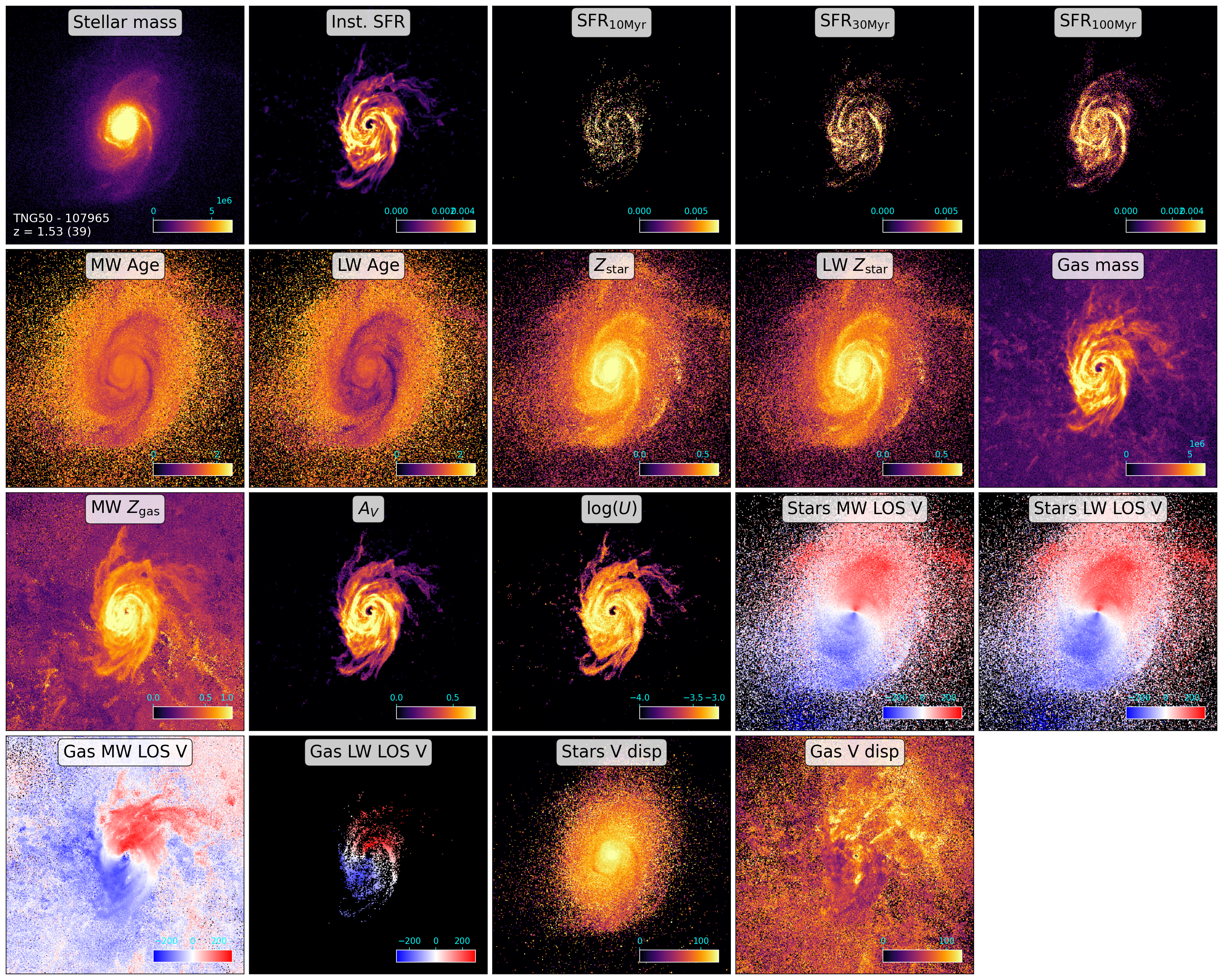}
\caption{Spatially resolved physical property maps of the galaxy TNG50-107965, the same galaxy shown in Figure~\ref{fig:idealized_images1}. These maps have the same spatial sampling as those of the synthetic images ($0.03"$ pixel$^{-1}$). This illustrates the comprehensive set of physical property maps generated by \galsyn.}
\label{fig:maps_props1}
\end{figure*}

Beyond creating synthetic images, \galsyn\ produces a comprehensive suite of 2D physical property maps of the simulated galaxies. These maps are generated by aggregating the properties of the constituent star and gas particles that fall within the line-of-sight of each pixel. The physical property maps are derived using three primary methods:
\begin{itemize}
\item \textbf{Summation}: This method involves summing a specific property from all relevant particles within a pixel. It is used for extensive quantities like total stellar or gas mass.
\item \textbf{Mass-Weighted Average}: This method is employed to calculate a physically representative mean value for a property such as age or metallicity. For a given property $P$, the average value $\langle P \rangle_{\text{mass}}$ for a pixel containing $N$ particles is calculated as:
\begin{equation}
\langle P \rangle_{\text{mass}} = \frac{\sum_{i=1}^{N} m_i P_i}{\sum_{i=1}^{N} m_i}
\end{equation}
where $m_i$ and $P_i$ are the mass and property of the $i$-th particle, respectively. This method ensures that more massive particles, which contribute more significantly to the system's total mass, have a proportionally larger impact on the average.
\item  \textbf{Light-Weighted Average}: This method is adopted to better represent what would be measured in an actual observation. This method weights each particle's contribution by its bolometric luminosity, $L_i$, which is calculated from its spectrum. Since young, massive stars are brighter than the old ones, this method provides an observationally-biased mean that reflects what we see in the observation. The light-weighted average, $\langle P \rangle_{\text{light}}$, is calculated as:
\begin{equation}
\langle P \rangle_{\text{light}} = \frac{\sum_{i=1}^{N} L_i P_i}{\sum_{i=1}^{N} L_i}.
\end{equation}
\end{itemize}
The list of generated physical property maps is given in Table~\ref{tab:physical_maps}, and an example of the maps for the galaxy TNG50-107965 is shown in Figure~\ref{fig:maps_props1}. 

\begin{figure}[ht]
\centering
\includegraphics[width=1.0\linewidth]{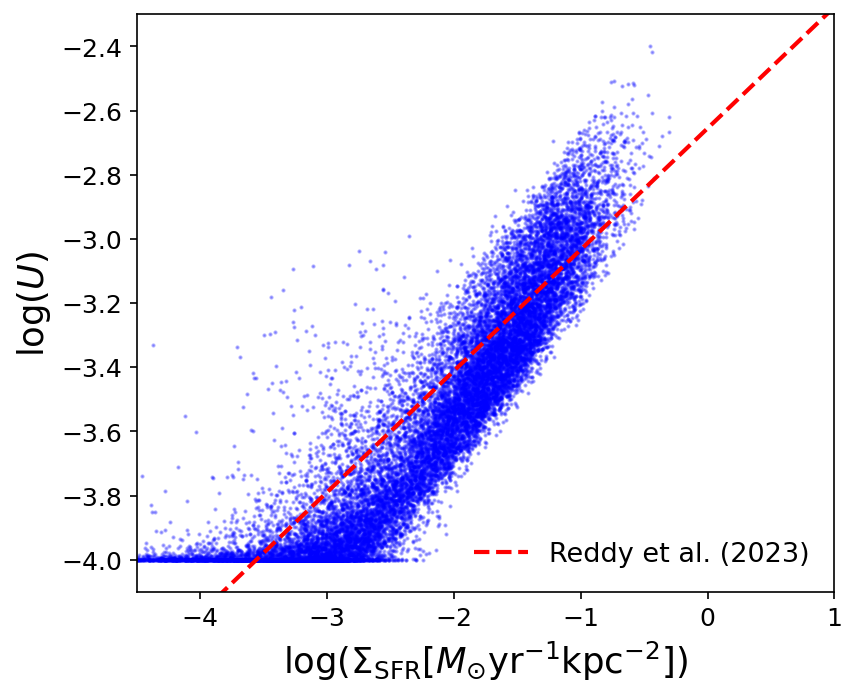}
\caption{Correlation between the spatially resolved ionization parameter and SFR surface density in the galaxy TNG50-107965 as derived using \galsyn\ (see Figure~\ref{fig:maps_props1}). Each data point represents a single pixel of size $0.03$ arcsec, corresponding to $0.26$ kpc. The dashed line shows the empirical correlation from \citet{Reddy2023a}. There is a good agreement between the predicted correlation from \galsyn\ and the empirical correlation. The calculated $U$ is capped to a physical range of $10^{-4}$ to $10^{-1}$ (see Section~\ref{sec:ionization_parameter}), which makes the artificial cut at the bottom of the correlation.}
\label{fig:plot_sfr_vs_logu}
\end{figure}

While an extensive analysis of property maps for a large galaxy sample is beyond the scope of this paper, we demonstrate here the robustness of the ionization parameter estimates formulated in Section~\ref{sec:ionization_parameter}. As shown in Figure~\ref{fig:plot_sfr_vs_logu}, the ionization parameter correlates strongly with SFR surface density on spatially resolved scales within the galaxy TNG50-107965. This aligns well with the empirical relation derived by \citet{Reddy2023a} for star-forming galaxies at $2.7 < z < 6.3$. 

\subsection{Spatially Resolved SFH Reconstruction} \label{sec:sfh_maps}

To understand how different regions of a galaxy were assembled over cosmic time, \galsyn\ includes functionality to reconstruct the spatially resolved SFH. This process analyzes the formation properties of star particles on a pixel-by-pixel basis, resulting in a comprehensive dataset that traces the growth and chemical evolution across the galaxy.

The reconstruction begins by preparing the simulation data for analysis. Essential properties, specifically the initial mass, formation time, metallicity, and 3D coordinates, are loaded for all star particles. The formation time of each particle is converted into a lookback time from the perspective of the galaxy's snapshot redshift. Subsequently, all star particles are projected onto a 2D pixel grid according to the specified viewing angle (see Section~\ref{sec:spatial_projection}). This step generates a membership list for each pixel, detailing exactly which star particles fall within its line-of-sight.

The core of the reconstruction is performed in parallel for each pixel on the grid to ensure computational efficiency. For every pixel, the associated star particles are gathered and binned into discrete time intervals based on their lookback time, with the user-defined bin width (with a default value of 50 Myr). Within each lookback time bin, several key physical properties are calculated:
\begin{itemize}
\item \textbf{Mass Formed}: The sum of the initial masses of all star particles in the bin.

\item \textbf{Star Formation Rate (SFR)}: The total mass formed in the bin divided by the time width of the bin, yielding an average SFR in M$_{\odot}$/yr.

\item \textbf{Mass-Weighted Metallicity}: The average metallicity of the stars in the bin, weighted by their initial mass.

\item \textbf{Cumulative Mass}: The running total of all stellar mass formed up to the end of that time bin
\end{itemize}

Once the SFH is constructed for a pixel, the cumulative mass history is used to derive key stages in its formation. By interpolating the cumulative mass array, the specific lookback time at which the pixel had formed a certain percentage of its final stellar mass is determined. This is calculated for several key fractions: 5\%, 10\%, 25\%, 50\%, 75\%, and 95\%. These mass assembly times can provide a useful insight on whether a region of the galaxy formed its stars early and rapidly or over a more extended period.

The final output is a single, comprehensive FITS file that aggregates the results from all pixels. This includes 3D data cubes for SFR, stellar mass formed, and metallicity, with axes corresponding to the two spatial dimensions and lookback time. Additionally, the derived mass assembly times (e.g., t${50}$, t${95}$) are saved as 2D maps. These are summarized in Table~\ref{tab:sfh_maps}.

\begin{figure*}
\centering
\includegraphics[width=0.65\textwidth]{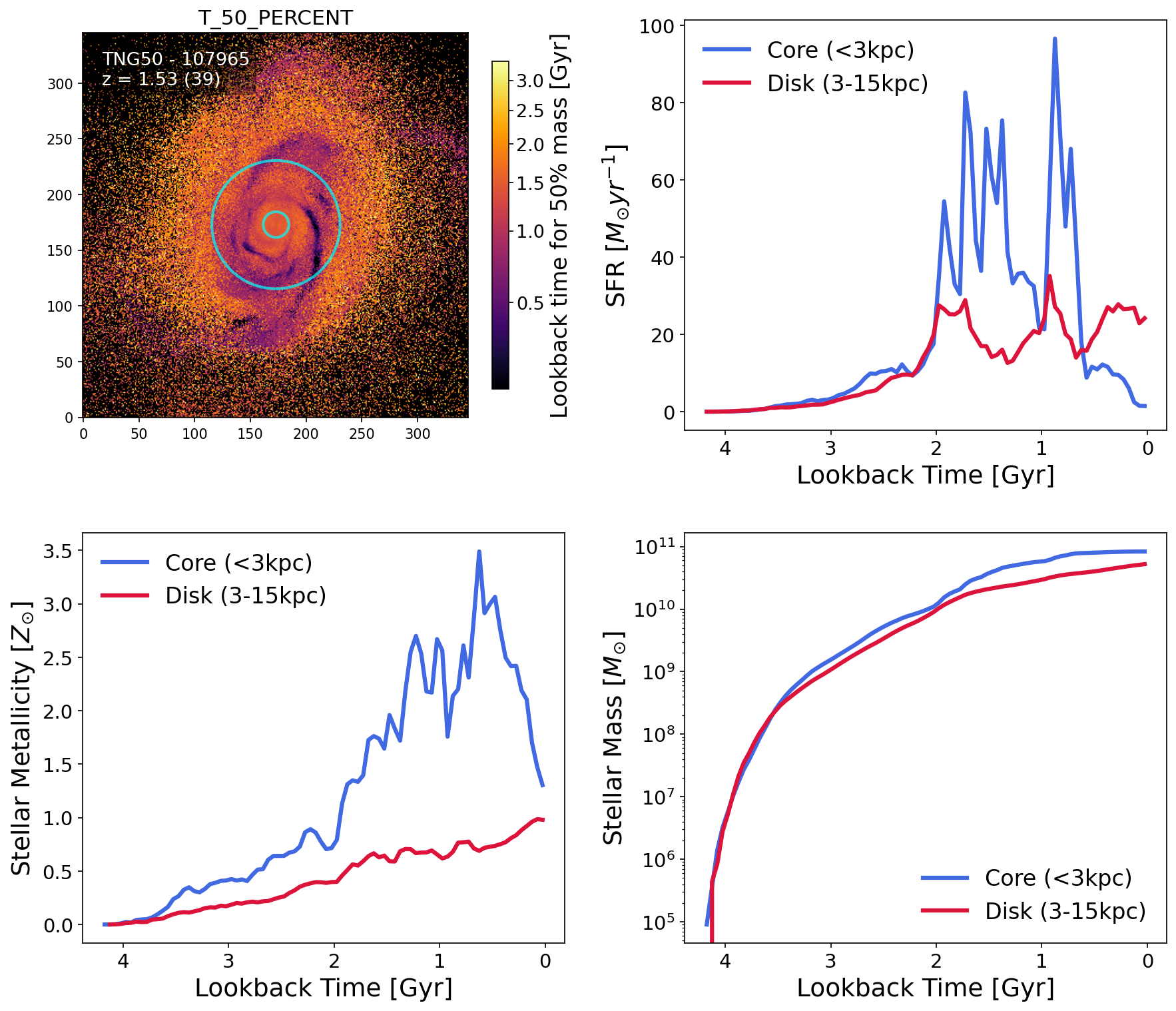}
\caption{Spatially resolved SFH of the galaxy TNG50-107965 as reconstructed by \galsyn. \textit{Top left}: the map of spatially resolved lookback time for 50\% mass assembly (\texttt{T\_50\_PERCENT}). \textit{Top right}: comparison between the integrated SFH of the central region ($r < 3$ kpc) and the disk ($3 < r < 15$ kpc). \textit{Bottom left}: comparison for metallicity history. \textit{Bottom right}: comparison for stellar mass growth history.}
\label{fig:sfh_maps}
\end{figure*}

\subsection{Regridding to Desired Pixel Scale} \label{sec:regridding}

As described in Section~\ref{sec:spatial_projection}, the initial 3D to 2D projection is performed at a high-resolution grid based on the smoothing length of the simulation. After the synthesis process, the synthetic images and property maps are resampled to a user-defined pixel scale using a spatial resampling process that ensures physical consistency. The regridding is handled differently depending on the type of data being processed. For images representing physical quantities like light (flux) or mass, the resampling is performed in a flux-conserving mode. In this case, the resampling process is designed to preserve the total integrated value across the grid. This ensures that even as the spatial resolution changes, the total physical quantity represented in the data cube remains consistent.

For maps that represent average characteristics, such as the stellar age, metallicity, or velocity, the module uses an averaging mode. This process is carefully managed to only include pixels that actually contain data, preventing empty space from ``diluting'' or artificially lowering the physical values of the galaxy properties.

The computational cost needed for synthesizing the data cube of a galaxy depends on the number of particles (stars and gas) in the galaxy. The typical computational cost for synthesizing imaging data (in 47 filters) of a galaxy with $5\times 10^{5}$ star particles and $7\times 10^{5}$ gas particles is $\sim 1$ CPU hour\footnote{This is $\sim 6$ minutes for a parallel computation using 10 cores.}.

\section{Synthetic Data Products and Observational Realism} \label{sec:datacube_observational_effects}

\subsection{Synthetic Data Cubes} \label{sec:data_product}

From the synthesis processes described in the previous section, \galsyn\ produces a comprehensive set of data products. The outputs can be categorized into four main types: synthetic imaging data cube, IFU data cube, physical property maps, and spatially resolved SFH maps. The imaging, IFU (when requested for output) data cubes, and physical property maps are stored in a single multi-extension FITS file. The resolved SFH maps are generated via a different module, and they are stored in a single multi-extension FITS file. 

The synthetic imaging and IFU data cubes contain dust-free and dust-attenuated data. The FITS file extensions for the synthetic data and the derived property maps are summarized in Table~\ref{tab:datacube_extensions} and~\ref{tab:physical_maps}, respectively, while the FITS file extension for the resolved SFH data product is listed in Table~\ref{tab:sfh_maps}.   

\begin{table*}[ht] 
\centering 
\caption{Synthetic Datacube FITS File Extensions} 
\label{tab:datacube_extensions} \begin{tabular}{l l l} 
\hline 
\hline 
\textbf{FITS Extension Name} & \textbf{Data Type} & \textbf{Description} \\ 
\hline \multicolumn{3}{c}{\textit{Imaging Data (2D Broadband Images)}} \\ 
\hline \texttt{NODUST\_[FILTER]} & 2D Image & Photometric flux map for a given filter without dust attenuation. \\ 
\texttt{DUST\_[FILTER]} & 2D Image & Photometric flux map for a given filter with dust attenuation. \\ 
\hline 
\multicolumn{3}{c}{\textit{IFU Data (3D Spectral Cubes)}} \\ 
\hline \texttt{OBS\_SPEC\_NODUST} & 3D Cube & Observed-frame spectra ($\lambda$, y, x) for each pixel without dust. \\ 
\texttt{OBS\_SPEC\_DUST} & 3D Cube & Observed-frame spectra ($\lambda$, y, x) for each pixel with dust. \\ 
\hline 
\multicolumn{3}{c}{\textit{Ancillary Data}} \\ 
\hline 
\texttt{WAVELENGTH\_GRID} & Binary Table & 1D array of observed-frame wavelengths for the spectral cubes. \\ 
\hline 
\end{tabular} 
\end{table*}

Figure~\ref{fig:idealized_images1} shows an example of synthetic images of a simulated galaxy from TNG50 (subhalo ID=107965) at $z=1.53$ across multiple observatories, including HST, JWST, Rubin/LSST, and the Roman Space Telescope. For this, we used SSP grids generated with FSPS and applied the ``line-of-sight'' dust method and the SKIRT-like dust attenuation curve with adaptive slope, bump amplitude, and bump width, as described in Section~\ref{sec:dust_attenuation}. The top-left panel displays a color composite image generated from JWST filters F115W (blue), F150W (green), and F200W (red) using the \texttt{make\_lupton\_rgb} function \citep{Lupton2004} from the \texttt{astropy} package \citep{Astropy2013,Astropy2018}. These images have a spatial sampling of $0.03$ arcsec pixel$^{-1}$ and do not contain simulated noise (to be described in the next section). The physical property maps of this galaxy are shown in Figure~\ref{fig:maps_props1}, while Figure~\ref{fig:sfh_maps} shows the galaxy's resolved SFH. In the latter, the top-left panel shows the map of the spatially resolved lookback time for 50\% mass assembly (\texttt{T\_50\_PERCENT}). The remaining panels illustrate the SFH (top right), metallicity history (bottom left), and stellar mass growth (bottom right), comparing the central region ($r < 3$ kpc) with the disk ($3 < r < 15$ kpc).

\begin{table*}[ht] 
\centering 
\caption{Physical Property Maps FITS File Extensions} 
\label{tab:physical_maps} 
\begin{tabular}{l l l} \hline \hline 
\textbf{FITS Extension Name} & \textbf{Property Description} & \textbf{Units} \\ 
\hline 
\multicolumn{3}{c}{\textit{Stellar Properties}} \\ 
\hline 
\texttt{STARS\_MASS} & Stellar mass & $M_{\odot}$ \\ 
\texttt{MW\_AGE} & Mass-weighted stellar age & Gyr \\ 
\texttt{STARS\_MW\_ZSOL} & Mass-weighted stellar metallicity & $Z_{\odot}$ \\ 
\texttt{LW\_AGE\_DUST} & Light-weighted stellar age with dust attenuation & Gyr \\ 
\texttt{LW\_AGE\_NODUST} & Light-weighted stellar age without dust attenuation & Gyr \\ 
\texttt{LW\_ZSOL\_DUST} & Light-weighted stellar metallicity with dust attenuation & $Z_{\odot}$ \\ 
\texttt{LW\_ZSOL\_NODUST} & Light-weighted stellar metallicity without dust attenuation & $Z_{\odot}$ \\ 
\texttt{SFR\_10MYR} & SFR averaged over the past 10 Myr & $M_{\odot}\,\mathrm{ yr}^{-1}$ \\ 
\texttt{SFR\_30MYR} & SFR averaged over the past 30 Myr & $M_{\odot}\,\mathrm{ yr}^{-1}$ \\  
\texttt{SFR\_100MYR} & SFR averaged over the past 100 Myr & $M_{\odot}\,\mathrm{ yr}^{-1}$ \\ 
\hline 
\multicolumn{3}{c}{\textit{Gas, Dust, \& Kinematic Properties}} \\ 
\hline 
\texttt{GAS\_MASS} & Gas mass & $M_{\odot}$ \\ 
\texttt{SFR\_INST} & Instantaneous SFR derived from gas particle & $M_{\odot}\,\mathrm{ yr}^{-1}$ \\ 
\texttt{GAS\_MW\_ZSOL} & Mass-weighted gas metallicity & $Z_{\odot}$ \\ 
\texttt{DUST\_MEAN\_AV} & Mean dust attenuation ($A_V$) among star particles & mag \\ 
\texttt{STARS\_MW\_VEL\_LOS} & Mass-weighted LOS stellar velocity & km s$^{-1}$ \\ 
\texttt{GAS\_MW\_VEL\_LOS} & Mass-weighted LOS gas velocity & km s$^{-1}$ \\ 
\texttt{STARS\_VEL\_DISP\_LOS} & LOS stellar velocity dispersion & km s$^{-1}$ \\ 
\texttt{GAS\_VEL\_DISP\_LOS} & LOS gas velocity dispersion & km s$^{-1}$ \\ 
\texttt{LW\_VEL\_LOS\_DUST} & LOS light-weighted stellar velocity with dust attenuation & km s$^{-1}$ \\ 
\texttt{LW\_VEL\_LOS\_NODUST} & LOS light-weighted stellar velocity without dust attenuation & km s$^{-1}$ \\ 
\texttt{LW\_VEL\_LOS\_NEBULAR} & LOS Light-weighted nebular velocity & km s$^{-1}$ \\ 
\texttt{GAS\_LOGU} & Ionization parameter in logarithmic scale & Dimensionless \\ \hline 
\end{tabular} 
\end{table*}

\subsection{Adding Observational Effects to the Synthetic Imaging Data Cube} \label{sec:obs_imaging}

\begin{table*}[ht] 
\centering 
\caption{Spatially Resolved Star Formation History Data Products} 
\label{tab:sfh_maps} 
\begin{tabular}{l l l l} 
\hline 
\hline 
\textbf{Property Description} & \textbf{FITS Extension Name} & \textbf{Description} & \textbf{Unit} \\ 
\hline 
\multicolumn{4}{c}{\textit{3D Data Cubes (y, x, lookback time)}} \\ 
\hline 
Star Formation Rate & \texttt{SFR} & Average SFR in each time bin & M$_{\odot}$ yr$^{-1}$ \\ 
Initial Stellar Mass Formed & \texttt{MASS} & Sum of initial stellar masses in bin & M$_{\odot}$ \\ 
Cumulative Stellar Mass & \texttt{CUMUL\_MASS} & Total stellar mass formed up to that time & M$_{\odot}$ \\ 
Mass-Weighted Metallicity & \texttt{METALLICITY} & Mass-weighted average metallicity &  Z$_{\odot}$ \\ 
Number of Stars & \texttt{N\_STARS} & Number of star particles in bin & count \\ 
\hline 
\multicolumn{4}{c}{\textit{2D Data Maps}} \\ 
\hline 
Mass Assembly Times & \texttt{T\_5\_PERCENT} & Lookback time for 5\% mass assembly & Gyr \\ 
& \texttt{T\_10\_PERCENT} & Lookback time for 10\% mass assembly & Gyr \\ 
& \texttt{T\_25\_PERCENT} & Lookback time for 25\% mass assembly & Gyr \\ 
& \texttt{T\_50\_PERCENT} & Lookback time for 50\% mass assembly & Gyr \\ 
& \texttt{T\_75\_PERCENT} & Lookback time for 75\% mass assembly & Gyr \\ 
& \texttt{T\_95\_PERCENT} & Lookback time for 95\% mass assembly & Gyr \\ 
\hline 
\multicolumn{4}{c}{\textit{1D Data Array}} \\ 
\hline 
Lookback Time Bins & \texttt{LOOKBACK\_TIME\_BINS} & Midpoints of the lookback time bins & Gyr \\ 
\hline 
\end{tabular} 
\end{table*}

To generate realistic synthetic data cubes, \galsyn\ is equipped with functionalities to add observational effects that include point spread function (PSF) convolution, noise simulation, and spatial resampling to a desired pixel scale. This can be performed using \texttt{observe} module. In the future update of the code, we plan to incorporate other public packages (e.g.,~\texttt{Galsim}, \citealt{Rowe2015}; \texttt{romanisim}; \texttt{MIRAGE}) for more realistic simulation of observational effects.  

The first step is to resample the synthetic images to a user-defined pixel scale, which can be set independently for each filter. This resampling process is done following procedures described in Section~\ref{sec:regridding}.

The next step is PSF convolution. A PSF is the characteristic blurred shape a telescope and camera system records when observing a point source of light, like a distant star. This PSF determines the image's fundamental resolution. In \galsyn\ processing, the PSF must be provided by the user as a 2D FITS image file. Each filter requires its own corresponding PSF file. This allows for wavelength-dependent variations in the blurring effect.

\begin{figure*}
\centering
\includegraphics[width=0.9\textwidth]{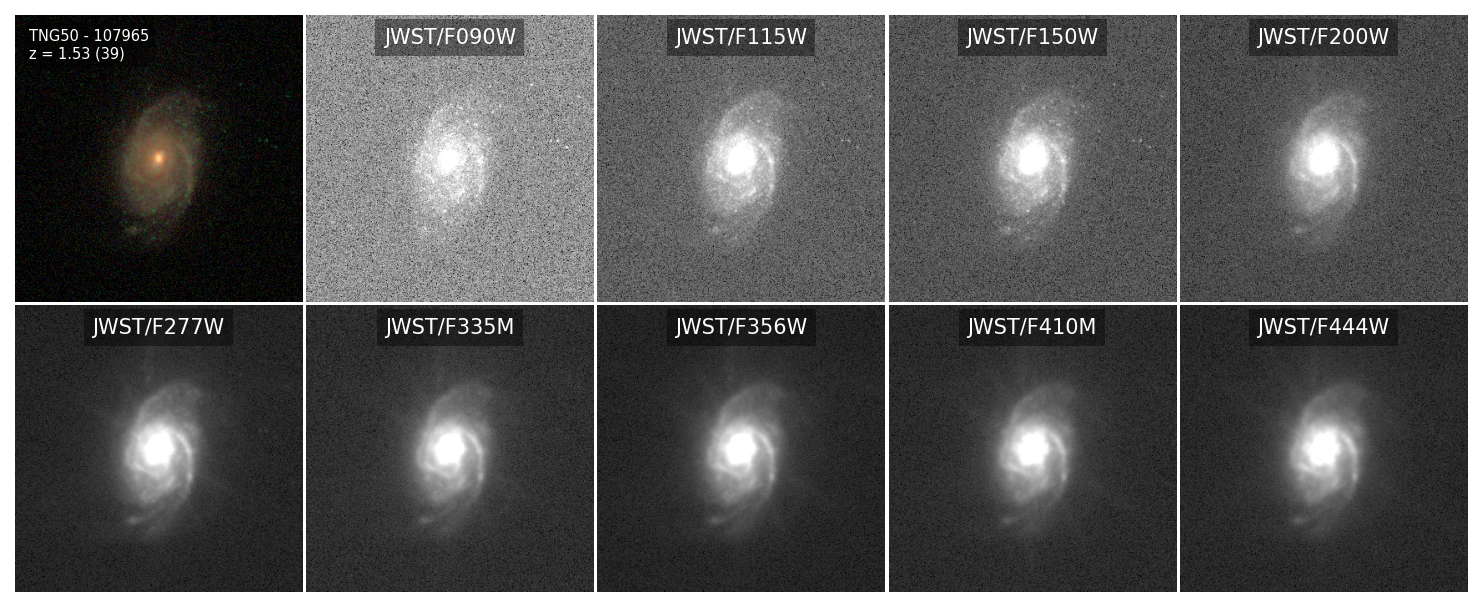}
\caption{Example of the post-processed images of the TNG50-107965 galaxy. The images are processed assuming the JADES \citep{Eisenstein2026} deep imaging observational characteristics and PSFs. More details about the observational parameters will be given in Section~\ref{sec:mock_fields}.}
\label{fig:mock_observed_images1}
\end{figure*}

In terms of noise simulation, \galsyn\ simulates the two primary sources of observational noise: photon shot noise from the astronomical source and background noise from the sky and instrument electronics. It achieves this by mimicking known (i.e.,~empirical) or expected noise characteristics rather than predicting them from detailed instrument parameters \citep[e.g.,][]{Merlin2023_euclid}. This approach simplifies the process, as users can input desired noise levels, whether determined empirically or from an Exposure Time Calculator (ETC), without needing complex or often unavailable instrument data. This method provides the flexibility to simulate noise for any instrument. The detailed description on the noise simulation is provided in Appendix~\ref{sec:technical_info_noise_simulation}.  

To illustrate the robustness of our noise simulation, we show in Figure~\ref{fig:mock_observed_images1} an example of the post-processing result for the TNG50-107965 galaxy. The images are processed assuming the JADES \citep{Eisenstein2026} deep imaging observational characteristics and PSFs, specifically for the subregion \texttt{jw011800-deep} from their DR 5. More details about the observational parameters will be given in Section~\ref{sec:mock_fields}. The F090W image, which has a limiting magnitude of $29.87$ (at S/N=5 measured on $0.15$ arcsec aperture), looks shallower than the other images, which reach deeper limiting magnitudes ($>30$). 

\begin{figure}
\centering
\includegraphics[width=1.0\linewidth]{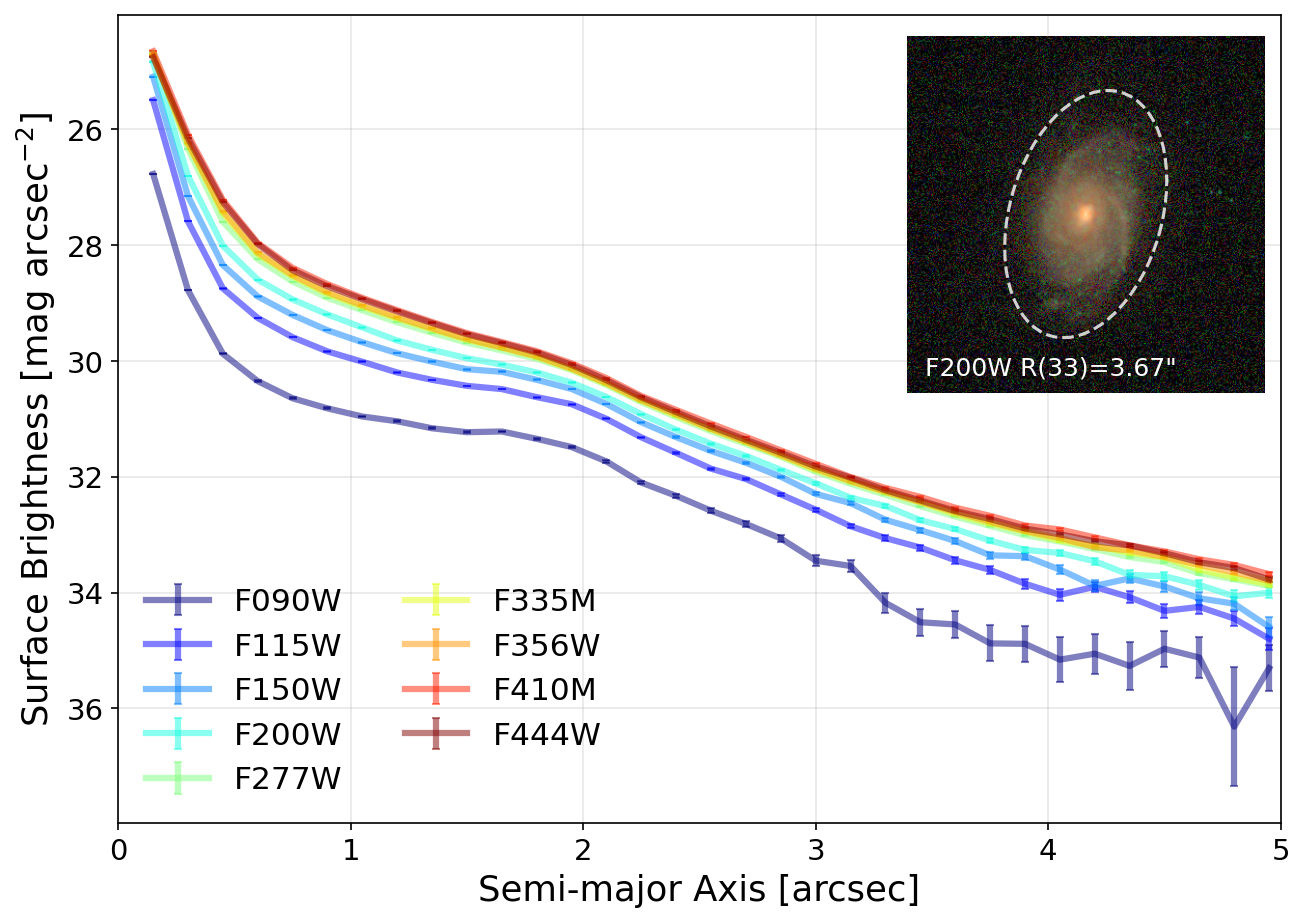}
\caption{Multi-wavelength surface brightness radial profiles of the TNG50-107965 galaxy. The profiles show a characteristic two-component structure: a steep central gradient ($R < 1$ arcsec) and a shallower exponential disk. The galaxy signal traced to extreme depths of $\sim 34\text{--}36 \text{ mag arcsec}^{-2}$. The white dashed line in the inset RGB map indicates the isophotal line of $33$ mag arcsec$^{-2}$ in F200W.}
\label{fig:plot_sb_profiles}
\end{figure}

Figure~\ref{fig:plot_sb_profiles} shows the surface brightness radial profiles across the filters. These profiles demonstrate a characteristic two-component structure across the wavelength. A steep central gradient ($R < 1$ arcsec) transitions into a shallower exponential disk, with the galaxy signal traced to extreme depths of $\sim 34\text{--}36 \text{ mag arcsec}^{-2}$. A consistent vertical offset is observed across the spectrum, where the long-wavelength bands (e.g., F444W) exhibit significantly higher surface brightness than the shorter-wavelength bands (e.g., F090W), indicating a globally red stellar population in the galaxy. The increasing error bars and surface-brightness fluctuations beyond $R \approx 4''$ in the F090W band mark the transition into the background-dominated regime. The white dashed line in the inset RGB image indicates the isophotal line of $33$ mag arcsec$^{-2}$ in F200W.

\subsection{Post-processing of IFU Data Cube} \label{sec:obs_ifu}

\begin{figure*}
\centering
\includegraphics[width=0.85\textwidth]{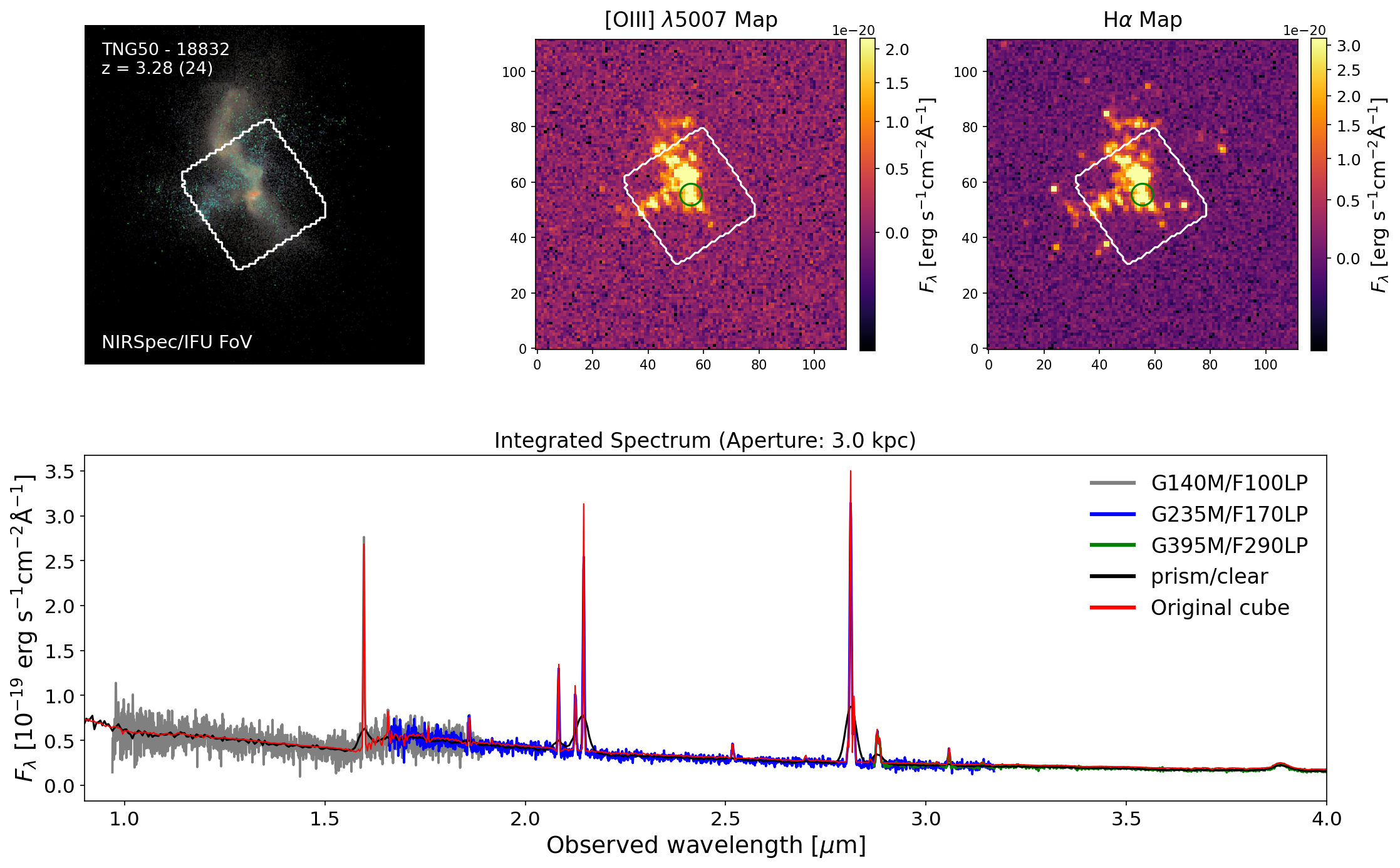}
\caption{Post-processed synthetic IFU datacube of galaxy TNG50-18832 at $z=3.28$, emulating JWST/NIRSpec IFU observations across four configurations: G140M/F100LP, G235M/F170LP, G395M/F290LP, and Prism/CLEAR. The noise is simulated based on the estimated $5\sigma$ limiting magnitudes of the instruments derived from the \texttt{Pandeia} ETC engine \citep{Pontoppidan2016}. The exposure time is set to 20 ks for Prism/CLEAR and 40 ks for the other instruments. The red curve in the bottom panel denotes the intrinsic spectrum without observational effects. The panels in the top row show, from the left to right: a color composite image, [OIII]$\lambda 5007$ flux map, and H$\alpha$ flux map. The white box in each panel shows the NIRSpec IFU field of view.}
\label{fig:plot_mockobs_ifu}
\end{figure*}

The post-processing of synthetic IFU data cubes in \galsyn\ begins with several spectral and spatial modifications to simulate realistic observational effects. First, the high-resolution synthetic cube is spectrally interpolated onto a user-defined wavelength grid, \texttt{desired\_wave\_grid}. The next step is to spatially resample the cube to a desired output pixel scale, which is performed using the same method as that for the synthetic images. Then, to simulate the effect of a finite instrumental resolution, the spectrum of each spaxel is convolved along the wavelength axis with a 1D Gaussian kernel. The width of this kernel is determined by a user-provided target spectral resolution, $R\equiv \lambda / \Delta\lambda$. The cube is then spatially convolved using a user-defined 3D PSF cube, a FITS file whose wavelength grid must exactly match the \texttt{desired\_wave\_grid}. The convolution is performed slice-by-slice, using the corresponding 2D PSF slice for each wavelength channel to model the wavelength-dependent blurring.

Following these spectral and spatial convolutions, observational noise is injected, also on a slice-by-slice basis. The process follows the same principles as those for the synthetic imaging data cube, but is performed independently for each wavelength slice of the data cube. The key difference is that the input parameters can be functions of wavelength: $m_{\text{zp}}(\lambda)$, $t_{\text{exp}}(\lambda)$, $m_{\text{lim}}(\lambda)$, and $\text{SNR}_{\text{lim}}(\lambda)$.

The same set of equations outlined in Appendix~\ref{sec:technical_info_noise_simulation} is used, but they are applied slice-by-slice, iterating through the wavelength axis of the cube. This ensures that the final 3D data cube and its corresponding 3D RMS error cube have a realistic, wavelength-dependent noise profile.

\begin{figure}
\centering
\includegraphics[width=0.47\textwidth]{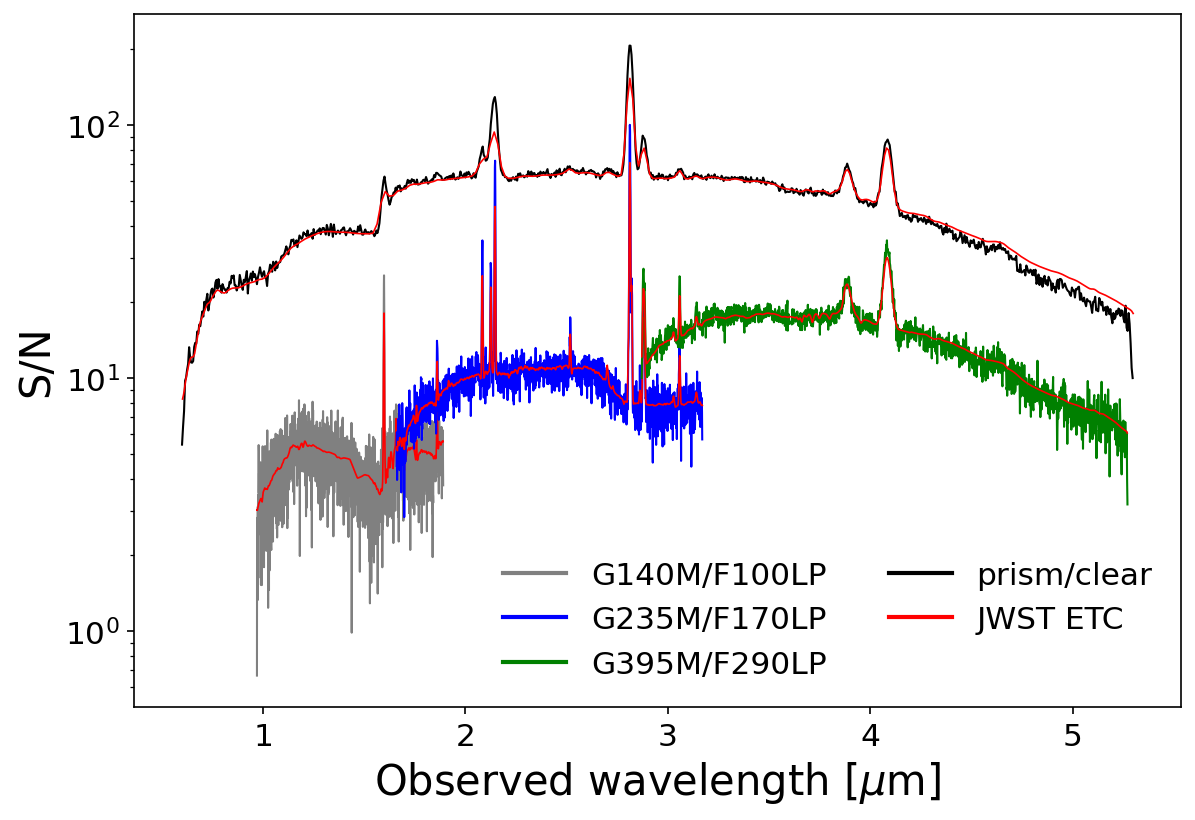}
\caption{S/N profiles of spectra across multiple instrument configurations, as shown in Figure~\ref{fig:plot_mockobs_ifu}, derived from \galsyn\ simulations and estimated using the \texttt{Pandeia} ETC engine (shown in red). The two methods show very good agreement, demonstrating the good recovery of observational parameters by \galsyn.}
\label{fig:comp_ifu_snr_vs_etc}
\end{figure}

Figure~\ref{fig:plot_mockobs_ifu} presents a synthetic IFU datacube of galaxy TNG50-18832 at $z=3.28$, post-processed to emulate JWST/NIRSpec IFU observations across four configurations: G140M/F100LP, G235M/F170LP, G395M/F290LP, and Prism/CLEAR. To determine the sensitivity limits, we performed S/N simulations using the \texttt{Pandeia} exposure time calculator (ETC) engine \citep{Pontoppidan2016}. We modeled a flat input spectrum in AB magnitudes across a grid of exposure times (10, 20, and 40 ks) and source magnitudes (18–28). For each configuration, the S/N was calculated per pixel on the native detector grid. We derived 5-$\sigma$ sensitivity limits by interpolating the magnitude–S/N relationship at each wavelength point, subsequently passing these limits to the \texttt{galsyn} module for noise simulation. A more detailed description of the procedure for estimating the sensitivity limits is given in Appendix~\ref{sec:mag_lims_nirspec_ifu}. 

The spectra shown in Figure~\ref{fig:plot_mockobs_ifu} correspond to a 40 ks exposure time (20 ks for Prism/CLEAR). For the PSF convolution, we modeled NIRSpec/IFU PSFs using \texttt{STPSF} \citep{Perrin2014}. In the figure, the red curve denotes the intrinsic spectrum without observational effects. The panels in the top row show [OIII]$\lambda 5007$ flux map, H$\alpha$ flux map, and a color composite image. The white box in each panel shows the NIRSpec IFU field of view.

To illustrate the robustness of the noise simulation in \galsyn, we compare the S/N profiles of the spectra across multiple instrument configurations with those derived using the \texttt{Pandeia} ETC engine. The comparison is shown in Figure~\ref{fig:comp_ifu_snr_vs_etc}. The red profiles are the S/N derived using the ETC, while the other profiles are the S/N recovered by \galsyn.

\section{First Data Release} \label{sec:data_release}

In this introductory paper, we present the first public data release of \galsyn\ to demonstrate its core capabilities. This release includes three different data cube categories generated from the IllustrisTNG simulation suites: four mock extragalactic survey fields, the progenitors of local massive galaxies $(\log(M_{*,z=0}/M_{\odot}) > 10.5)$, and a sample of major-merging systems. For each galaxy, the data cubes consist of synthetic imaging across 47 filters (spanning JWST, HST, Euclid, Rubin/LSST, and the Roman Space Telescope), a comprehensive suite of physical property maps (see Section~\ref{sec:property_maps}), and spatially resolved SFHs. 

We implement the ``line-of-sight'' dust method with three different dust attenuation curves: (1) dust0-S18K13: the option 0 dust model with an adaptive slope following the $\delta$-$A_{V}$ relation from \citet{Salim2018} and bump amplitude following the $B$-$\delta$ relation from \citet{Kriek2013}; (2) dust0-skirt-like: the option 0 dust model with an adaptive slope, bump amplitude, and bump width following the $\delta$-$B$-$\gamma$-$A_{V}$ correlations inferred from the NIHAO-SKIRT synthetic spectroscopic cubes \citep{Faucher2023}; and (3) the option 6 dust model: Milky Way dust extinction law from \citet{Fitzpatrick1999}. 

To capture a representative range of viewing geometries, we produce these data cubes at four distinct viewing angles: A1 ($\theta=45$, $\phi=45$), A2 ($\theta=0$, $\phi=0$), A3 ($\theta=90$, $\phi=90$), and A4 ($\theta=90$, $\phi=0$). These datasets are publicly available at \url{https://github.com/aabdurrouf/GalSyn_dr1} (\dataset[doi: 10.5281/zenodo.21498294]{https://doi.org/10.5281/zenodo.21498294}).

\begin{figure*}[ht]
\centering
\includegraphics[width=1\textwidth]{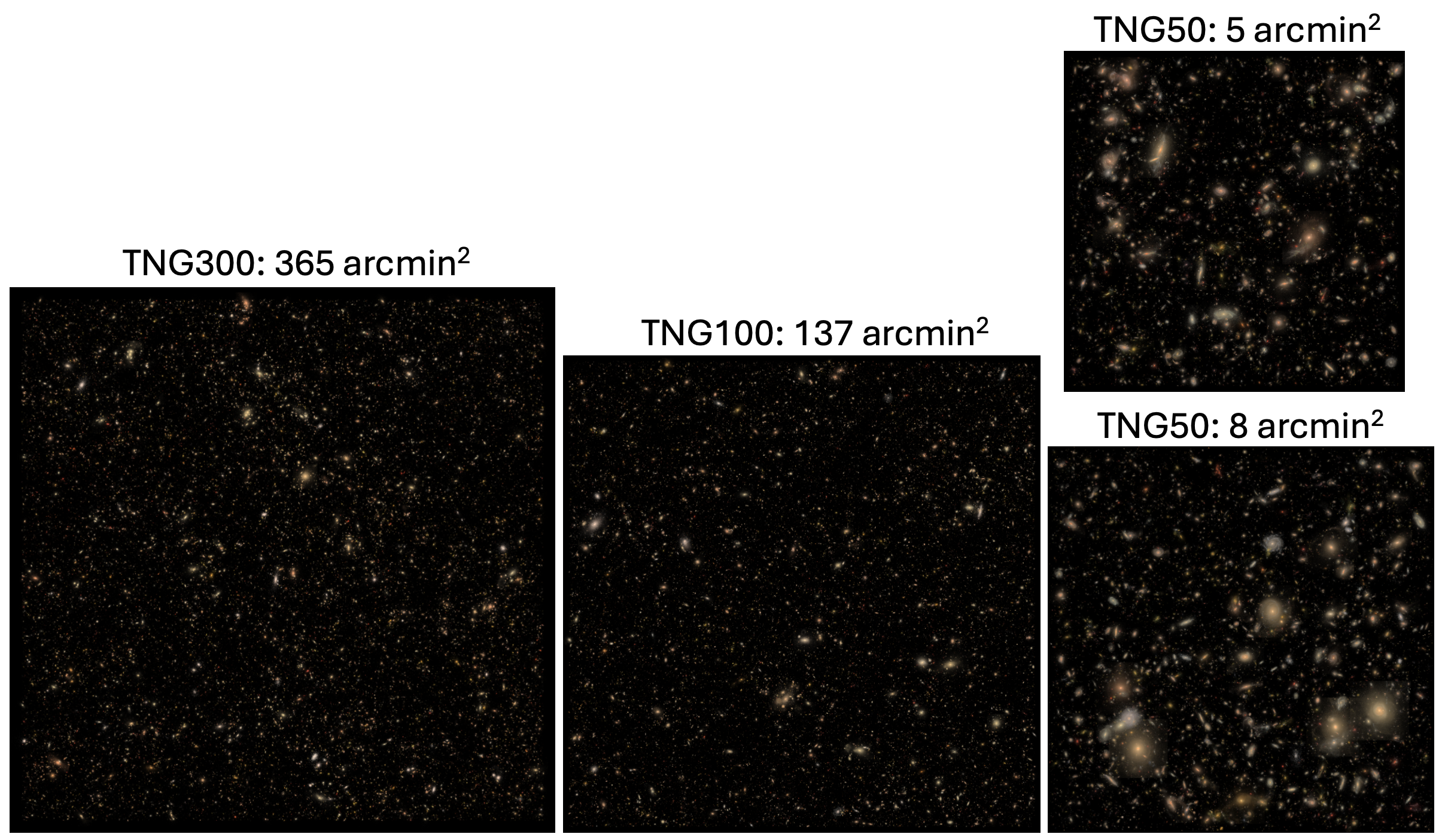}
\caption{Mock extragalactic survey fields generated from the three IllustrisTNG simulation suites (TNG50-1, TNG100-1, and TNG300-1) using the lightcone catalogs from \citet{Snyder2023}. These fields cover areas of 5, 8, 137, and 365 arcmin$^{2}$. These idealized mosaics implement the LOS dust method with the dust0-skirt-like prescription at viewing angle A1, excluding observational noise or PSF effects. The color composite images are created using JWST images: F115W (blue), F150W (green), and F200W (red).}
\label{fig:plot_mosaics}
\end{figure*}

\subsection{Mock Extragalactic Survey Fields} \label{sec:mock_fields}

The current rapid expansion of galaxy redshift surveys necessitates high-fidelity simulations to aid in observational design, pipeline testing, and the interpretation of complex datasets. Previous studies have produced synthetic survey imaging for these purposes \citep[e.g.,][]{Williams2018, Kauffmann2020, Merlin2023_euclid, Fortuni2023}. While the primary objective of \galsyn\ is to produce synthetic data cubes of individual galaxies from hydrodynamical simulation cutouts, we present these mock survey fields to demonstrate one of the applications: combining individual galaxy data cubes into large-scale synthetic mosaics using lightcone catalogs. 

Here, we produce four mock survey fields using the lightcone catalogs from \citet{Snyder2023}:
\begin{itemize}
    \item 5 arcmin$^{2}$ (TNG50) and 8 arcmin$^{2}$ fields generated from TNG50-1
    \item 137 arcmin$^{2}$ field generated from TNG100-1
    \item 365 arcmin$^{2}$ field generated from TNG300-1
\end{itemize}
These mosaics utilize the snapshots, subhalo IDs, and celestial coordinates (RA and DEC) assigned by the lightcone catalogs from \citet{Snyder2023}. Although \citet{Snyder2023} produced synthetic \textit{JWST} and \textit{HST} imaging fields, they did not incorporate dust attenuation modeling.

To create the mosaics, we generate a synthetic imaging data cube for every galaxy in the lightcone catalogs. The image size is set to encompass at least 99\% of the total stellar mass of the subhalo associated with the galaxy. Mosaics are constructed by creating a blank FITS array using the \texttt{Astropy} package \citep{Astropy2013,Astropy2018}, defining a World Coordinate System (WCS) centered at $(\text{RA, DEC}) = (0,0)$, and populating the field with the individual galaxy synthetic images. The mock survey fields are visualized in Figure~\ref{fig:plot_mosaics}.

\begin{figure*}[ht]
\centering
\includegraphics[width=1\textwidth]{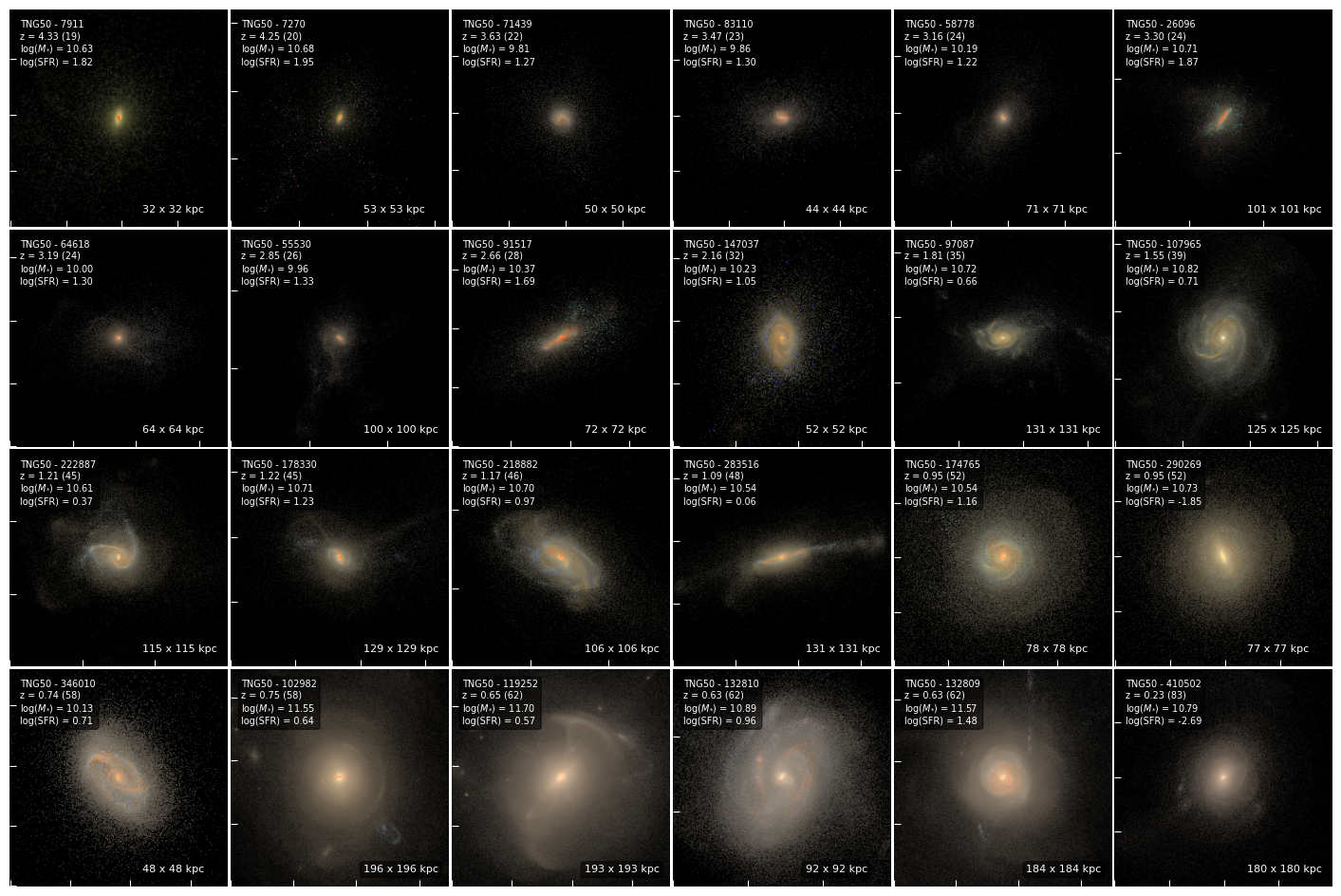}
\caption{Representative image stamps of 24 galaxies selected from the TNG50-11-10 field. Extracted from TNG subhalo cutouts, these data cubes provide an isolated view of each system. In this example, galaxies are seen from the A1 viewing angle ($\theta=45^\circ$, $\phi=45^\circ$). The physical scale of each stamp is indicated in the bottom-right corner of the respective panels. The color composite images are created using JWST filters: $z<0.2$ (F115W, F150W, F200W), $0.2<z<1.2$ (F150W, F200W, F277W), and $>1.2$ (F200W, F277W, F356W).}
\label{fig:stamp_images_mosaics}
\end{figure*}

In addition to the mosaics, we provide isolated data cubes for each constituent galaxy. These cutouts represent the isolated subhalos, free from the crowded mosaic environment. Each cube contains 47-filter imaging, physical property maps, and spatially resolved SFHs. We also include catalogs of integrated physical properties, including stellar, gas, dark matter, and black hole masses, as well as SFRs, half-stellar-mass radii, and metallicities. Figure \ref{fig:stamp_images_mosaics} shows examples of 24 galaxies from the TNG50-11-10 field, covering a redshift range of $0.2 < z < 4.5$ and stellar masses of $M_{*} > 10^{9.5} M_\odot$.

These individual cubes enable diverse scientific analyses, such as systematic comparisons with observations to gain physical insights. While a full assessment of \texttt{GalSyn}'s physical modeling, especially the dust attenuation implementations, is deferred to a forthcoming paper, we present a preliminary evaluation using color-color diagnostics. Specifically, we employ observed-frame JWST filter combinations that mimic the rest-frame $UVJ$ diagram \citep[e.g.,][]{Williams2009}. This diagnostic is widely used to identify quiescent galaxies due to its effectiveness in breaking the degeneracy between stellar aging and dust attenuation. To approximate the rest-frame $U-V$ color, we select filter pairs that bracket the rest-frame 4000 \AA\ break.

\begin{figure*}[ht]
\centering
\includegraphics[width=1\textwidth]{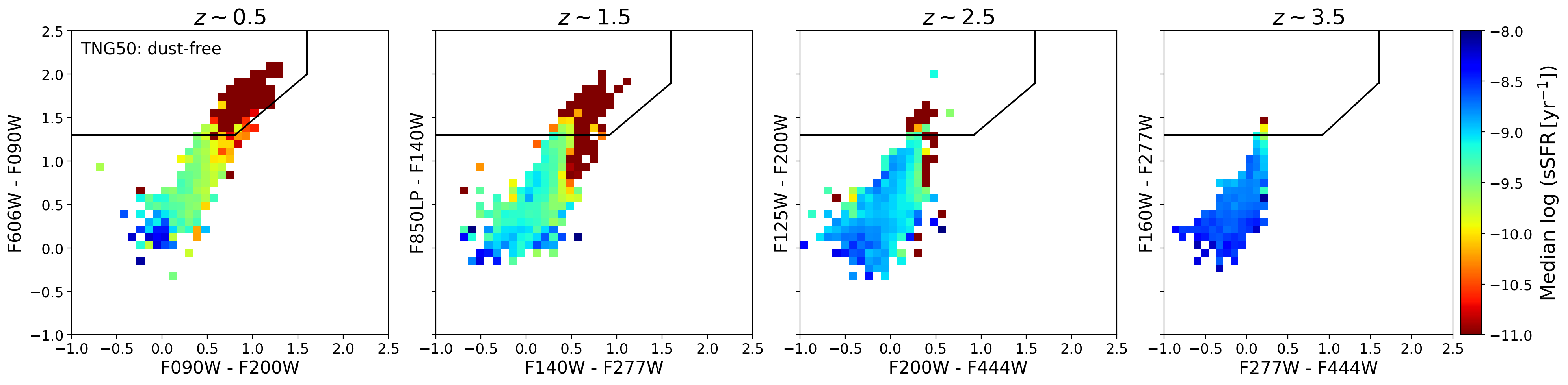}
\includegraphics[width=1\textwidth]{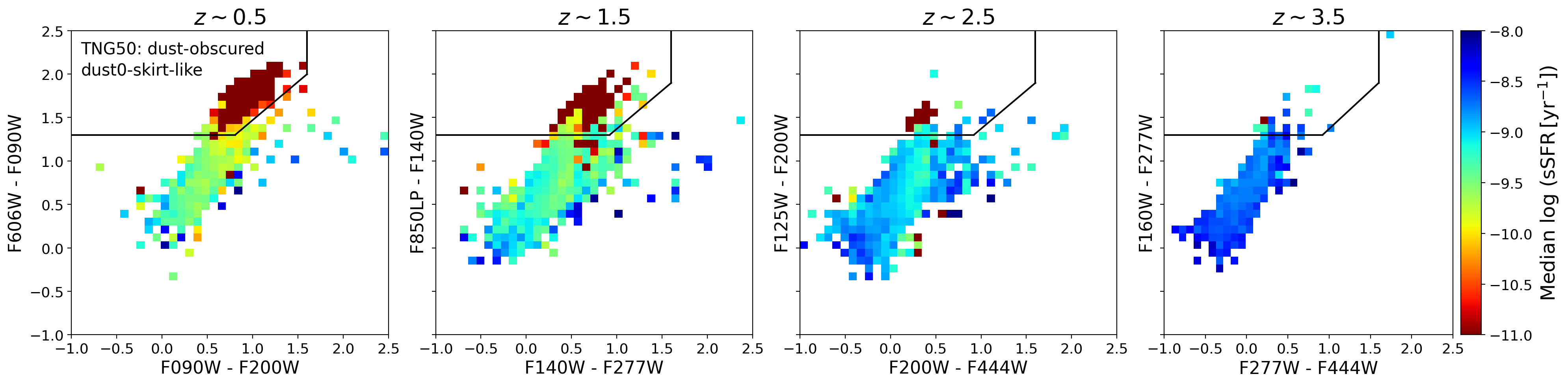}
\includegraphics[width=1\textwidth]{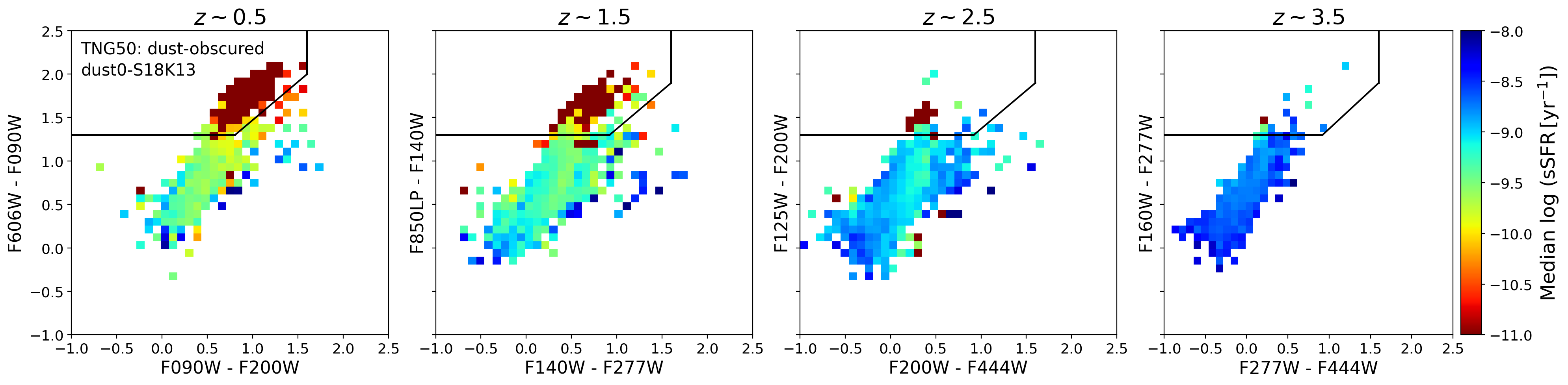}
\caption{JWST color-color diagrams across four redshift bins, designed to mimic the rest-frame $UVJ$ diagram. The galaxies shown here are selected from the TNG50-11-10 and TNG50-12-11 fields (viewing angle A1) with $M_* > 10^7 M_\odot$. The photometry is derived from idealized data cubes without noise. The top, middle, and bottom rows show results for no dust, dust0-skirt-like, and dust0-S18K13 models, respectively. Bins are color-coded by median sSFR of their constituent galaxies. The black line indicates the quiescent selection box from \citet{Williams2009}. The inclusion of dust (middle and bottom) significantly improves the alignment of quiescent galaxies with empirical selection boundaries.}
\label{fig:uvj_tng50_dust0skirt}
\end{figure*}

We select galaxies with $M_* > 10^7 M_\odot$ from the TNG50-11-10 and TNG50-12-11 fields (viewing angle A1) and group them in four redshift bins ($z \approx 0.5, 1.5, 2.5, 3.5$) with a width of $\Delta z = \pm 0.4$. We then calculate photometry of each galaxy within a circular aperture of twice the half-stellar mass radius applied on idealized data cube without noise. Figure \ref{fig:uvj_tng50_dust0skirt} shows the resulting color-color distributions: the top row displays results without dust attenuation, while the middle and bottom rows show the results that implement the dust0-skirt and dust0-S18K13 models, respectively. The color coding represents the median sSFR of galaxies in the bins. At $z \approx 0.5$ and $1.5$, models including dust attenuation show a clear separation for quiescent galaxies ($\text{sSFR} \lesssim 10^{-10.5} \text{ yr}^{-1}$), consistent with the \citet{Williams2009} selection box. In contrast, dust-free models result in a more scattered vertical distribution, with many quiescent galaxies falling below the selection region. From the color-color distribution, we see a good consistency between the results obtained with the dust0-skirt-like and dust0-S18K13.

Finally, we provide versions of these mosaics that include realistic observational effects (Section~\ref{sec:obs_imaging}). These are designed to emulate JADES deep imaging \citep{Eisenstein2026}, specifically the \texttt{jw011800-deep} subregion. We adopted empirical photometric characteristics of this field, including PSF images, limiting magnitudes ($S/N=5$ at $0.15''$), and an exposure time of 87 hours \citep{Johnson2026}.  

\begin{figure*}[ht]
\centering
\includegraphics[width=0.45\textwidth]{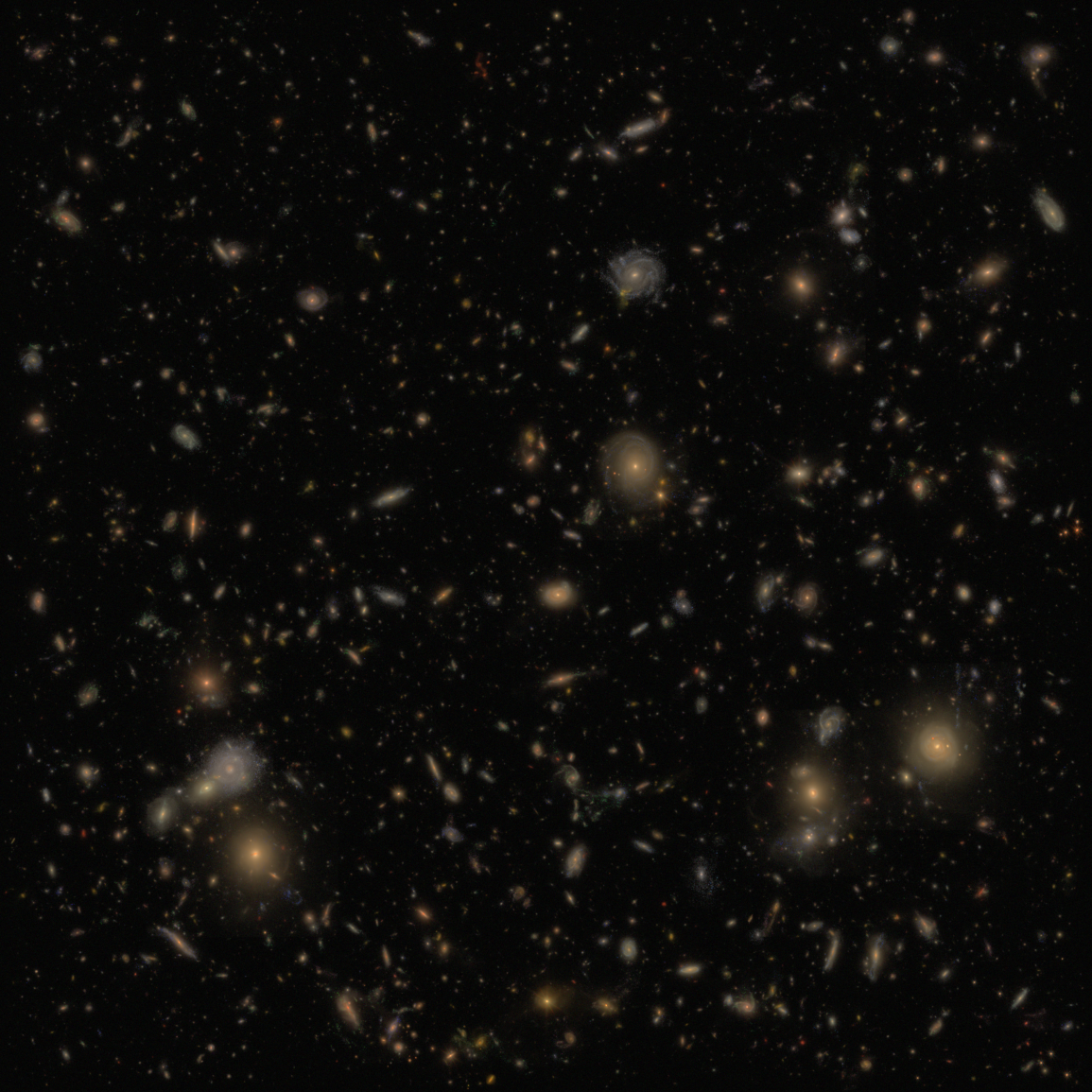}
\includegraphics[width=0.4\linewidth]{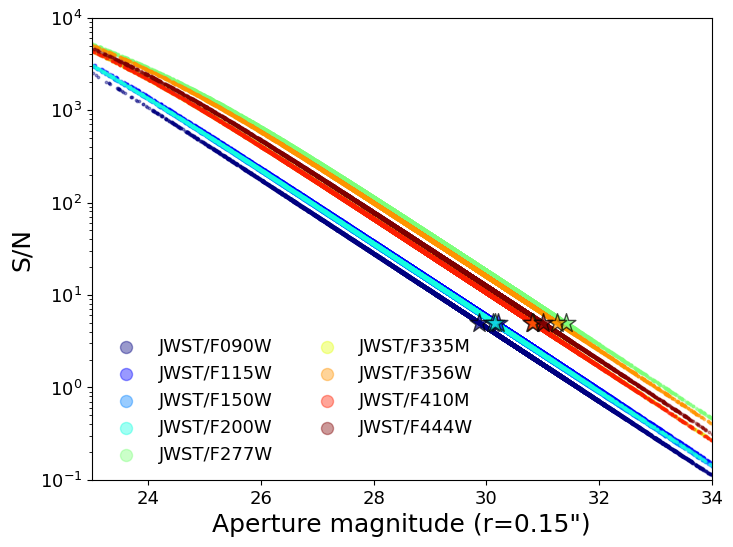}
\caption{\textit{Left}: Color composite of the TNG50-11-10 field with added observational effects emulating the JADES deep survey (jw011800-deep). \textit{Right}: S/N versus on-source magnitude within a $0.15''$ radius aperture. Star symbols denote the target empirical limiting magnitudes ($S/N=5$), which are precisely recovered by the \galsyn\ noise simulation.}
\label{fig:rgb_obsimg_tng50-11-10_8sqarcmin}
\end{figure*}

An example of a post-processed mosaic for TNG50-11-10 is shown in the left panel of Figure~\ref{fig:rgb_obsimg_tng50-11-10_8sqarcmin}. The right panel shows the correlation between the S/N and on-source magnitudes measured within a $0.15''$ radius aperture. The star symbols denote the empirical limiting magnitudes used as input parameters for \galsyn; as shown, these values are perfectly recovered in the simulated data. Further details regarding the JADES deep imaging characteristics and the fidelity with which \galsyn\ reproduces them are provided in Appendix~\ref{sec:determine_limit_mags}.

Given the wide range of potential observational depths, we do not include observational effects in the individual galaxy data cubes. Instead, we provide these effects only for the mosaics, specifically tailored to the characteristics of the JADES deep subregion. However, users can easily extend this analysis to other observational configurations by using function \texttt{GalSynMockObservation\_mosaic} in \texttt{observe} module.

\subsection{Progenitors of Massive Galaxies} \label{sec:progenitors_massive_gal}

The recent discovery of massive, mature (i.e.,~quiescent) galaxies at high redshifts ($z > 3$) by JWST has challenged existing galaxy formation models, raising fundamental questions about how such systems assembled substantial stellar mass so early in cosmic time \citep[e.g.,][]{Carnall2023, Haryana2025, Baker2025}. To facilitate a systematic study of these ``early bloomers'' and their evolutionary pathways, we provide synthetic data cubes for the progenitors of 290 massive galaxies identified in the TNG50 $z=0$ snapshot with stellar masses $\log(M_{*,z=0}/M_{\odot}) > 10.5$. By focusing on the $z=0$ descendants, this sample allows users to perform ``galaxy archaeology'' in reverse, tracing the structural and photometric properties of today's most massive systems back to their infancy in the early universe.

\begin{figure*}[ht]
\centering
\includegraphics[width=1.0\linewidth]{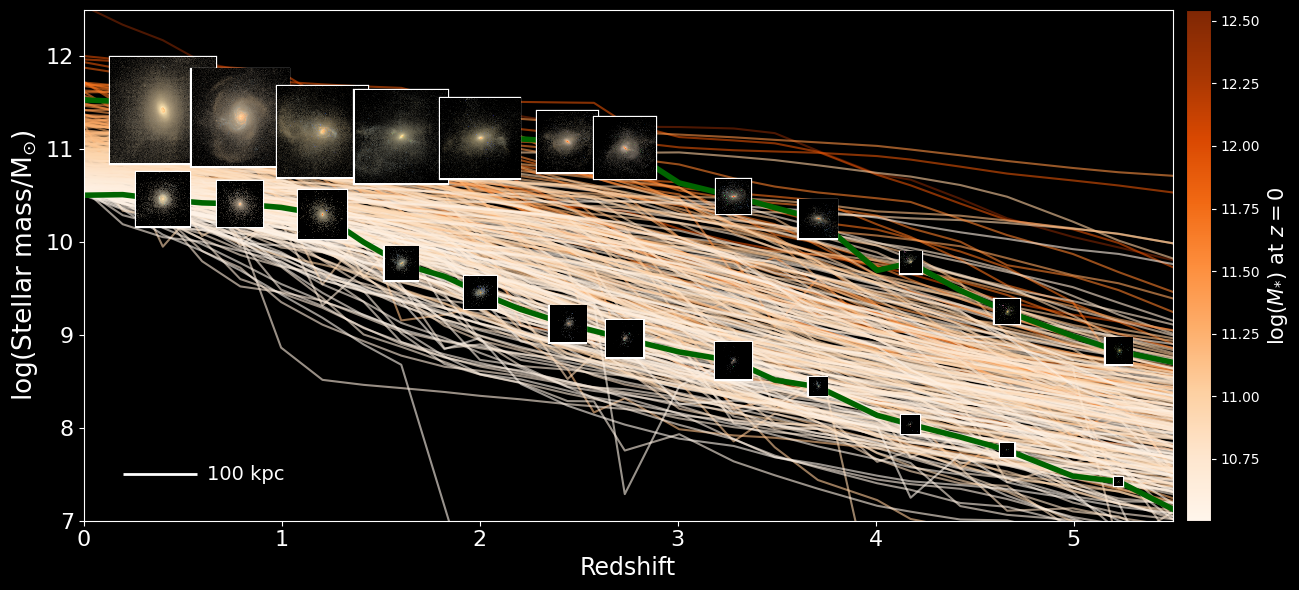}
\caption{Stellar mass assembly histories for the progenitors of 290 massive galaxies ($\log(M_{*,z=0}/M_{\odot}) > 10.5$) across the redshift range $0 < z < 5$. The tracks are color-coded by their $z=0$ descendant mass, illustrating the ``early-and-fast'' growth characteristic of the most massive systems compared to the more gradual assembly of lower-mass galaxies. Representative JWST color-composite stamps are overlaid to show the morphological and photometric evolution along two example tracks.}
\label{fig:plot_z_mass}
\end{figure*}

The progenitor histories are tracked using the \texttt{SubLink} algorithm \citep{Rodriguez-Gomez2015}, which connects subhalos across snapshots by mapping unique particle IDs to determine robust descendant-progenitor relationships. We track these lineages from the local universe out to $z=5$ at intervals of approximately $\Delta z \approx 0.2$. This fine temporal resolution results in 26 snapshots per lineage and a total sample of 7,540 individual galaxies. This systematic time-stepping is crucial for capturing short-lived evolutionary phases, such as rapid starbursts or AGN-driven quenching events, which might be missed in broader redshift samplings.

Figure \ref{fig:plot_z_mass} shows the stellar mass assembly histories (stellar mass vs. redshift) directly extracted from the TNG50 \texttt{SubLink} merger trees \citep{Rodriguez-Gomez2015}. We emphasize that the mass-growth tracks are intrinsic features of the TNG50 simulation rather than results of \galsyn's forward modeling. The primary contribution added by \galsyn\ is providing the spatially resolved synthetic observations of the progenitors along the evolutionary pathways.

Overlaid on the evolutionary tracks are representative postage stamps depicting the morphological evolution of two specific progenitors: one focusing on a lower-mass descendant $(\sim 10^{10.5} M_{\odot})$ and another reaching a final mass of $\sim 10^{11.5} M_{\odot}$. Qualitatively, the figure reveals a clear divergence in growth modes: the most massive systems tend to assemble a significant fraction of their stellar mass early on ($z \gtrsim 2.5$), followed by a phase of relatively modest accretion or quiescent evolution. In contrast, less massive systems exhibit a more continuous, gradual growth trend. This difference in mass assembly aligns with the ``Downsizing'' established by previous observational studies \citep[e.g.,][]{Cowie1996, Thomas2005, Perez-Gonzalez2008, Muzzin2013}.

These data cubes could serve as a useful laboratory for studying the spatially resolved evolution of massive galaxy progenitors. They enable us to move beyond integrated properties and examine the spatial distribution of star formation, the buildup of stellar disks and bulges, and the specific physical mechanisms, such as environmental effects or internal feedback, that lead to the cessation of star formation. Ultimately, this dataset provides the community with a self-consistent bridge between the high-redshift ``progenitors'' seen by JWST and the well-studied massive ``descendants'' in the local universe, offering a unique resource for testing the physics of galaxy evolution.

\subsection{Major-merger Systems} \label{sec:major_mergers}

Galaxy mergers are among the most transformative events in the cosmic life cycle, capable of triggering intense starbursts, fueling rapid supermassive black hole growth, and fundamentally reshaping galaxy morphology. To facilitate the study of these violent transitions, we have curated a specialized sample of major merger events ($M_{*,1}/M_{*,2} \geq 1/4$) occurring within the lineages of the massive galaxy progenitors described in Section \ref{sec:progenitors_massive_gal}. Drawing from the IllustrisTNG merger history catalogs \citep{Rodriguez-Gomez2017, Eisert2023}, we identified 259 unique major mergers spanning the redshift range $0 < z < 5$.

\begin{figure*}[ht]
\centering
\includegraphics[width=1\textwidth]{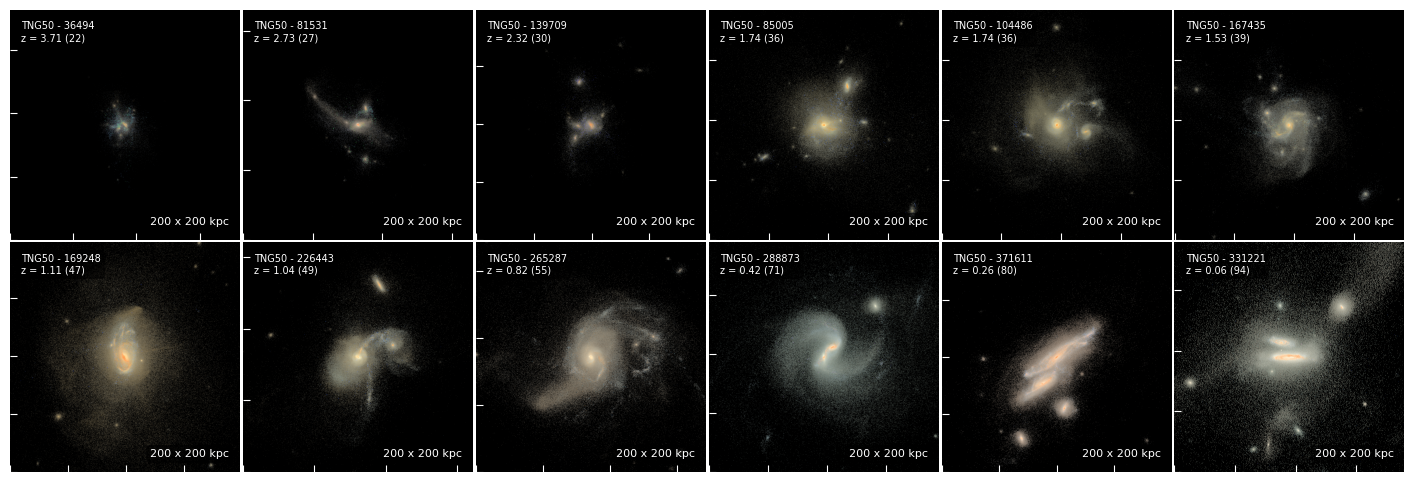}
\caption{A collage of color-composite stamps showcasing examples of major-merging systems. Unlike isolated subhalo cutouts, these images are generated from parent halo data to capture the full spatial extent of tidal features, bridges, and multiple interacting nuclei. The color composite images are created using the same JWST filter combinations as those used for Figure~\ref{fig:stamp_images_mosaics}.}
\label{fig:stamp_images_mergers}
\end{figure*}

Unlike the progenitor sample, which follows a rigid temporal grid, this dataset is event-driven; we extract snapshots that specifically capture the major-merger event, based on the merger catalogs. A key technical distinction in this release is that these synthetic data cubes are generated using parent halo cutouts rather than isolated subhalos. By including all particles associated with the parent Friends-of-Friends (FoF) group, we ensure that both interacting progenitors, as well as any tidal bridges, tails, or diffuse intra-halo light, are preserved in the synthetic observations. This is essential for capturing the complex environments that define merging systems. Figure~\ref{fig:stamp_images_mergers} shows color-composite stamps of some examples of major-merging systems.

Beyond the raw photometry, the data cube includes the physical property maps (see Section~\ref{sec:property_maps}). These sets provide a valuable resource for a variety of scientific analyses, such as developing and training machine learning algorithms for merger identification. Traditional methods often rely on simple morphological asymmetries or non-parametric statistics (e.g., Gini, $M_{20}$); however, our datasets provide a multi-wavelength training set that includes realistic synthetic images and spatially resolved physical property maps. These can be used for investigating how merger signatures vary across different filters. Furthermore, the spatially resolved SFHs provide mapping exactly where star formation is enhanced, whether it is concentrated in the nuclei due to gas inflows or triggered in the tidal tails. 

Finally, the inclusion of multiple viewing angles (A1–A4) for each merger event addresses the significant challenge of projection effects in observational studies. A system that appears as a late-stage merger from one perspective may look like a projected pair from another. By providing these multi-angle synthetic cubes, this data release enables a statistical assessment of how orientation influences the measured merger rates and the derived physical properties of interacting systems. This dataset thus serves as a comprehensive bridge between theoretical predictions of merger-driven evolution and the diverse morphologies observed in deep extragalactic fields. Currently, the data release focuses exclusively on the major mergers associated with massive galaxy progenitors. Future releases will expand this catalog to include mergers across a broader range of descendant masses, as well as minor merger events. 

\section{Summary and Outlooks} \label{sec:summary}

In this paper, we introduce \galsyn\ (Galaxy Synthesizer), a modular and flexible Python package for generating synthetic spectrophotometric observations from hydrodynamical galaxy simulations. \galsyn\ bridges the gap between numerical simulations and real observations by implementing an efficient particle-by-particle spectral modeling approach, which enables the rapid production of large synthetic datasets well-suited to population studies and statistical analyses of galaxies across a wide redshift range.

The core capabilities of \galsyn\ can be summarized as follows. The package accepts a simulation-agnostic, standardized HDF5 input format, enabling its application to a wide range of hydrodynamical simulations. It provides users with unprecedented control over the spectral synthesis process, including the choice of SPS engine (\texttt{FSPS} or \texttt{bagpipes}), stellar isochrone sets, spectral libraries, and initial mass functions. A spatially varying ionization parameter is computed self-consistently from the local gas properties of each pixel and is used to model the nebular emission from young stellar populations. \galsyn\ implements a decoupled kinematics model that independently Doppler-shifts the stellar continuum and nebular emission components, reproducing the observational signatures of distinct gas and stellar kinematic fields that are critical for realistic synthetic IFU data. Dust attenuation is modeled at the spatially resolved level through either a line-of-sight gas column density method or an SFR-surface-density-based approach. A comprehensive suite of dust attenuation laws is provided, including both fixed and adaptive prescriptions that allow the shape of the attenuation curve to vary with the local $A_V$. \galsyn\ also provides post-processing features to add observational realism, including PSF convolution, noise simulation, and spatial resampling. Beyond synthetic imaging and IFU data cubes, \galsyn\ reconstructs spatially resolved physical property maps and star formation histories, enabling direct spatially resolved comparison with observational data.

Together with this paper, we present the first public data release of synthetic galaxy observations generated from the IllustrisTNG simulation suites. This release comprises four mock extragalactic survey fields covering areas of 5, 8, 137, and 365 arcmin$^{2}$ drawn from TNG50-1, TNG100-1, and TNG300-1, respectively, synthetic data cubes for the progenitors of 290 local massive galaxies ($\log(M_{*,z=0}/M_{\odot}) > 10.5$) tracked from $z = 0$ to $5$, and 259 major-merger systems identified within those progenitor lineages. For each galaxy, the data products include synthetic imaging in 47 filters spanning HST, JWST, Euclid, Rubin/LSST, and the Roman Space Telescope, a comprehensive suite of spatially resolved physical property maps, and spatially resolved star formation histories. Datasets are provided at four distinct viewing angles and under three dust attenuation prescriptions.

Looking ahead, several extensions to \galsyn\ are planned for future work. First, modules for natively handling simulation data from EAGLE and other hydrodynamical simulations (such as COLIBRE, \citealt{Schaye2025}) will be added. Second, the current \texttt{GalSyn} framework models dust attenuation but defers dust emission; the incorporation of far-infrared and submillimeter emission from heated dust grains is a natural next step, which will extend the utility of \galsyn\ to long-wavelength observatories such as ALMA and \textit{Herschel}. Third, a forthcoming companion paper will present a systematic investigation of how different choices of dust attenuation law impact the observable photometric and spectroscopic properties of simulated galaxies, directly exploiting the flexibility of \texttt{GalSyn}'s modular design. Fourth, the first data release presented here focuses on IllustrisTNG; future releases will incorporate galaxies from complementary simulation suites to enable broader cross-simulation comparisons and to expand the available training sets for machine learning applications.

Finally, a particularly promising avenue for future development is the integration of \galsyn\ into parameter estimation and statistical inference frameworks. The computational efficiency of \galsyn\ makes it well-suited as a forward model within simulation-based inference (SBI) frameworks (also known as likelihood-free inference). In this framework, \galsyn\ would serve as the simulator, generating large ensembles of synthetic observations across a broad range of physical parameter combinations. It is possible to further accelerate the synthesis process in \galsyn\ by re-implementing key components of \galsyn\ using \texttt{JAX}. 


\begin{acknowledgments}

The authors acknowledge the Indiana University Pervasive Technology Institute (PTI) and UITS for providing supercomputing and visualization resources on Big Red 200 and Quartz that have contributed to the research results reported within this paper. The computations reported in this paper were (in part) performed using resources made available by the Flatiron Institute. The Flatiron Institute is a division of the Simons Foundation. The IllustrisTNG simulations were undertaken with compute time awarded by the Gauss Centre for Supercomputing (GCS) under GCS Large-Scale Projects GCS-ILLU and GCS-DWAR on the GCS share of the supercomputer Hazel Hen at the High Performance Computing Center Stuttgart (HLRS), as well as on the machines of the Max Planck Computing and Data Facility (MPCDF) in Garching, Germany.
\end{acknowledgments}

\begin{contribution}


All authors contributed to the work that resulted in this paper.


\end{contribution}

%


\software{FSPS \citep{Conroy2009,Conroy2010}, Python FSPS \citep{ben_johnson_2024_fsps}, Bagpipes \citep{Carnall2018}, piXedfit \citep{Abdurrouf2021,Abdurrouf2022_pixedfit_code}}


\appendix

\section{Robustness of SSP interpolation methods} \label{sec:ssp_interpolation_robustness}

The core of \texttt{GalSyn}'s efficiency lies in its ability to map physical properties from simulation particles to synthetic spectra using pre-computed SSP and nebular emission grids. As described in Sections~\ref{sec:ssp_grids} and ~\ref{sec:ssp_assignment}, these grids are discretized across age, metallicity, and ionization parameter. However, because simulation particles possess a continuous range of physical properties, the accuracy of the resulting synthetic spectra depends on the interpolation scheme used to traverse these grids. 

\begin{figure*}[ht]
\centering
\includegraphics[width=1.0\textwidth]{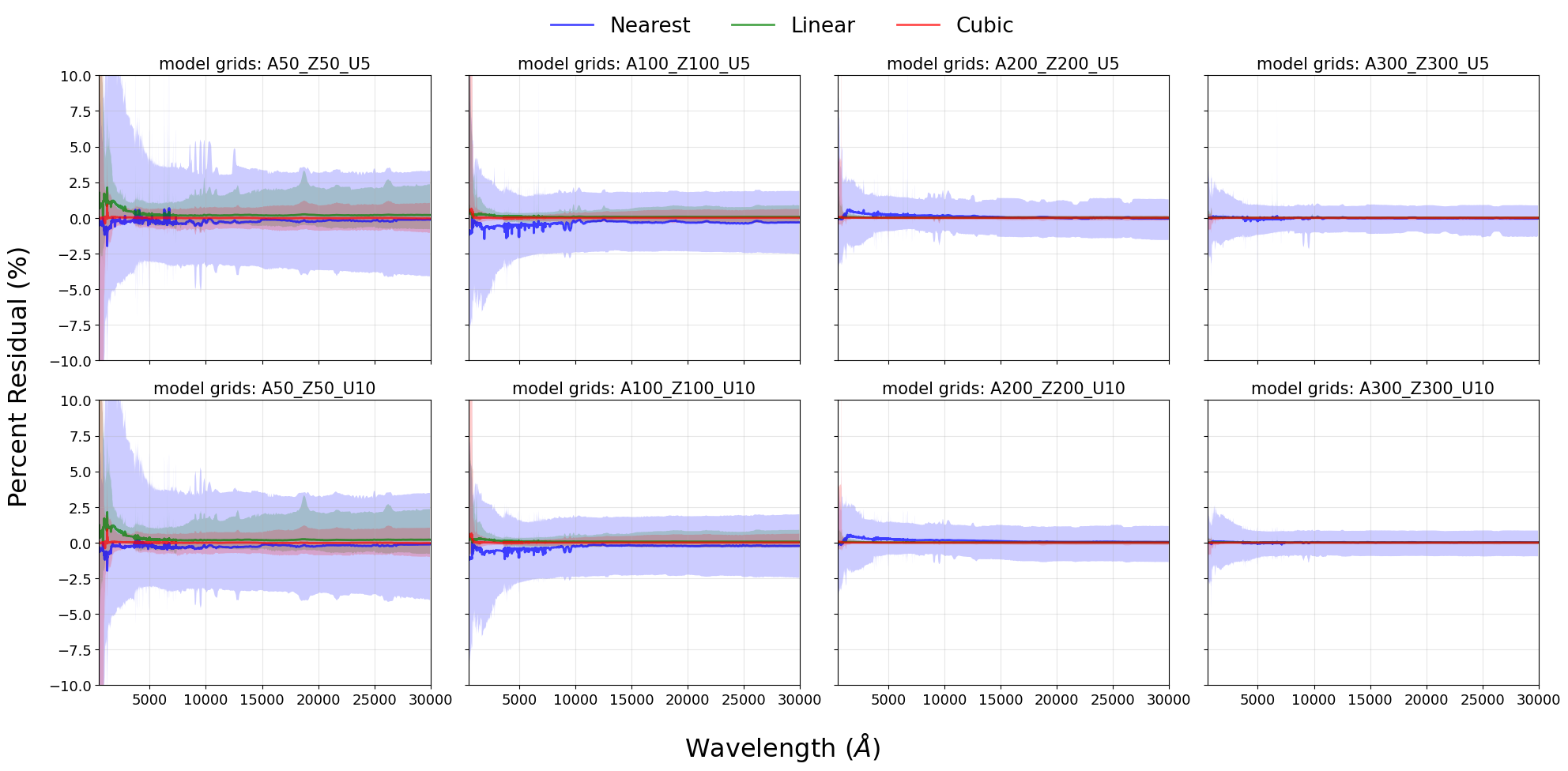}
\caption{Percentage residuals between interpolated SSP+nebular spectra and ground-truth models across various grid configurations. The top row utilizes 5 ionization parameter bins ($U5$), while the bottom row utilizes 10 ($U10$). Different columns represent grid densities in age and metallicity ($A$ and $Z$). Different colors represent interpolation methods. The comparison demonstrates that while cubic interpolation (red) is the most precise, linear interpolation (green) on a $100 \times 100 \times 10$ grid provides a high-fidelity recovery (residuals $\lesssim 1\%$) at a fraction of the computational cost.}
\label{fig:plot_interp_grids}
\end{figure*}

Here, we quantify the trade-offs between various interpolation methods and grid densities to provide users with a baseline for model reliability. To assess the robustness of our spectral synthesis, we experiment by comparing interpolated spectra against ``ground truth'' models (spectra generated directly by the SPS engine without interpolation). We tested three interpolation methods: nearest neighbor, linear, and cubic spline. These methods are applied across eight SSP grid configurations of increasing density, labeled by the number of bins in Age ($A$), Metallicity ($Z$), and Ionization ($U$): A50\_Z50\_U5, A50\_Z50\_U10, A100\_Z100\_U5, A100\_Z100\_U10, A200\_Z200\_U5, A200\_Z200\_U10, A300\_Z300\_U5, and A300\_Z300\_U10. The accuracy was measured via the percentage residual, defined as $100 \times (F_{\text{interp}} - F_{\text{truth}}) / F_{\text{truth}}$, across the full wavelength range.

The results of this comparison are summarized in Figure \ref{fig:plot_interp_grids}. The nearest neighbor method consistently shows the highest residuals ($\gtrsim 3\%$) across the whole wavelength range. Conversely, the linear and cubic methods significantly reduce these errors. For the lower-density A50\_Z50 grids, even higher-order interpolation struggles to capture sharp spectral features (e.g., the 4000\AA\ break or emission lines) accurately. However, once the grid density reaches A100\_Z100 or higher, the residuals drop below $\sim 1-2\%$ for both linear and cubic methods. Increasing the $U$ resolution from 5 to 10 bins (bottom panels vs. top panels) provides a marginal but noticeable improvement in the recovery of nebular-dominated spectra.

While the cubic method provides the most robust recovery with minimal residuals, it is the most computationally expensive, significantly increasing the processing time per galaxy. The nearest neighbor method, while fast, is insufficient for high-precision spectrophotometric analysis. We find that the linear interpolation method applied to the A100\_Z100\_U10 grid offers the optimal balance between accuracy and computational efficiency.

\section{Technical information on noise simulation} \label{sec:technical_info_noise_simulation}

In the noise simulation, \galsyn\ simulates both photon shot noise and background noise by mimicking known (i.e.,~empirical or expected) noise characteristics of an instrument. Here, we provide technical information about the noise simulation.  

For the synthetic imaging, noise is calculated for each filter based on user-defined observational magnitude limits at given S/N thresholds. The process converts the ideal, noise-free image into a realistic mock observation. The ideal flux of each pixel is first converted into $f_{\lambda}$ units of erg s$^{-1}$ cm$^{-2}$ \AA$^{-1}$ if it is not the original unit, and then converted into the expected number of photon counts that the detector would register. First, the flux per pixel is converted to an AB magnitude:
\begin{equation}
m_{\text{AB}} = -2.5 \cdot \log_{10}(f_{\nu}) - 48.6
\end{equation}
where $f_{\nu} = f_{\lambda} \cdot \frac{\lambda_{p}^2}{c}$ is the flux density in erg s$^{-1}$ cm$^{-2}$ Hz$^{-1}$. Then, the expected source counts are calculated using the instrumental zero-point magnitude ($m_{\text{zp}}$) and the exposure time ($t_{\text{exp}}$):
\begin{equation}
C_{\text{src}} = t_{\text{exp}} \, 10^{0.4 \, (m_{\text{zp}} - m_{\text{AB}})}.
\end{equation}

The background noise is derived from the specified limiting magnitude ($m_{\text{lim}}$) and the signal-to-noise ratio ($\text{SNR}_{\text{lim}}$) achieved at that limit within a given aperture. First, the total counts from a source at the limiting magnitude ($C_{\text{lim}}$) are calculated as:
\begin{equation}
C_{\text{lim}} = t_{\text{exp}} \cdot 10^{0.4 \cdot (m_{\text{zp}} - m_{\text{lim}})}
\end{equation}
The SNR is defined as $\text{SNR} = \frac{C_{\text{src}}}{\sqrt{C_{\text{src}} + C_{\text{bg}}}}$, where $C_{\text{bg}}$ is counts associated with the background. Rearranging to solve for the variance of the background counts ($\sigma_{\text{bg, apert}}^2 = C_{\text{bg}}$) within the aperture gives:
\begin{equation}
\sigma_{\text{bg, apert}}^2 = \left(\frac{C_{\text{lim}}}{\text{SNR}_{\text{lim}}}\right)^2 - C_{\text{lim}}
\end{equation}
The standard deviation of the background counts in the aperture is $\sigma_{\text{bg,apert}}$. This is then converted to a per-pixel standard deviation by dividing by the square root of the aperture area in pixels ($A_{\text{apert,pix}}$):
\begin{equation}
\sigma_{\text{bg,pix}} = \frac{\sigma_{\text{bg,apert}}}{\sqrt{A_{\text{apert,pix}}}}
\end{equation}

The final noisy image is created by combining the simulated noise components. The source noise is simulated by drawing from a Poisson distribution with the expected source counts as the mean:
\begin{equation}
N_{\text{src}} \sim \text{Poisson}(\lambda = C_{\text{src}}),
\end{equation}
whereas the background noise is simulated by drawing from a Gaussian (i.e.,~normal) distribution:
\begin{equation}
N_{\text{bg}} \sim \mathcal{N}(\mu=0, \sigma=\sigma_{\text{bg,pix}}).
\end{equation}
The final noisy image, in counts, is then the sum of these two components:
\begin{equation}
C_{\text{noisy}} = N_{\text{src}} + N_{\text{bg}}.
\end{equation}
This count value is then converted back into physical flux units. The corresponding error (i.e., Root Mean Square, RMS) map is calculated by adding the variances in quadrature, where the variance of the source's shot noise is equal to the expected source counts:
\begin{equation}
\sigma_{\text{total,pix}} = \sqrt{\sigma_{\text{src}}^2 + \sigma_{\text{bg,pix}}^2} = \sqrt{C_{\text{src}} + \sigma_{\text{bg,pix}}^2}
\end{equation}
This RMS value in counts is also converted back to flux units to provide the final error map.

\section{Emulating empirical magnitude limits of an imaging observation} \label{sec:determine_limit_mags}

\begin{figure*}[ht]
\centering
\includegraphics[width=1.0\textwidth]{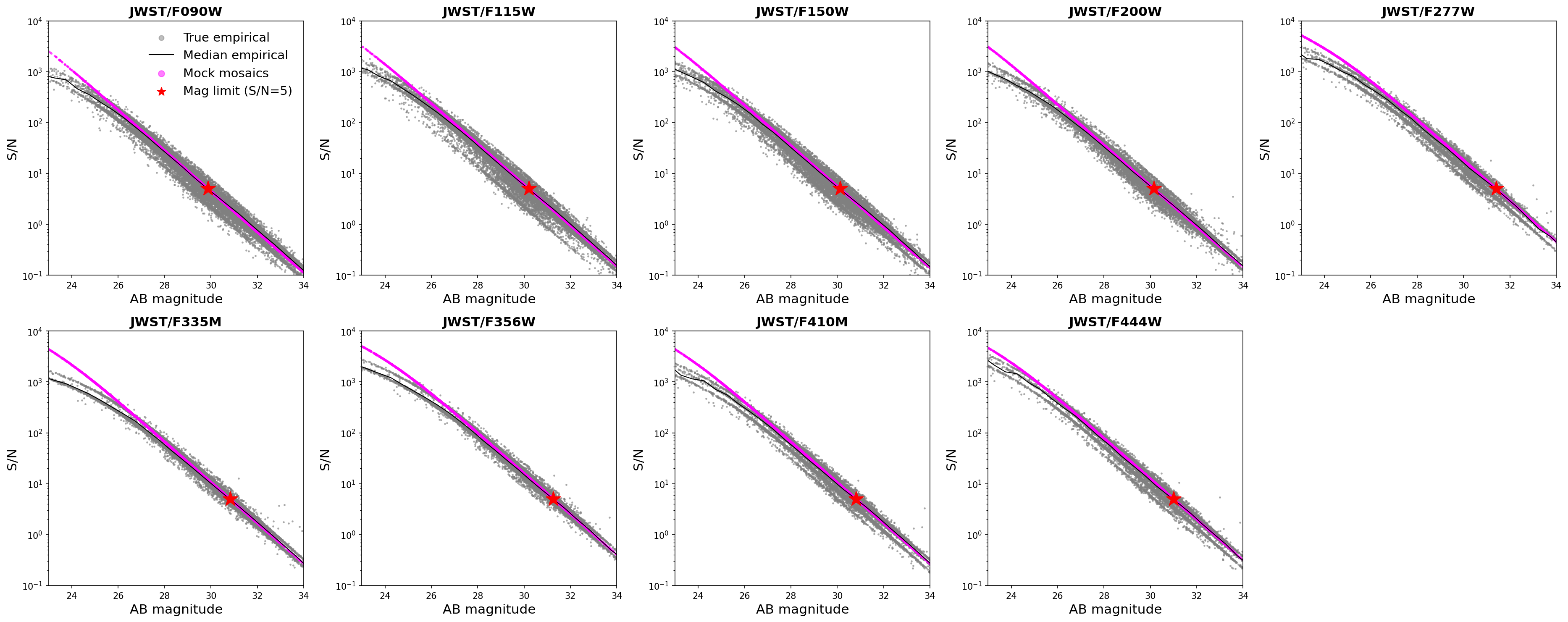}
\caption{Comparison between the empirical and simulated S/N as a function of AB magnitude across nine NIRCam filters. The grey data points represent the empirical relation derived from the JADES deep imaging observation of jw011800-deep field (87-hour exposure). The black line denotes the median relation. The magenta data points represent the S/N relation recovered by \galsyn. The red star symbols indicate the empirical $5-\sigma$ limiting magnitude. \galsyn\ shows high fidelity in emulating the $5\sigma$ limiting magnitudes across all bands, with a slight overestimation of S/N for the brightest sources.}
\label{fig:tng50-11-10_mag_snr_9panels}
\end{figure*}

We validate \texttt{GalSyn}'s noise simulation framework against empirical data from the JADES survey. Specifically, we utilize a subregion of the JADES deep imaging, jw011800-deep field, with a total exposure time of approximately 87 hours \citep{Johnson2026}. This deep baseline allows us to test the accuracy of our noise injection over a wide dynamic range of magnitudes across nine NIRCam filters.

The noise simulation is designed to match the $5\sigma$ limiting magnitudes of the reference observation. Using the empirical limits (e.g., $m_{\text{lim}} \sim 29.88$ in F090W and $m_{\text{lim}} \sim 31.00$ in F444W), we inject noise into the TNG50-11-10 mosaic (see Figure~\ref{fig:rgb_obsimg_tng50-11-10_8sqarcmin}) to recover the S/N vs. magnitude relation. As shown in Figure~\ref{fig:tng50-11-10_mag_snr_9panels}, \galsyn\ effectively recovers the median S/N-magnitude relation and the $5\sigma$ depth across all filters, confirming that the tool can reliably emulate the sensitivity of imaging surveys.

We note a slight discrepancy at the brightest end, where \galsyn\ tends to overestimate the S/N compared to the empirical JADES data. This offset likely stems from imperfect treatment of pixel-level saturation and non-linear noise components in the observational data, which become more prominent for bright sources and are difficult to reproduce with our current noise model. However, for the bulk of the galaxy population near the detection limits, which is the primary focus for high-redshift studies, the agreement is excellent, demonstrating the robustness of the noise simulation.

\section{Estimating magnitude limits of NIRSpec IFU with JWST ETC}
\label{sec:mag_lims_nirspec_ifu}

To simulate realistic observational noise in synthetic JWST/NIRSpec IFU datacubes, we conducted systematic S/N simulations using the \texttt{Pandeia engine} \citep{Pontoppidan2016}, the official exposure time calculator for JWST. We modeled a ``flat'' input spectrum (i.e.,~constant in AB magnitude) across a grid of source magnitudes ranging from 18 to 28 and for effective exposure times of 10, 20, and 40 ks. This experiment was carried out for nine primary instrument configurations: the high-resolution gratings (G140H/F070LP, G140H/F100LP, G235H/F170LP, and G395H/F290LP), the medium-resolution gratings (G140M/F070LP, G140M/F100LP, G235M/F170LP, and G395M/F290LP), and the low-resolution Prism/CLEAR mode. By calculating the S/N per pixel on the native detector grid, our model accounts for critical wavelength-dependent factors including instrument throughput, thermal background, and detector noise.

\begin{figure*}[ht]
\centering
\includegraphics[width=0.47\textwidth]{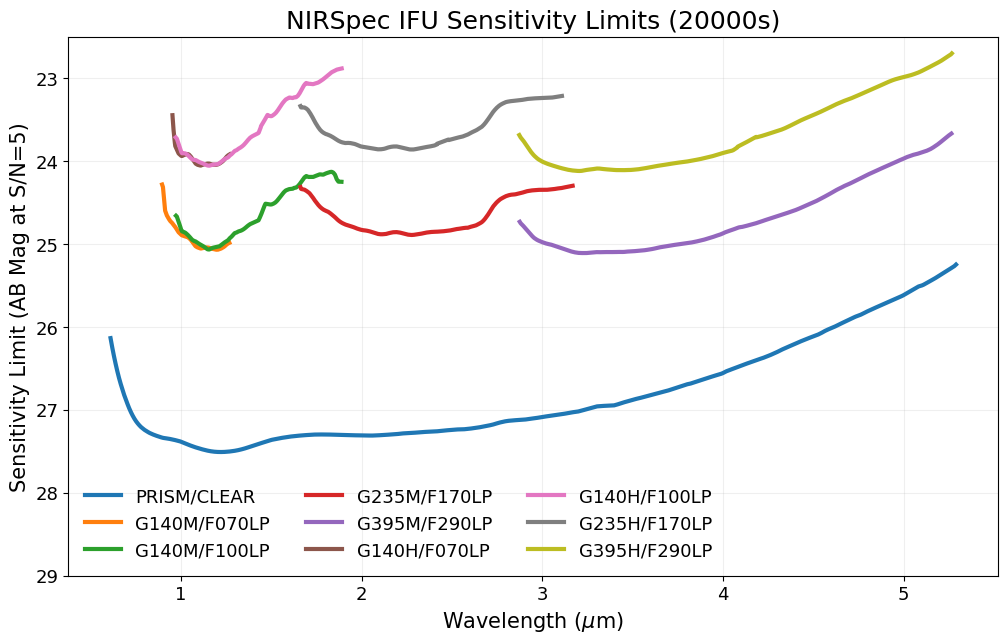}
\includegraphics[width=0.47\textwidth]{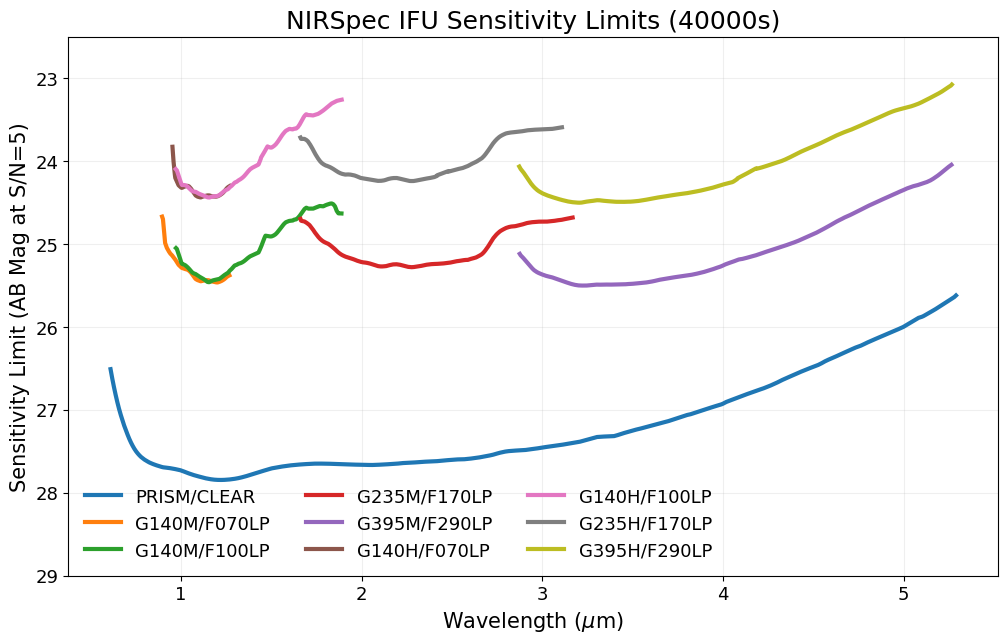}
\caption{Estimated $5\sigma$ limiting magnitudes for JWST/NIRSpec IFU observations as a function of wavelength, derived from the \texttt{Pandeia} ETC engine \citep{Pontoppidan2016}. The left and right panels display the sensitivity limits for exposure times of 20 ks and 40 ks, respectively. The colored lines indicate the performance of the nine primary instrument configurations: the high-resolution gratings (G140H/F070LP, G140H/F100LP, G235H/F170LP, and G395H/F290LP), the medium-resolution gratings (G140M/F070LP, G140M/F100LP, G235M/F170LP, and G395M/F290LP), and the low-resolution Prism/CLEAR mode.}
\label{fig:nirspec_ifu_sensitivity}
\end{figure*}

The resulting $5\sigma$ sensitivity limits, illustrated in Figure~\ref{fig:nirspec_ifu_sensitivity}, reveal the depth achievable across the NIRSpec wavelength range. For a 40 ks exposure, the medium-resolution gratings typically reach limits between 24 and 26 mag. In contrast, the low-resolution Prism mode, even with a shorter 20 ks exposure, achieves significantly greater depth—exceeding 27 mag, due to its lower spectral resolution, spreading the signal over fewer pixels.

These calculated limits are integrated into the \galsyn\ post-processing suite to ensure observational realism in our mock datasets. We interpolate the magnitude–S/N relationship at every wavelength point to derive a continuous, wavelength-dependent sensitivity curve. This information is then passed to the \galsyn\ noise simulation module (Section~\ref{sec:obs_ifu}). By scaling the noise variance according to the interpolated \texttt{Pandeia} results, we produce synthetic IFU observations that are statistically consistent with the expected performance of JWST. An example of a post-processed result is shown in Figure~\ref{fig:plot_mockobs_ifu}. 

\bibliography{galsyn}{}

\begin{thebibliography}{}
\expandafter\ifx\csname natexlab\endcsname\relax\def\natexlab#1{#1}\fi
\providecommand{\url}[1]{\href{#1}{#1}}
\providecommand{\dodoi}[1]{doi:~\href{http://doi.org/#1}{\nolinkurl{#1}}}
\providecommand{\doeprint}[1]{\href{http://ascl.net/#1}{\nolinkurl{http://ascl.net/#1}}}
\providecommand{\doarXiv}[1]{\href{https://arxiv.org/abs/#1}{\nolinkurl{https://arxiv.org/abs/#1}}}

\bibitem[{ {Abdurro'uf} {et~al.}(2021){Abdurro'uf}, {Lin}, {Wu}, \& {Akiyama}}]{Abdurrouf2021}
{Abdurro'uf}, {Lin}, Y.-T., {Wu}, P.-F., \& {Akiyama}, M. 2021, \bibinfo{title}{{Introducing piXedfit: A Spectral Energy Distribution Fitting Code Designed for Resolved Sources},} \apjs, 254, 15, \dodoi{10.3847/1538-4365/abebe2}

\bibitem[{ {Abdurro'uf} {et~al.}(2022){Abdurro'uf}, {Lin}, {Wu}, \& {Akiyama}}]{Abdurrouf2022_pixedfit_code}
{Abdurro'uf}, {Lin}, Y.-T., {Wu}, P.-F., \& {Akiyama}, M. 2022, {piXedfit: Analyze spatially resolved SEDs of galaxies},, Astrophysics Source Code Library, record ascl:2207.033

\bibitem[{ {Astropy Collaboration} {et~al.}(2013){Astropy Collaboration}, {Robitaille}, {Tollerud}, {Greenfield}, {Droettboom}, {Bray}, {Aldcroft}, {Davis}, {Ginsburg}, {Price-Whelan}, {Kerzendorf}, {Conley}, {Crighton}, {Barbary}, {Muna}, {Ferguson}, {Grollier}, {Parikh}, {Nair}, {Unther}, {Deil}, {Woillez}, {Conseil}, {Kramer}, {Turner}, {Singer}, {Fox}, {Weaver}, {Zabalza}, {Edwards}, {Azalee Bostroem}, {Burke}, {Casey}, {Crawford}, {Dencheva}, {Ely}, {Jenness}, {Labrie}, {Lim}, {Pierfederici}, {Pontzen}, {Ptak}, {Refsdal}, {Servillat}, \& {Streicher}}]{Astropy2013}
{Astropy Collaboration}, {Robitaille}, T.~P., {Tollerud}, E.~J., {et~al.} 2013, \bibinfo{title}{{Astropy: A community Python package for astronomy},} \aap, 558, A33, \dodoi{10.1051/0004-6361/201322068}

\bibitem[{ {Astropy Collaboration} {et~al.}(2018){Astropy Collaboration}, {Price-Whelan}, {Sip{\H{o}}cz}, {G{\"u}nther}, {Lim}, {Crawford}, {Conseil}, {Shupe}, {Craig}, {Dencheva}, {Ginsburg}, {VanderPlas}, {Bradley}, {P{\'e}rez-Su{\'a}rez}, {de Val-Borro}, {Aldcroft}, {Cruz}, {Robitaille}, {Tollerud}, {Ardelean}, {Babej}, {Bach}, {Bachetti}, {Bakanov}, {Bamford}, {Barentsen}, {Barmby}, {Baumbach}, {Berry}, {Biscani}, {Boquien}, {Bostroem}, {Bouma}, {Brammer}, {Bray}, {Breytenbach}, {Buddelmeijer}, {Burke}, {Calderone}, {Cano Rodr{\'\i}guez}, {Cara}, {Cardoso}, {Cheedella}, {Copin}, {Corrales}, {Crichton}, {D'Avella}, {Deil}, {Depagne}, {Dietrich}, {Donath}, {Droettboom}, {Earl}, {Erben}, {Fabbro}, {Ferreira}, {Finethy}, {Fox}, {Garrison}, {Gibbons}, {Goldstein}, {Gommers}, {Greco}, {Greenfield}, {Groener}, {Grollier}, {Hagen}, {Hirst}, {Homeier}, {Horton}, {Hosseinzadeh}, {Hu}, {Hunkeler}, {Ivezi{\'c}}, {Jain}, {Jenness}, {Kanarek}, {Kendrew}, {Kern}, {Kerzendorf}, {Khvalko}, {King}, {Kirkby}, {Kulkarni},
  {Kumar}, {Lee}, {Lenz}, {Littlefair}, {Ma}, {Macleod}, {Mastropietro}, {McCully}, {Montagnac}, {Morris}, {Mueller}, {Mumford}, {Muna}, {Murphy}, {Nelson}, {Nguyen}, {Ninan}, {N{\"o}the}, {Ogaz}, {Oh}, {Parejko}, {Parley}, {Pascual}, {Patil}, {Patil}, {Plunkett}, {Prochaska}, {Rastogi}, {Reddy Janga}, {Sabater}, {Sakurikar}, {Seifert}, {Sherbert}, {Sherwood-Taylor}, {Shih}, {Sick}, {Silbiger}, {Singanamalla}, {Singer}, {Sladen}, {Sooley}, {Sornarajah}, {Streicher}, {Teuben}, {Thomas}, {Tremblay}, {Turner}, {Terr{\'o}n}, {van Kerkwijk}, {de la Vega}, {Watkins}, {Weaver}, {Whitmore}, {Woillez}, {Zabalza}, \& {Astropy Contributors}}]{Astropy2018}
{Astropy Collaboration}, {Price-Whelan}, A.~M., {Sip{\H{o}}cz}, B.~M., {et~al.} 2018, \bibinfo{title}{{The Astropy Project: Building an Open-science Project and Status of the v2.0 Core Package},} \aj, 156, 123, \dodoi{10.3847/1538-3881/aabc4f}

\bibitem[{W.~M. {Baker} {et~al.}(2025){Baker}, {Valentino}, {Lagos}, {Ito}, {Jespersen}, {Gottumukkala}, {Hjorth}, {Langeroodi}, \& {Sedgewick}}]{Baker2025}
{Baker}, W.~M., {Valentino}, F., {Lagos}, C. d.~P., {et~al.} 2025, \bibinfo{title}{{Exploring over 700 massive quiescent galaxies at z = 2─7: Demographics and stellar mass functions},} \aap, 702, A270, \dodoi{10.1051/0004-6361/202555829}

\bibitem[{A.~J. {Battisti} {et~al.}(2020){Battisti}, {Cunha}, {Shivaei}, \& {Calzetti}}]{Battisti2020}
{Battisti}, A.~J., {Cunha}, E.~d., {Shivaei}, I., \& {Calzetti}, D. 2020, \bibinfo{title}{{The Strength of the 2175 {\r{A}} Feature in the Attenuation Curves of Galaxies at 0.1 < z {\ensuremath{\lesssim}} 3},} \apj, 888, 108, \dodoi{10.3847/1538-4357/ab5fdd}

\bibitem[{A. {Bressan} {et~al.}(2012){Bressan}, {Marigo}, {Girardi}, {Salasnich}, {Dal Cero}, {Rubele}, \& {Nanni}}]{Bressan2012}
{Bressan}, A., {Marigo}, P., {Girardi}, L., {et~al.} 2012, \bibinfo{title}{{PARSEC: stellar tracks and isochrones with the PAdova and TRieste Stellar Evolution Code},} \mnras, 427, 127, \dodoi{10.1111/j.1365-2966.2012.21948.x}

\bibitem[{G. {Bruzual} \& S. {Charlot}(2003){Bruzual} \& {Charlot}}]{Bruzual2003}
{Bruzual}, G., \& {Charlot}, S. 2003, \bibinfo{title}{{Stellar population synthesis at the resolution of 2003},} \mnras, 344, 1000, \dodoi{10.1046/j.1365-8711.2003.06897.x}

\bibitem[{D. {Calzetti} {et~al.}(2000){Calzetti}, {Armus}, {Bohlin}, {Kinney}, {Koornneef}, \& {Storchi-Bergmann}}]{Calzetti2000}
{Calzetti}, D., {Armus}, L., {Bohlin}, R.~C., {et~al.} 2000, \bibinfo{title}{{The Dust Content and Opacity of Actively Star-forming Galaxies},} \apj, 533, 682, \dodoi{10.1086/308692}

\bibitem[{D. {Calzetti} {et~al.}(1994){Calzetti}, {Kinney}, \& {Storchi-Bergmann}}]{Calzetti1994}
{Calzetti}, D., {Kinney}, A.~L., \& {Storchi-Bergmann}, T. 1994, \bibinfo{title}{{Dust Extinction of the Stellar Continua in Starburst Galaxies: The Ultraviolet and Optical Extinction Law},} \apj, 429, 582, \dodoi{10.1086/174346}

\bibitem[{P. {Camps} \& M. {Baes}(2015){Camps} \& {Baes}}]{Camps2015}
{Camps}, P., \& {Baes}, M. 2015, \bibinfo{title}{{SKIRT: An advanced dust radiative transfer code with a user-friendly architecture},} Astronomy and Computing, 9, 20, \dodoi{10.1016/j.ascom.2014.10.004}

\bibitem[{P. {Camps} \& M. {Baes}(2020){Camps} \& {Baes}}]{Camps2020}
{Camps}, P., \& {Baes}, M. 2020, \bibinfo{title}{{SKIRT 9: Redesigning an advanced dust radiative transfer code to allow kinematics, line transfer and polarization by aligned dust grains},} Astronomy and Computing, 31, 100381, \dodoi{10.1016/j.ascom.2020.100381}

\bibitem[{J.~A. {Cardelli} {et~al.}(1989){Cardelli}, {Clayton}, \& {Mathis}}]{Cardelli1989}
{Cardelli}, J.~A., {Clayton}, G.~C., \& {Mathis}, J.~S. 1989, \bibinfo{title}{{The Relationship between Infrared, Optical, and Ultraviolet Extinction},} \apj, 345, 245, \dodoi{10.1086/167900}

\bibitem[{A.~C. {Carnall} {et~al.}(2018){Carnall}, {McLure}, {Dunlop}, \& {Dav{\'e}}}]{Carnall2018}
{Carnall}, A.~C., {McLure}, R.~J., {Dunlop}, J.~S., \& {Dav{\'e}}, R. 2018, \bibinfo{title}{{Inferring the star formation histories of massive quiescent galaxies with BAGPIPES: evidence for multiple quenching mechanisms},} \mnras, 480, 4379, \dodoi{10.1093/mnras/sty2169}

\bibitem[{A.~C. {Carnall} {et~al.}(2023){Carnall}, {McLure}, {Dunlop}, {McLeod}, {Wild}, {Cullen}, {Magee}, {Begley}, {Cimatti}, {Donnan}, {Hamadouche}, {Jewell}, \& {Walker}}]{Carnall2023}
{Carnall}, A.~C., {McLure}, R.~J., {Dunlop}, J.~S., {et~al.} 2023, \bibinfo{title}{{A massive quiescent galaxy at redshift 4.658},} \nat, 619, 716, \dodoi{10.1038/s41586-023-06158-6}

\bibitem[{G. {Chabrier}(2003){Chabrier}}]{2003Chabrier}
{Chabrier}, G. 2003, \bibinfo{title}{{Galactic Stellar and Substellar Initial Mass Function},} \pasp, 115, 763, \dodoi{10.1086/376392}

\bibitem[{J. {Chevallard} {et~al.}(2013){Chevallard}, {Charlot}, {Wandelt}, \& {Wild}}]{Chevallard2013}
{Chevallard}, J., {Charlot}, S., {Wandelt}, B., \& {Wild}, V. 2013, \bibinfo{title}{{Insights into the content and spatial distribution of dust from the integrated spectral properties of galaxies},} \mnras, 432, 2061, \dodoi{10.1093/mnras/stt523}

\bibitem[{J. {Choi} {et~al.}(2016){Choi}, {Dotter}, {Conroy}, {Cantiello}, {Paxton}, \& {Johnson}}]{Choi2016}
{Choi}, J., {Dotter}, A., {Conroy}, C., {et~al.} 2016, \bibinfo{title}{{Mesa Isochrones and Stellar Tracks (MIST). I. Solar-scaled Models},} \apj, 823, 102, \dodoi{10.3847/0004-637X/823/2/102}

\bibitem[{G.~C. {Clayton} \& P.~G. {Martin}(1985){Clayton} \& {Martin}}]{Clayton1985}
{Clayton}, G.~C., \& {Martin}, P.~G. 1985, \bibinfo{title}{{Interstellar dust in the Large Magellanic Cloud.},} \apj, 288, 558, \dodoi{10.1086/162821}

\bibitem[{C. {Conroy} \& J.~E. {Gunn}(2010){Conroy} \& {Gunn}}]{Conroy2010}
{Conroy}, C., \& {Gunn}, J.~E. 2010, \bibinfo{title}{{The Propagation of Uncertainties in Stellar Population Synthesis Modeling. III. Model Calibration, Comparison, and Evaluation},} \apj, 712, 833, \dodoi{10.1088/0004-637X/712/2/833}

\bibitem[{C. {Conroy} {et~al.}(2009){Conroy}, {Gunn}, \& {White}}]{Conroy2009}
{Conroy}, C., {Gunn}, J.~E., \& {White}, M. 2009, \bibinfo{title}{{The Propagation of Uncertainties in Stellar Population Synthesis Modeling. I. The Relevance of Uncertain Aspects of Stellar Evolution and the Initial Mass Function to the Derived Physical Properties of Galaxies},} \apj, 699, 486, \dodoi{10.1088/0004-637X/699/1/486}

\bibitem[{L.~L. {Cowie} {et~al.}(1996){Cowie}, {Songaila}, {Hu}, \& {Cohen}}]{Cowie1996}
{Cowie}, L.~L., {Songaila}, A., {Hu}, E.~M., \& {Cohen}, J.~G. 1996, \bibinfo{title}{{New Insight on Galaxy Formation and Evolution From Keck Spectroscopy of the Hawaii Deep Fields},} \aj, 112, 839, \dodoi{10.1086/118058}

\bibitem[{R. {Dav{\'e}}(2008){Dav{\'e}}}]{Dave2008}
{Dav{\'e}}, R. 2008, \bibinfo{title}{{The galaxy stellar mass-star formation rate relation: evidence for an evolving stellar initial mass function?},} \mnras, 385, 147, \dodoi{10.1111/j.1365-2966.2008.12866.x}

\bibitem[{D.~J. {Eisenstein} {et~al.}(2026){Eisenstein}, {Willott}, {Alberts}, {Arribas}, {Bonaventura}, {Bunker}, {Cameron}, {Carniani}, {Charlot}, {Curtis-Lake}, {D'Eugenio}, {Ferruit}, {Giardino}, {Hainline}, {Hausen}, {Jakobsen}, {Johnson}, {Maiolino}, {Rauscher}, {Rieke}, {Rieke}, {Rix}, {Robertson}, {Stark}, {Tacchella}, {Williams}, {Willmer}, {Baker}, {Baum}, {Bhatawdekar}, {Boyett}, {Chen}, {Chevallard}, {Circosta}, {Curti}, {Danhaive}, {DeCoursey}, {Endsley}, {de Graaff}, {Dressler}, {Egami}, {Helton}, {Hviding}, {Ji}, {Jones}, {Kumari}, {L{\"u}tzgendorf}, {Laseter}, {Looser}, {Lyu}, {Maseda}, {Nelson}, {Parlanti}, {Perna}, {Pusk{\'a}s}, {Rawle}, {Rodr{\'\i}guez Del Pino}, {Rujopakarn}, {Sandles}, {Saxena}, {Scholtz}, {Sharpe}, {Shivaei}, {Silcock}, {Simmonds}, {Skarbinski}, {Smit}, {Stone}, {Suess}, {Sun}, {Tang}, {Topping}, {{\"U}bler}, {Villanueva}, {Wallace}, {Whitler}, {Witstok}, \& {Woodrum}}]{Eisenstein2026}
{Eisenstein}, D.~J., {Willott}, C., {Alberts}, S., {et~al.} 2026, \bibinfo{title}{{Overview of the JWST Advanced Deep Extragalactic Survey (JADES)},} \apjs, 283, 6, \dodoi{10.3847/1538-4365/ae3163}

\bibitem[{L. {Eisert} {et~al.}(2023){Eisert}, {Pillepich}, {Nelson}, {Klessen}, {Huertas-Company}, \& {Rodriguez-Gomez}}]{Eisert2023}
{Eisert}, L., {Pillepich}, A., {Nelson}, D., {et~al.} 2023, \bibinfo{title}{{ERGO-ML I: inferring the assembly histories of IllustrisTNG galaxies from integral observable properties via invertible neural networks},} \mnras, 519, 2199, \dodoi{10.1093/mnras/stac3295}

\bibitem[{J.~J. {Eldridge} {et~al.}(2017){Eldridge}, {Stanway}, {Xiao}, {McClelland}, {Taylor}, {Ng}, {Greis}, \& {Bray}}]{Eldridge2017}
{Eldridge}, J.~J., {Stanway}, E.~R., {Xiao}, L., {et~al.} 2017, \bibinfo{title}{{Binary Population and Spectral Synthesis Version 2.1: Construction, Observational Verification, and New Results},} \pasa, 34, e058, \dodoi{10.1017/pasa.2017.51}

\bibitem[{Y. {Euclid Collaboration: Mellier} {et~al.}(2025){Euclid Collaboration: Mellier}, {Abdurro'uf}, {Acevedo~Barroso}, {et~al.}}]{EuclidSkyOverview}
{Euclid Collaboration: Mellier}, Y., {Abdurro'uf}, {Acevedo~Barroso}, J., {et~al.} 2025, \bibinfo{title}{Euclid - I. Overview of the Euclid mission,} A\&A, 697, A1, \dodoi{10.1051/0004-6361/202450810}

\bibitem[{E. {Euclid Collaboration: Merlin} {et~al.}(2023){Euclid Collaboration: Merlin}, {Castellano}, {Bretonni{\`e}re}, {et~al.}}]{Merlin2023_euclid}
{Euclid Collaboration: Merlin}, E., {Castellano}, M., {Bretonni{\`e}re}, H., {et~al.} 2023, \bibinfo{title}{{Euclid preparation. XXV. The Euclid Morphology Challenge: Towards model-fitting photometry for billions of galaxies},} \aap, 671, A101, \dodoi{10.1051/0004-6361/202245041}

\bibitem[{N. {Faucher} {et~al.}(2023){Faucher}, {Blanton}, \& {Macci{\`o}}}]{Faucher2023}
{Faucher}, N., {Blanton}, M.~R., \& {Macci{\`o}}, A.~V. 2023, \bibinfo{title}{{Panchromatic Simulated Galaxy Observations from the NIHAO Project},} \apj, 957, 7, \dodoi{10.3847/1538-4357/acf9f0}

\bibitem[{G.~J. {Ferland} {et~al.}(1998){Ferland}, {Korista}, {Verner}, {Ferguson}, {Kingdon}, \& {Verner}}]{Ferland1998}
{Ferland}, G.~J., {Korista}, K.~T., {Verner}, D.~A., {et~al.} 1998, \bibinfo{title}{{CLOUDY 90: Numerical Simulation of Plasmas and Their Spectra},} \pasp, 110, 761, \dodoi{10.1086/316190}

\bibitem[{G.~J. {Ferland} {et~al.}(2013){Ferland}, {Porter}, {van Hoof}, {Williams}, {Abel}, {Lykins}, {Shaw}, {Henney}, \& {Stancil}}]{Ferland2013}
{Ferland}, G.~J., {Porter}, R.~L., {van Hoof}, P.~A.~M., {et~al.} 2013, \bibinfo{title}{{The 2013 Release of Cloudy},} \rmxaa, 49, 137.
\newblock \doarXiv{1302.4485}

\bibitem[{E.~L. {Fitzpatrick}(1999){Fitzpatrick}}]{Fitzpatrick1999}
{Fitzpatrick}, E.~L. 1999, \bibinfo{title}{{Correcting for the Effects of Interstellar Extinction},} \pasp, 111, 63, \dodoi{10.1086/316293}

\bibitem[{F. {Fortuni} {et~al.}(2023){Fortuni}, {Merlin}, {Fontana}, {Giocoli}, {Romelli}, {Graziani}, {Santini}, {Castellano}, {Charlot}, \& {Chevallard}}]{Fortuni2023}
{Fortuni}, F., {Merlin}, E., {Fontana}, A., {et~al.} 2023, \bibinfo{title}{{FORECAST: A flexible software to forward model cosmological hydrodynamical simulations mimicking real observations},} \aap, 677, A102, \dodoi{10.1051/0004-6361/202346725}

\bibitem[{J.~P. {Gardner} {et~al.}(2023){Gardner}, {Mather}, {Abbott}, {Abell}, {Abernathy}, {Abney}, {Abraham}, {Abraham}, {Abul-Huda}, {Acton}, {Adams}, {Adams}, {Adler}, {Adriaensen}, {Aguilar}, {Ahmed}, {Ahmed}, {Ahmed}, {Albat}, {Albert}, {Alberts}, {Aldridge}, {Allen}, {Allen}, {Altenburg}, {Altunc}, {Alvarez}, {{\'A}lvarez-M{\'a}rquez}, {Alves de Oliveira}, {Ambrose}, {Anandakrishnan}, {Andersen}, {Anderson}, {Anderson}, {Anderson}, {Anderson}, {Aprea}, {Archer}, {Arenberg}, {Argyriou}, {Arribas}, {Artigau}, {Arvai}, {Atcheson}, {Atkinson}, {Averbukh}, {Aymergen}, {Bacinski}, {Baggett}, {Bagnasco}, {Baker}, {Balzano}, {Banks}, {Baran}, {Barker}, {Barrett}, {Barringer}, {Barto}, {Bast}, {Baudoz}, {Baum}, {Beatty}, {Beaulieu}, {Bechtold}, {Beck}, {Beddard}, {Beichman}, {Bellagama}, {Bely}, {Berger}, {Bergeron}, {Bernier}, {Bertch}, {Beskow}, {Betz}, {Biagetti}, {Birkmann}, {Bjorklund}, {Blackwood}, {Blazek}, {Blossfeld}, {Bluth}, {Boccaletti}, {Boegner}, {Bohlin}, {Boia}, {B{\"o}ker}, {Bonaventura},
  {Bond}, {Bosley}, {Boucarut}, {Bouchet}, {Bouwman}, {Bower}, {Bowers}, {Bowers}, {Boyce}, {Boyer}, {Boyer}, {Boyer}, {Boyer}, {Bradley}, {Brady}, {Brandl}, {Brannen}, {Breda}, {Bremmer}, {Brennan}, {Bresnahan}, {Bright}, {Broiles}, {Bromenschenkel}, {Brooks}, {Brooks}, {Brown}, {Brown}, {Brown}, {Bruce}, {Bryson}, {Bujanda}, {Bullock}, {Bunker}, {Bureo}, {Burt}, {Bush}, {Bushouse}, {Bussman}, {Cabaud}, {Cale}, {Calhoon}, {Calvani}, {Canipe}, {Caputo}, {Cara}, {Carey}, {Case}, {Cesari}, {Cetorelli}, {Chance}, {Chandler}, {Chaney}, {Chapman}, {Charlot}, {Chayer}, {Cheezum}, {Chen}, {Chen}, {Cherinka}, {Chichester}, {Chilton}, {Chittiraibalan}, {Clampin}, {Clark}, {Clark}, {Clark}, {Claybrooks}, {Cleveland}, {Cohen}, {Cohen}, {Col{\'o}n}, {Coleman}, {Colina}, {Comber}, {Comeau}, {Comer}, {Conde Reis}, {Connolly}, {Conroy}, {Contos}, {Contreras}, {Cook}, {Cooper}, {Cooper}, {Correia}, {Correnti}, {Cossou}, {Costanza}, {Coulais}, {Cox}, {Coyle}, {Cracraft}, {Crew}, {Curtis}, {Cusveller}, {Da Costa Maciel},
  {Dailey}, {Daugeron}, {Davidson}, {Davies}, {Davis}, {Davis}, {Day}, {de Chambure}, {de Jong}, {De Marchi}, {Dean}, {Decker}, {Delisa}, {Dell}, \& {Dellagatta}}]{Gardner2023}
{Gardner}, J.~P., {Mather}, J.~C., {Abbott}, R., {et~al.} 2023, \bibinfo{title}{{The James Webb Space Telescope Mission},} \pasp, 135, 068001, \dodoi{10.1088/1538-3873/acd1b5}

\bibitem[{L. {Girardi} {et~al.}(2000){Girardi}, {Bressan}, {Bertelli}, \& {Chiosi}}]{Girardi2000}
{Girardi}, L., {Bressan}, A., {Bertelli}, G., \& {Chiosi}, C. 2000, \bibinfo{title}{{Evolutionary tracks and isochrones for low- and intermediate-mass stars: From 0.15 to 7 M$_{sun}$, and from Z=0.0004 to 0.03},} \aaps, 141, 371, \dodoi{10.1051/aas:2000126}

\bibitem[{K.~D. {Gordon} {et~al.}(2003){Gordon}, {Clayton}, {Misselt}, {Landolt}, \& {Wolff}}]{Gordon2003}
{Gordon}, K.~D., {Clayton}, G.~C., {Misselt}, K.~A., {Landolt}, A.~U., \& {Wolff}, M.~J. 2003, \bibinfo{title}{{A Quantitative Comparison of the Small Magellanic Cloud, Large Magellanic Cloud, and Milky Way Ultraviolet to Near-Infrared Extinction Curves},} \apj, 594, 279, \dodoi{10.1086/376774}

\bibitem[{T. {Harvey} {et~al.}(2026){Harvey}, {Lovell}, {Newman}, {Conselice}, {Austin}, {Roper}, {Vijayan}, {Wilkins}, {Iglesias-Navarro}, {Rusakov}, {Li}, {Adams}, {Magdwick}, {Goolsby}, {Huertas-Company}, \& {Ho}}]{Harvey2026}
{Harvey}, T., {Lovell}, C.~C., {Newman}, S., {et~al.} 2026, \bibinfo{title}{{Flexible simulation-based inference for galaxy photometric fitting with synthesizer},} \mnras, 547, stag282, \dodoi{10.1093/mnras/stag282}

\bibitem[{N.~S. {Haryana} {et~al.}(2025){Haryana}, {Akiyama}, {Abdurro'uf}, {Wulandari}, {Alfonzo}, {Lee}, {Matsumoto}, {Sutanto}, {Effendi}, {Fitriana}, {Huda}, {Jaelani}, {Kusuma}, {Puspitarini}, \& {Triani}}]{Haryana2025}
{Haryana}, N.~S., {Akiyama}, M., {Abdurro'uf}, {et~al.} 2025, \bibinfo{title}{{Stellar Mass Assembly History of Massive Quiescent Galaxies since z {\ensuremath{\sim}} 4: Insights from Spatially Resolved Spectral Energy Distribution Fitting with JWST Data},} \apj, 994, 215, \dodoi{10.3847/1538-4357/ae03ad}

\bibitem[{A.~K. {Inoue} {et~al.}(2014){Inoue}, {Shimizu}, {Iwata}, \& {Tanaka}}]{Inoue2014}
{Inoue}, A.~K., {Shimizu}, I., {Iwata}, I., \& {Tanaka}, M. 2014, \bibinfo{title}{{An updated analytic model for attenuation by the intergalactic medium},} \mnras, 442, 1805, \dodoi{10.1093/mnras/stu936}

\bibitem[{{\v{Z}}. {Ivezi{\'c}} {et~al.}(2019){Ivezi{\'c}}, {Kahn}, {Tyson}, {Abel}, {Acosta}, {Allsman}, {Alonso}, {AlSayyad}, {Anderson}, {Andrew}, {Angel}, {Angeli}, {Ansari}, {Antilogus}, {Araujo}, {Armstrong}, {Arndt}, {Astier}, {Aubourg}, {Auza}, {Axelrod}, {Bard}, {Barr}, {Barrau}, {Bartlett}, {Bauer}, {Bauman}, {Baumont}, {Bechtol}, {Bechtol}, {Becker}, {Becla}, {Beldica}, {Bellavia}, {Bianco}, {Biswas}, {Blanc}, {Blazek}, {Blandford}, {Bloom}, {Bogart}, {Bond}, {Booth}, {Borgland}, {Borne}, {Bosch}, {Boutigny}, {Brackett}, {Bradshaw}, {Brandt}, {Brown}, {Bullock}, {Burchat}, {Burke}, {Cagnoli}, {Calabrese}, {Callahan}, {Callen}, {Carlin}, {Carlson}, {Chandrasekharan}, {Charles-Emerson}, {Chesley}, {Cheu}, {Chiang}, {Chiang}, {Chirino}, {Chow}, {Ciardi}, {Claver}, {Cohen-Tanugi}, {Cockrum}, {Coles}, {Connolly}, {Cook}, {Cooray}, {Covey}, {Cribbs}, {Cui}, {Cutri}, {Daly}, {Daniel}, {Daruich}, {Daubard}, {Daues}, {Dawson}, {Delgado}, {Dellapenna}, {de Peyster}, {de Val-Borro}, {Digel}, {Doherty},
  {Dubois}, {Dubois-Felsmann}, {Durech}, {Economou}, {Eifler}, {Eracleous}, {Emmons}, {Fausti Neto}, {Ferguson}, {Figueroa}, {Fisher-Levine}, {Focke}, {Foss}, {Frank}, {Freemon}, {Gangler}, {Gawiser}, {Geary}, {Gee}, {Geha}, {Gessner}, {Gibson}, {Gilmore}, {Glanzman}, {Glick}, {Goldina}, {Goldstein}, {Goodenow}, {Graham}, {Gressler}, {Gris}, {Guy}, {Guyonnet}, {Haller}, {Harris}, {Hascall}, {Haupt}, {Hernandez}, {Herrmann}, {Hileman}, {Hoblitt}, {Hodgson}, {Hogan}, {Howard}, {Huang}, {Huffer}, {Ingraham}, {Innes}, {Jacoby}, {Jain}, {Jammes}, {Jee}, {Jenness}, {Jernigan}, {Jevremovi{\'c}}, {Johns}, {Johnson}, {Johnson}, {Jones}, {Juramy-Gilles}, {Juri{\'c}}, {Kalirai}, {Kallivayalil}, {Kalmbach}, {Kantor}, {Karst}, {Kasliwal}, {Kelly}, {Kessler}, {Kinnison}, {Kirkby}, {Knox}, {Kotov}, {Krabbendam}, {Krughoff}, {Kub{\'a}nek}, {Kuczewski}, {Kulkarni}, {Ku}, {Kurita}, {Lage}, {Lambert}, {Lange}, {Langton}, {Le Guillou}, {Levine}, {Liang}, {Lim}, {Lintott}, {Long}, {Lopez}, {Lotz}, {Lupton}, {Lust}, {MacArthur},
  {Mahabal}, {Mandelbaum}, {Markiewicz}, {Marsh}, {Marshall}, {Marshall}, {May}, {McKercher}, {McQueen}, {Meyers}, {Migliore}, {Miller}, \& {Mills}}]{Ivezic2019}
{Ivezi{\'c}}, {\v{Z}}., {Kahn}, S.~M., {Tyson}, J.~A., {et~al.} 2019, \bibinfo{title}{{LSST: From Science Drivers to Reference Design and Anticipated Data Products},} \apj, 873, 111, \dodoi{10.3847/1538-4357/ab042c}

\bibitem[{P. {Janulewicz} \& W. {Cui}(2025){Janulewicz} \& {Cui}}]{Janulewicz2025}
{Janulewicz}, P., \& {Cui}, W. 2025, \bibinfo{title}{{PYMGAL: a python package for generating optical mock observations from hydrodynamical simulations},} RAS Techniques and Instruments, 4, rzaf027, \dodoi{10.1093/rasti/rzaf027}

\bibitem[{B. Johnson {et~al.}(2024)Johnson, Foreman-Mackey, Sick, Leja, Walmsley, Tollerud, Leung, Scott, \& Park}]{ben_johnson_2024_fsps}
Johnson, B., Foreman-Mackey, D., Sick, J., {et~al.} 2024, dfm/python-fsps: v0.4.7, v0.4.7 Zenodo, \dodoi{10.5281/zenodo.12447779}

\bibitem[{B.~D. {Johnson} {et~al.}(2026){Johnson}, {Robertson}, {Eisenstein}, {Tacchella}, {Pusk{\'a}s}, {Duan}, {Wu}, {Hainline}, {Rieke}, {Willott}, {Willmer}, {Trussler}, {Alberts}, {Arribas}, {Baker}, {Bunker}, {Cameron}, {Carniani}, {Carreira}, {Cargile}, {Curtis-Lake}, {Egami}, {Hausen}, {Helton}, {Ji}, {Maiolino}, {P{\'e}rez-Gonz{\'a}lez}, {Rinaldi}, {Sun}, {Sun}, {Villanueva}, {Williams}, \& {Zhu}}]{Johnson2026}
{Johnson}, B.~D., {Robertson}, B.~E., {Eisenstein}, D.~J., {et~al.} 2026, \bibinfo{title}{{JWST Advanced Deep Extragalactic Survey (JADES) Data Release 5: NIRCam Imaging in GOODS-S and GOODS-N},} arXiv e-prints, arXiv:2601.15954, \dodoi{10.48550/arXiv.2601.15954}

\bibitem[{P. {Jonsson}(2006){Jonsson}}]{Jonsson2006}
{Jonsson}, P. 2006, \bibinfo{title}{{SUNRISE: polychromatic dust radiative transfer in arbitrary geometries},} \mnras, 372, 2, \dodoi{10.1111/j.1365-2966.2006.10884.x}

\bibitem[{O.~B. {Kauffmann} {et~al.}(2020){Kauffmann}, {Le F{\`e}vre}, {Ilbert}, {Chevallard}, {Williams}, {Curtis-Lake}, {Colina}, {P{\'e}rez-Gonz{\'a}lez}, {Pye}, \& {Caputi}}]{Kauffmann2020}
{Kauffmann}, O.~B., {Le F{\`e}vre}, O., {Ilbert}, O., {et~al.} 2020, \bibinfo{title}{{Simulating JWST deep extragalactic imaging surveys and physical parameter recovery},} \aap, 640, A67, \dodoi{10.1051/0004-6361/202037450}

\bibitem[{R.~C. {Kennicutt}(1998){Kennicutt}}]{Kennicutt1998}
{Kennicutt}, Jr., R.~C. 1998, \bibinfo{title}{{Star Formation in Galaxies Along the Hubble Sequence},} \araa, 36, 189, \dodoi{10.1146/annurev.astro.36.1.189}

\bibitem[{M. {Kriek} \& C. {Conroy}(2013){Kriek} \& {Conroy}}]{Kriek2013}
{Kriek}, M., \& {Conroy}, C. 2013, \bibinfo{title}{{The Dust Attenuation Law in Distant Galaxies: Evidence for Variation with Spectral Type},} \apjl, 775, L16, \dodoi{10.1088/2041-8205/775/1/L16}

\bibitem[{P. {Kroupa}(2001){Kroupa}}]{Kroupa2001}
{Kroupa}, P. 2001, \bibinfo{title}{{On the variation of the initial mass function},} \mnras, 322, 231, \dodoi{10.1046/j.1365-8711.2001.04022.x}

\bibitem[{P. {Kroupa} \& C.~M. {Boily}(2002){Kroupa} \& {Boily}}]{Kroupa2002}
{Kroupa}, P., \& {Boily}, C.~M. 2002, \bibinfo{title}{{On the mass function of star clusters},} \mnras, 336, 1188, \dodoi{10.1046/j.1365-8711.2002.05848.x}

\bibitem[{T. {Lejeune} {et~al.}(1997){Lejeune}, {Cuisinier}, \& {Buser}}]{Lejeune1997}
{Lejeune}, T., {Cuisinier}, F., \& {Buser}, R. 1997, \bibinfo{title}{{Standard stellar library for evolutionary synthesis. I. Calibration of theoretical spectra},} \aaps, 125, 229, \dodoi{10.1051/aas:1997373}

\bibitem[{C.~C. {Lovell} {et~al.}(2025){Lovell}, {Roper}, {Vijayan}, {Wilkins}, {Newman}, \& {Seeyave}}]{Lovell2025}
{Lovell}, C.~C., {Roper}, W.~J., {Vijayan}, A.~P., {et~al.} 2025, \bibinfo{title}{{Synthesizer: a Software Package for Synthetic Astronomical Observables},} arXiv e-prints, arXiv:2508.03888, \dodoi{10.48550/arXiv.2508.03888}

\bibitem[{R. {Lupton} {et~al.}(2004){Lupton}, {Blanton}, {Fekete}, {Hogg}, {O'Mullane}, {Szalay}, \& {Wherry}}]{Lupton2004}
{Lupton}, R., {Blanton}, M.~R., {Fekete}, G., {et~al.} 2004, \bibinfo{title}{{Preparing Red-Green-Blue Images from CCD Data},} \pasp, 116, 133, \dodoi{10.1086/382245}

\bibitem[{P. {Madau}(1995){Madau}}]{Madau1995}
{Madau}, P. 1995, \bibinfo{title}{{Radiative Transfer in a Clumpy Universe: The Colors of High-Redshift Galaxies},} \apj, 441, 18, \dodoi{10.1086/175332}

\bibitem[{F. {Marinacci} {et~al.}(2018){Marinacci}, {Vogelsberger}, {Pakmor}, {Torrey}, {Springel}, {Hernquist}, {Nelson}, {Weinberger}, {Pillepich}, {Naiman}, \& {Genel}}]{Marinacci2018}
{Marinacci}, F., {Vogelsberger}, M., {Pakmor}, R., {et~al.} 2018, \bibinfo{title}{{First results from the IllustrisTNG simulations: radio haloes and magnetic fields},} \mnras, 480, 5113, \dodoi{10.1093/mnras/sty2206}

\bibitem[{A. {Muzzin} {et~al.}(2013){Muzzin}, {Marchesini}, {Stefanon}, {Franx}, {McCracken}, {Milvang-Jensen}, {Dunlop}, {Fynbo}, {Brammer}, {Labb{\'e}}, \& {van Dokkum}}]{Muzzin2013}
{Muzzin}, A., {Marchesini}, D., {Stefanon}, M., {et~al.} 2013, \bibinfo{title}{{The Evolution of the Stellar Mass Functions of Star-forming and Quiescent Galaxies to z = 4 from the COSMOS/UltraVISTA Survey},} \apj, 777, 18, \dodoi{10.1088/0004-637X/777/1/18}

\bibitem[{G. {Nagaraj} {et~al.}(2022){Nagaraj}, {Forbes}, {Leja}, {Foreman-Mackey}, \& {Hayward}}]{Nagaraj2022}
{Nagaraj}, G., {Forbes}, J.~C., {Leja}, J., {Foreman-Mackey}, D., \& {Hayward}, C.~C. 2022, \bibinfo{title}{{A Bayesian Population Model for the Observed Dust Attenuation in Galaxies},} \apj, 932, 54, \dodoi{10.3847/1538-4357/ac6c80}

\bibitem[{J.~P. {Naiman} {et~al.}(2018){Naiman}, {Pillepich}, {Springel}, {Ramirez-Ruiz}, {Torrey}, {Vogelsberger}, {Pakmor}, {Nelson}, {Marinacci}, {Hernquist}, {Weinberger}, \& {Genel}}]{Naiman2018}
{Naiman}, J.~P., {Pillepich}, A., {Springel}, V., {et~al.} 2018, \bibinfo{title}{{First results from the IllustrisTNG simulations: a tale of two elements - chemical evolution of magnesium and europium},} \mnras, 477, 1206, \dodoi{10.1093/mnras/sty618}

\bibitem[{K. {Nandy} {et~al.}(1981){Nandy}, {Morgan}, {Willis}, {Wilson}, \& {Gondhalekar}}]{Nandy1981}
{Nandy}, K., {Morgan}, D.~H., {Willis}, A.~J., {Wilson}, R., \& {Gondhalekar}, P.~M. 1981, \bibinfo{title}{{Interstellar extinction in the Large Magellanic Cloud.},} \mnras, 196, 955, \dodoi{10.1093/mnras/196.4.955}

\bibitem[{D. {Narayanan} {et~al.}(2021){Narayanan}, {Turk}, {Robitaille}, {Kelly}, {McClellan}, {Sharma}, {Garg}, {Abruzzo}, {Choi}, {Conroy}, {Johnson}, {Kimock}, {Li}, {Lovell}, {Lower}, {Privon}, {Roberts}, {Sethuram}, {Snyder}, {Thompson}, \& {Wise}}]{Narayanan2021}
{Narayanan}, D., {Turk}, M.~J., {Robitaille}, T., {et~al.} 2021, \bibinfo{title}{{POWDERDAY: Dust Radiative Transfer for Galaxy Simulations},} \apjs, 252, 12, \dodoi{10.3847/1538-4365/abc487}

\bibitem[{D. {Nelson} {et~al.}(2018){Nelson}, {Pillepich}, {Springel}, {Weinberger}, {Hernquist}, {Pakmor}, {Genel}, {Torrey}, {Vogelsberger}, {Kauffmann}, {Marinacci}, \& {Naiman}}]{Nelson2018}
{Nelson}, D., {Pillepich}, A., {Springel}, V., {et~al.} 2018, \bibinfo{title}{{First results from the IllustrisTNG simulations: the galaxy colour bimodality},} \mnras, 475, 624, \dodoi{10.1093/mnras/stx3040}

\bibitem[{D. {Nelson} {et~al.}(2019{\natexlab{a}}){Nelson}, {Springel}, {Pillepich}, {Rodriguez-Gomez}, {Torrey}, {Genel}, {Vogelsberger}, {Pakmor}, {Marinacci}, {Weinberger}, {Kelley}, {Lovell}, {Diemer}, \& {Hernquist}}]{Nelson2019}
{Nelson}, D., {Springel}, V., {Pillepich}, A., {et~al.} 2019{\natexlab{a}}, \bibinfo{title}{{The IllustrisTNG simulations: public data release},} Computational Astrophysics and Cosmology, 6, 2, \dodoi{10.1186/s40668-019-0028-x}

\bibitem[{D. {Nelson} {et~al.}(2019{\natexlab{b}}){Nelson}, {Pillepich}, {Springel}, {Pakmor}, {Weinberger}, {Genel}, {Torrey}, {Vogelsberger}, {Marinacci}, \& {Hernquist}}]{Nelson2019_b}
{Nelson}, D., {Pillepich}, A., {Springel}, V., {et~al.} 2019{\natexlab{b}}, \bibinfo{title}{{First results from the TNG50 simulation: galactic outflows driven by supernovae and black hole feedback},} \mnras, 490, 3234, \dodoi{10.1093/mnras/stz2306}

\bibitem[{S. {Noll} {et~al.}(2009){Noll}, {Pierini}, {Cimatti}, {Daddi}, {Kurk}, {Bolzonella}, {Cassata}, {Halliday}, {Mignoli}, {Pozzetti}, {Renzini}, {Berta}, {Dickinson}, {Franceschini}, {Rodighiero}, {Rosati}, \& {Zamorani}}]{Noll2009}
{Noll}, S., {Pierini}, D., {Cimatti}, A., {et~al.} 2009, \bibinfo{title}{{GMASS ultradeep spectroscopy of galaxies at z \raisebox{-0.5ex}\textasciitilde 2. IV. The variety of dust populations},} \aap, 499, 69, \dodoi{10.1051/0004-6361/200811526}

\bibitem[{P.~G. {P{\'e}rez-Gonz{\'a}lez} {et~al.}(2008){P{\'e}rez-Gonz{\'a}lez}, {Rieke}, {Villar}, {Barro}, {Blaylock}, {Egami}, {Gallego}, {Gil de Paz}, {Pascual}, {Zamorano}, \& {Donley}}]{Perez-Gonzalez2008}
{P{\'e}rez-Gonz{\'a}lez}, P.~G., {Rieke}, G.~H., {Villar}, V., {et~al.} 2008, \bibinfo{title}{{The Stellar Mass Assembly of Galaxies from z = 0 to z = 4: Analysis of a Sample Selected in the Rest-Frame Near-Infrared with Spitzer},} \apj, 675, 234, \dodoi{10.1086/523690}

\bibitem[{M.~D. {Perrin} {et~al.}(2014){Perrin}, {Sivaramakrishnan}, {Lajoie}, {Elliott}, {Pueyo}, {Ravindranath}, \& {Albert}}]{Perrin2014}
{Perrin}, M.~D., {Sivaramakrishnan}, A., {Lajoie}, C.-P., {et~al.} 2014, \bibinfo{title}{{Updated point spread function simulations for JWST with WebbPSF},} in Society of Photo-Optical Instrumentation Engineers (SPIE) Conference Series, Vol. 9143, Space Telescopes and Instrumentation 2014: Optical, Infrared, and Millimeter Wave, ed. J.~M. {Oschmann}, Jr., M.~{Clampin}, G.~G. {Fazio}, \& H.~A. {MacEwen}, 91433X, \dodoi{10.1117/12.2056689}

\bibitem[{A. {Pietrinferni} {et~al.}(2004){Pietrinferni}, {Cassisi}, {Salaris}, \& {Castelli}}]{Pietrinferni2004}
{Pietrinferni}, A., {Cassisi}, S., {Salaris}, M., \& {Castelli}, F. 2004, \bibinfo{title}{{A Large Stellar Evolution Database for Population Synthesis Studies. I. Scaled Solar Models and Isochrones},} \apj, 612, 168, \dodoi{10.1086/422498}

\bibitem[{A. {Pillepich} {et~al.}(2018){Pillepich}, {Nelson}, {Hernquist}, {Springel}, {Pakmor}, {Torrey}, {Weinberger}, {Genel}, {Naiman}, {Marinacci}, \& {Vogelsberger}}]{Pillepich2018}
{Pillepich}, A., {Nelson}, D., {Hernquist}, L., {et~al.} 2018, \bibinfo{title}{{First results from the IllustrisTNG simulations: the stellar mass content of groups and clusters of galaxies},} \mnras, 475, 648, \dodoi{10.1093/mnras/stx3112}

\bibitem[{A. {Pillepich} {et~al.}(2019){Pillepich}, {Nelson}, {Springel}, {Pakmor}, {Torrey}, {Weinberger}, {Vogelsberger}, {Marinacci}, {Genel}, {van der Wel}, \& {Hernquist}}]{Pillepich2019}
{Pillepich}, A., {Nelson}, D., {Springel}, V., {et~al.} 2019, \bibinfo{title}{{First results from the TNG50 simulation: the evolution of stellar and gaseous discs across cosmic time},} \mnras, 490, 3196, \dodoi{10.1093/mnras/stz2338}

\bibitem[{K.~M. {Pontoppidan} {et~al.}(2016){Pontoppidan}, {Pickering}, {Laidler}, {Gilbert}, {Sontag}, {Slocum}, {Sienkiewicz}, {Hanley}, {Earl}, {Pueyo}, {Ravindranath}, {Karakla}, {Robberto}, {Noriega-Crespo}, \& {Barker}}]{Pontoppidan2016}
{Pontoppidan}, K.~M., {Pickering}, T.~E., {Laidler}, V.~G., {et~al.} 2016, \bibinfo{title}{{Pandeia: a multi-mission exposure time calculator for JWST and WFIRST},} in Society of Photo-Optical Instrumentation Engineers (SPIE) Conference Series, Vol. 9910, Observatory Operations: Strategies, Processes, and Systems VI, ed. A.~B. {Peck}, R.~L. {Seaman}, \& C.~R. {Benn}, 991016, \dodoi{10.1117/12.2231768}

\bibitem[{M.~L. {Prevot} {et~al.}(1984){Prevot}, {Lequeux}, {Maurice}, {Prevot}, \& {Rocca-Volmerange}}]{Prevot1984}
{Prevot}, M.~L., {Lequeux}, J., {Maurice}, E., {Prevot}, L., \& {Rocca-Volmerange}, B. 1984, \bibinfo{title}{{The typical interstellar extinction in the Small Magellanic Cloud.},} \aap, 132, 389

\bibitem[{N.~A. {Reddy} {et~al.}(2023{\natexlab{a}}){Reddy}, {Topping}, {Sanders}, {Shapley}, \& {Brammer}}]{Reddy2023a}
{Reddy}, N.~A., {Topping}, M.~W., {Sanders}, R.~L., {Shapley}, A.~E., \& {Brammer}, G. 2023{\natexlab{a}}, \bibinfo{title}{{A JWST/NIRSpec Exploration of the Connection between Ionization Parameter, Electron Density, and Star-formation-rate Surface Density in z = 2.7-6.3 Galaxies},} \apj, 952, 167, \dodoi{10.3847/1538-4357/acd754}

\bibitem[{N.~A. {Reddy} {et~al.}(2023{\natexlab{b}}){Reddy}, {Sanders}, {Shapley}, {Topping}, {Kriek}, {Coil}, {Mobasher}, {Siana}, \& {Rezaee}}]{Reddy2023b}
{Reddy}, N.~A., {Sanders}, R.~L., {Shapley}, A.~E., {et~al.} 2023{\natexlab{b}}, \bibinfo{title}{{The Impact of Star-formation-rate Surface Density on the Electron Density and Ionization Parameter of High-redshift Galaxies},} \apj, 951, 56, \dodoi{10.3847/1538-4357/acd0b1}

\bibitem[{J. {Rigby} {et~al.}(2023){Rigby}, {Perrin}, {McElwain}, {Kimble}, {Friedman}, {Lallo}, {Doyon}, {Feinberg}, {Ferruit}, {Glasse}, {Rieke}, {Rieke}, {Wright}, {Willott}, {Colon}, {Milam}, {Neff}, {Stark}, {Valenti}, {Abell}, {Abney}, {Abul-Huda}, {Acton}, {Adams}, {Adler}, {Aguilar}, {Ahmed}, {Albert}, {Alberts}, {Aldridge}, {Allen}, {Altenburg}, {{\'A}lvarez-M{\'a}rquez}, {Alves de Oliveira}, {Andersen}, {Anderson}, {Anderson}, {Argyriou}, {Armstrong}, {Arribas}, {Artigau}, {Arvai}, {Atkinson}, {Bacon}, {Bair}, {Banks}, {Barrientes}, {Barringer}, {Bartosik}, {Bast}, {Baudoz}, {Beatty}, {Bechtold}, {Beck}, {Bergeron}, {Bergkoetter}, {Bhatawdekar}, {Birkmann}, {Blazek}, {Blome}, {Boccaletti}, {B{\"o}ker}, {Boia}, {Bonaventura}, {Bond}, {Bosley}, {Boucarut}, {Bourque}, {Bouwman}, {Bower}, {Bowers}, {Boyer}, {Bradley}, {Brady}, {Braun}, {Breda}, {Bresnahan}, {Bright}, {Britt}, {Bromenschenkel}, {Brooks}, {Brooks}, {Brown}, {Brown}, {Brown}, {Bunker}, {Burger}, {Bushouse}, {Cale}, {Cameron}, {Cameron},
  {Canipe}, {Caplinger}, {Caputo}, {Cara}, {Carey}, {Carniani}, {Carrasquilla}, {Carruthers}, {Case}, {Catherine}, {Chance}, {Chapman}, {Charlot}, {Charlow}, {Chayer}, {Chen}, {Cherinka}, {Chichester}, {Chilton}, {Chonis}, {Clampin}, {Clark}, {Clark}, {Coe}, {Coleman}, {Comber}, {Comeau}, {Connolly}, {Cooper}, {Cooper}, {Coppock}, {Correnti}, {Cossou}, {Coulais}, {Coyle}, {Cracraft}, {Curti}, {Cuturic}, {Davis}, {Davis}, {Dean}, {DeLisa}, {deMeester}, {Dencheva}, {Dencheva}, {DePasquale}, {Deschenes}, {Hunor Detre}, {Diaz}, {Dicken}, {DiFelice}, {Dillman}, {Dixon}, {Doggett}, {Donaldson}, {Douglas}, {DuPrie}, {Dupuis}, {Durning}, {Easmin}, {Eck}, {Edeani}, {Egami}, {Ehrenwinkler}, {Eisenhamer}, {Eisenhower}, {Elie}, {Elliott}, {Elliott}, {Ellis}, {Engesser}, {Espinoza}, {Etienne}, {Etxaluze}, {Falini}, {Feeney}, {Ferry}, {Filippazzo}, {Fincham}, {Fix}, {Flagey}, {Florian}, {Flynn}, {Fontanella}, {Ford}, {Forshay}, {Fox}, {Franz}, {Fu}, {Fullerton}, {Galkin}, {Galyer}, {Garc{\'\i}a Mar{\'\i}n}, {Gardner},
  {Gardner}, {Garland}, {Garrett}, {Gasman}, {Gaspar}, {Gaudreau}, {Gauthier}, {Geers}, {Geithner}, {Gennaro}, {Giardino}, {Girard}, {Giuliano}, {Glassmire}, \& {Glauser}}]{Rigby2023}
{Rigby}, J., {Perrin}, M., {McElwain}, M., {et~al.} 2023, \bibinfo{title}{{The Science Performance of JWST as Characterized in Commissioning},} \pasp, 135, 048001, \dodoi{10.1088/1538-3873/acb293}

\bibitem[{T.~P. {Robitaille}(2011){Robitaille}}]{Robitaille2011}
{Robitaille}, T.~P. 2011, \bibinfo{title}{{HYPERION: an open-source parallelized three-dimensional dust continuum radiative transfer code},} \aap, 536, A79, \dodoi{10.1051/0004-6361/201117150}

\bibitem[{V. {Rodriguez-Gomez} {et~al.}(2015){Rodriguez-Gomez}, {Genel}, {Vogelsberger}, {Sijacki}, {Pillepich}, {Sales}, {Torrey}, {Snyder}, {Nelson}, {Springel}, {Ma}, \& {Hernquist}}]{Rodriguez-Gomez2015}
{Rodriguez-Gomez}, V., {Genel}, S., {Vogelsberger}, M., {et~al.} 2015, \bibinfo{title}{{The merger rate of galaxies in the Illustris simulation: a comparison with observations and semi-empirical models},} \mnras, 449, 49, \dodoi{10.1093/mnras/stv264}

\bibitem[{V. {Rodriguez-Gomez} {et~al.}(2017){Rodriguez-Gomez}, {Sales}, {Genel}, {Pillepich}, {Zjupa}, {Nelson}, {Griffen}, {Torrey}, {Snyder}, {Vogelsberger}, {Springel}, {Ma}, \& {Hernquist}}]{Rodriguez-Gomez2017}
{Rodriguez-Gomez}, V., {Sales}, L.~V., {Genel}, S., {et~al.} 2017, \bibinfo{title}{{The role of mergers and halo spin in shaping galaxy morphology},} \mnras, 467, 3083, \dodoi{10.1093/mnras/stx305}

\bibitem[{W.~J. Roper {et~al.}(2026)Roper, Lovell, Vijayan, Wilkins, Akins, Berger, Sant~Fournier, Harvey, Iyer, Leonardi, Newman, Pautasso, Perry, Seeyave, Sommovigo, Punyasheel, d'Hautefort, \& Rawlings}]{Roper2025}
Roper, W.~J., Lovell, C.~C., Vijayan, A., {et~al.} 2026, \bibinfo{title}{Synthesizer: Synthetic Observables for Modern Astronomy,} Journal of Open Source Software, 11, 9436, \dodoi{10.21105/joss.09436}

\bibitem[{B.~T.~P. {Rowe} {et~al.}(2015){Rowe}, {Jarvis}, {Mandelbaum}, {Bernstein}, {Bosch}, {Simet}, {Meyers}, {Kacprzak}, {Nakajima}, {Zuntz}, {Miyatake}, {Dietrich}, {Armstrong}, {Melchior}, \& {Gill}}]{Rowe2015}
{Rowe}, B.~T.~P., {Jarvis}, M., {Mandelbaum}, R., {et~al.} 2015, \bibinfo{title}{{GALSIM: The modular galaxy image simulation toolkit},} Astronomy and Computing, 10, 121, \dodoi{10.1016/j.ascom.2015.02.002}

\bibitem[{S. {Salim} {et~al.}(2018){Salim}, {Boquien}, \& {Lee}}]{Salim2018}
{Salim}, S., {Boquien}, M., \& {Lee}, J.~C. 2018, \bibinfo{title}{{Dust Attenuation Curves in the Local Universe: Demographics and New Laws for Star-forming Galaxies and High-redshift Analogs},} \apj, 859, 11, \dodoi{10.3847/1538-4357/aabf3c}

\bibitem[{S. {Salim} \& D. {Narayanan}(2020){Salim} \& {Narayanan}}]{Salim2020}
{Salim}, S., \& {Narayanan}, D. 2020, \bibinfo{title}{{The Dust Attenuation Law in Galaxies},} \araa, 58, 529, \dodoi{10.1146/annurev-astro-032620-021933}

\bibitem[{B. {Salmon} {et~al.}(2016){Salmon}, {Papovich}, {Long}, {Willner}, {Finkelstein}, {Ferguson}, {Dickinson}, {Duncan}, {Faber}, {Hathi}, {Koekemoer}, {Kurczynski}, {Newman}, {Pacifici}, {P{\'e}rez-Gonz{\'a}lez}, \& {Pforr}}]{Salmon2016}
{Salmon}, B., {Papovich}, C., {Long}, J., {et~al.} 2016, \bibinfo{title}{{Breaking the Curve with CANDELS: A Bayesian Approach to Reveal the Non-Universality of the Dust-Attenuation Law at High Redshift},} \apj, 827, 20, \dodoi{10.3847/0004-637X/827/1/20}

\bibitem[{E.~E. {Salpeter}(1955){Salpeter}}]{Salpeter1955}
{Salpeter}, E.~E. 1955, \bibinfo{title}{{The Luminosity Function and Stellar Evolution.},} \apj, 121, 161, \dodoi{10.1086/145971}

\bibitem[{P. {S{\'a}nchez-Bl{\'a}zquez} {et~al.}(2006){S{\'a}nchez-Bl{\'a}zquez}, {Peletier}, {Jim{\'e}nez-Vicente}, {Cardiel}, {Cenarro}, {Falc{\'o}n-Barroso}, {Gorgas}, {Selam}, \& {Vazdekis}}]{Sanchez-Blazquez2006}
{S{\'a}nchez-Bl{\'a}zquez}, P., {Peletier}, R.~F., {Jim{\'e}nez-Vicente}, J., {et~al.} 2006, \bibinfo{title}{{Medium-resolution Isaac Newton Telescope library of empirical spectra},} \mnras, 371, 703, \dodoi{10.1111/j.1365-2966.2006.10699.x}

\bibitem[{G. {Schaller} {et~al.}(1992){Schaller}, {Schaerer}, {Meynet}, \& {Maeder}}]{Schaller1992}
{Schaller}, G., {Schaerer}, D., {Meynet}, G., \& {Maeder}, A. 1992, \bibinfo{title}{{New Grids of Stellar Models from 0.8-SOLAR-MASS to 120-SOLAR-MASSES at Z=0.020 and Z=0.001},} \aaps, 96, 269

\bibitem[{J. {Schaye} {et~al.}(2015){Schaye}, {Crain}, {Bower}, {Furlong}, {Schaller}, {Theuns}, {Dalla Vecchia}, {Frenk}, {McCarthy}, {Helly}, {Jenkins}, {Rosas-Guevara}, {White}, {Baes}, {Booth}, {Camps}, {Navarro}, {Qu}, {Rahmati}, {Sawala}, {Thomas}, \& {Trayford}}]{Schaye2015}
{Schaye}, J., {Crain}, R.~A., {Bower}, R.~G., {et~al.} 2015, \bibinfo{title}{{The EAGLE project: simulating the evolution and assembly of galaxies and their environments},} \mnras, 446, 521, \dodoi{10.1093/mnras/stu2058}

\bibitem[{J. {Schaye} {et~al.}(2025){Schaye}, {Chaikin}, {Schaller}, {Ploeckinger}, {Hu{\v{s}}ko}, {McGibbon}, {Trayford}, {Ben{\'\i}tez-Llambay}, {Correa}, {Frenk}, {Richings}, {Forouhar Moreno}, {Bah{\'e}}, {Borrow}, {Durrant}, {Gebek}, {Helly}, {Jenkins}, {Lacey}, {Ludlow}, \& {Nobels}}]{Schaye2025}
{Schaye}, J., {Chaikin}, E., {Schaller}, M., {et~al.} 2025, \bibinfo{title}{{The COLIBRE project: cosmological hydrodynamical simulations of galaxy formation and evolution},} arXiv e-prints, arXiv:2508.21126, \dodoi{10.48550/arXiv.2508.21126}

\bibitem[{I. {Smail} {et~al.}(2021){Smail}, {Dudzevi{\v{c}}i{\={u}}t{\.{e}}}, {Stach}, {Almaini}, {Birkin}, {Chapman}, {Chen}, {Geach}, {Gullberg}, {Hodge}, {Ikarashi}, {Ivison}, {Scott}, {Simpson}, {Swinbank}, {Thomson}, {Walter}, {Wardlow}, \& {van der Werf}}]{Smail2021}
{Smail}, I., {Dudzevi{\v{c}}i{\={u}}t{\.{e}}}, U., {Stach}, S.~M., {et~al.} 2021, \bibinfo{title}{{An ALMA survey of the S2CLS UDS field: optically invisible submillimetre galaxies},} \mnras, 502, 3426, \dodoi{10.1093/mnras/stab283}

\bibitem[{G.~F. {Snyder} {et~al.}(2023){Snyder}, {Pe{\~n}a}, {Yung}, {Rose}, {Kartaltepe}, \& {Ferguson}}]{Snyder2023}
{Snyder}, G.~F., {Pe{\~n}a}, T., {Yung}, L.~Y.~A., {et~al.} 2023, \bibinfo{title}{{Mock galaxy surveys for HST and JWST from the IllustrisTNG simulations},} \mnras, 518, 6318, \dodoi{10.1093/mnras/stac3397}

\bibitem[{R.~S. {Somerville} {et~al.}(2012){Somerville}, {Gilmore}, {Primack}, \& {Dom{\'\i}nguez}}]{Somerville2012}
{Somerville}, R.~S., {Gilmore}, R.~C., {Primack}, J.~R., \& {Dom{\'\i}nguez}, A. 2012, \bibinfo{title}{{Galaxy properties from the ultraviolet to the far-infrared: {\ensuremath{\Lambda}} cold dark matter models confront observations},} \mnras, 423, 1992, \dodoi{10.1111/j.1365-2966.2012.20490.x}

\bibitem[{L. {Sommovigo} {et~al.}(2025){Sommovigo}, {Cochrane}, {Somerville}, {Hayward}, {Lovell}, {Starkenburg}, {Popping}, {Iyer}, {Gabrielpillai}, {Ho}, {Steinwandel}, \& {Perez}}]{Sommovigo2025}
{Sommovigo}, L., {Cochrane}, R.~K., {Somerville}, R.~S., {et~al.} 2025, \bibinfo{title}{{Learning the Universe: Physically Motivated Priors for Dust Attenuation Curves},} \apj, 990, 114, \dodoi{10.3847/1538-4357/addec1}

\bibitem[{D. {Spergel} {et~al.}(2015){Spergel}, {Gehrels}, {Baltay}, {Bennett}, {Breckinridge}, {Donahue}, {Dressler}, {Gaudi}, {Greene}, {Guyon}, {Hirata}, {Kalirai}, {Kasdin}, {Macintosh}, {Moos}, {Perlmutter}, {Postman}, {Rauscher}, {Rhodes}, {Wang}, {Weinberg}, {Benford}, {Hudson}, {Jeong}, {Mellier}, {Traub}, {Yamada}, {Capak}, {Colbert}, {Masters}, {Penny}, {Savransky}, {Stern}, {Zimmerman}, {Barry}, {Bartusek}, {Carpenter}, {Cheng}, {Content}, {Dekens}, {Demers}, {Grady}, {Jackson}, {Kuan}, {Kruk}, {Melton}, {Nemati}, {Parvin}, {Poberezhskiy}, {Peddie}, {Ruffa}, {Wallace}, {Whipple}, {Wollack}, \& {Zhao}}]{2015Spergel_roman}
{Spergel}, D., {Gehrels}, N., {Baltay}, C., {et~al.} 2015, \bibinfo{title}{{Wide-Field InfrarRed Survey Telescope-Astrophysics Focused Telescope Assets WFIRST-AFTA 2015 Report},} arXiv e-prints, arXiv:1503.03757, \dodoi{10.48550/arXiv.1503.03757}

\bibitem[{V. {Springel} \& L. {Hernquist}(2003){Springel} \& {Hernquist}}]{Springel2003}
{Springel}, V., \& {Hernquist}, L. 2003, \bibinfo{title}{{Cosmological smoothed particle hydrodynamics simulations: a hybrid multiphase model for star formation},} \mnras, 339, 289, \dodoi{10.1046/j.1365-8711.2003.06206.x}

\bibitem[{V. {Springel} {et~al.}(2018){Springel}, {Pakmor}, {Pillepich}, {Weinberger}, {Nelson}, {Hernquist}, {Vogelsberger}, {Genel}, {Torrey}, {Marinacci}, \& {Naiman}}]{Springel2018}
{Springel}, V., {Pakmor}, R., {Pillepich}, A., {et~al.} 2018, \bibinfo{title}{{First results from the IllustrisTNG simulations: matter and galaxy clustering},} \mnras, 475, 676, \dodoi{10.1093/mnras/stx3304}

\bibitem[{D. {Thomas} {et~al.}(2005){Thomas}, {Maraston}, {Bender}, \& {Mendes de Oliveira}}]{Thomas2005}
{Thomas}, D., {Maraston}, C., {Bender}, R., \& {Mendes de Oliveira}, C. 2005, \bibinfo{title}{{The Epochs of Early-Type Galaxy Formation as a Function of Environment},} \apj, 621, 673, \dodoi{10.1086/426932}

\bibitem[{P.~G. {van Dokkum}(2008){van Dokkum}}]{vanDokkum2008}
{van Dokkum}, P.~G. 2008, \bibinfo{title}{{Evidence of Cosmic Evolution of the Stellar Initial Mass Function},} \apj, 674, 29, \dodoi{10.1086/525014}

\bibitem[{M. {Vogelsberger} {et~al.}(2020){Vogelsberger}, {Nelson}, {Pillepich}, {Shen}, {Marinacci}, {Springel}, {Pakmor}, {Tacchella}, {Weinberger}, {Torrey}, \& {Hernquist}}]{Vogelsberger2020}
{Vogelsberger}, M., {Nelson}, D., {Pillepich}, A., {et~al.} 2020, \bibinfo{title}{{High-redshift JWST predictions from IllustrisTNG: dust modelling and galaxy luminosity functions},} \mnras, 492, 5167, \dodoi{10.1093/mnras/staa137}

\bibitem[{C.~C. {Williams} {et~al.}(2018){Williams}, {Curtis-Lake}, {Hainline}, {Chevallard}, {Robertson}, {Charlot}, {Endsley}, {Stark}, {Willmer}, {Alberts}, {Amorin}, {Arribas}, {Baum}, {Bunker}, {Carniani}, {Crandall}, {Egami}, {Eisenstein}, {Ferruit}, {Husemann}, {Maseda}, {Maiolino}, {Rawle}, {Rieke}, {Smit}, {Tacchella}, \& {Willott}}]{Williams2018}
{Williams}, C.~C., {Curtis-Lake}, E., {Hainline}, K.~N., {et~al.} 2018, \bibinfo{title}{{The JWST Extragalactic Mock Catalog: Modeling Galaxy Populations from the UV through the Near-IR over 13 Billion Years of Cosmic History},} \apjs, 236, 33, \dodoi{10.3847/1538-4365/aabcbb}

\bibitem[{R.~J. {Williams} {et~al.}(2009){Williams}, {Quadri}, {Franx}, {van Dokkum}, \& {Labb{\'e}}}]{Williams2009}
{Williams}, R.~J., {Quadri}, R.~F., {Franx}, M., {van Dokkum}, P., \& {Labb{\'e}}, I. 2009, \bibinfo{title}{{Detection of Quiescent Galaxies in a Bicolor Sequence from Z = 0-2},} \apj, 691, 1879, \dodoi{10.1088/0004-637X/691/2/1879}

\bibitem[{L.~Y.~A. {Yung} {et~al.}(2019){Yung}, {Somerville}, {Finkelstein}, {Popping}, \& {Dav{\'e}}}]{Yung2019}
{Yung}, L.~Y.~A., {Somerville}, R.~S., {Finkelstein}, S.~L., {Popping}, G., \& {Dav{\'e}}, R. 2019, \bibinfo{title}{{Semi-analytic forecasts for JWST - I. UV luminosity functions at z = 4-10},} \mnras, 483, 2983, \dodoi{10.1093/mnras/sty3241}

\end{thebibliography}
\bibliographystyle{aasjournalv7}



\end{document}